\documentclass[graybox,envcountchap]{svmono}
\pdfoutput=1
\usepackage{float}
\usepackage{hyperref}
\usepackage{bm}
\usepackage{mathtools}
\usepackage{tcolorbox}
\usepackage[utf8]{inputenc}
\usepackage{slashed,verbatim}
\usepackage{epigraph,lipsum}
\usepackage{fancyhdr}
\usepackage{commath}
\usepackage{epigraph}
\usepackage[final]{pdfpages}
\usepackage{cancel}
\usepackage{graphicx}
\makeatletter
\def\@fpheader{\relax}
\makeatother
\usepackage[normalem]{ulem}

\usepackage{xcolor}

\newcommand*{\colorboxed}{}
\def\colorboxed#1#{%
  \colorboxedAux{#1}%
}

\usepackage{epigraph}
\usepackage{booktabs}

\newcommand{\MB}[1]{{\textcolor{red}{MB: #1}}}

\newcommand{\be}{\begin{equation}}
\newcommand{\ee}{\end{equation}}
\newcommand{\bea}{\begin{eqnarray}}
\newcommand{\eea}{\end{eqnarray}}
\usepackage{gensymb}

\usepackage[framemethod=tikz]{mdframed}

\usepackage{amsmath,epsfig}
\usepackage{amssymb,amsfonts}
\usepackage{latexsym}
\usepackage{graphicx}
\usepackage{commath}
 \usepackage{enumitem}

\usepackage{lipsum}
\mdfdefinestyle{MyFrame}{%
    linecolor=blue,
    outerlinewidth=0.1pt,
    roundcorner=10pt,
    innertopmargin=\baselineskip,
    innerbottommargin=\baselineskip,
    innerrightmargin=20pt,
    innerleftmargin=20pt,
    backgroundcolor=gray!20!white}

\mdfdefinestyle{MyFrame2}{%
    linecolor=orange,
    outerlinewidth=0.1pt,
    roundcorner=10pt,
    innertopmargin=\baselineskip,
    innerbottommargin=\baselineskip,
    innerrightmargin=20pt,
    innerleftmargin=20pt,
    backgroundcolor=red!20!white}
    
    \mdfdefinestyle{MyFrame3}{%
    linecolor=green,
    outerlinewidth=0.1pt,
    roundcorner=10pt,
    innertopmargin=\baselineskip,
    innerbottommargin=\baselineskip,
    innerrightmargin=20pt,
    innerleftmargin=20pt,
    backgroundcolor=green!20!white}
    
     \mdfdefinestyle{MyFrame4}{%
    linecolor=green,
    outerlinewidth=0.1pt,
    roundcorner=10pt,
    innertopmargin=\baselineskip,
    innerbottommargin=\baselineskip,
    innerrightmargin=20pt,
    innerleftmargin=20pt,
    backgroundcolor=yellow!60!white}
\usepackage{subeqnarray}
\usepackage{ulem}

\newcommand*{\colorboxedAux}[3]{%
  \begingroup
    \colorlet{cb@saved}{.}%
    \color#1{#2}%
    \boxed{%
      \color{cb@saved}%
      #3%
    }%
  \endgroup
}
\usepackage{lipsum,calc}
\usepackage{enumitem}

\def\be{\begin{equation}}
\def\ee{\end{equation}}
\def\bea{\begin{eqnarray}}
\def\eea{\end{eqnarray}}

\newcommand\fverb{\setbox\pippobox=\hbox\bgroup\verb}
\newcommand\fverbdo{\egroup\medskip\noindent%
                        \fbox{\unhbox\pippobox}\ }
\newcommand\fverbit{\egroup\item[\fbox{\unhbox\pippobox}]}

\newcommand{\bear}{\begin{eqnarray}}

\newcommand{\eear}{\end{eqnarray}}

\newcommand{\bsea}{\begin{subeqnarray}}
\newcommand{\esea}{\end{subeqnarray}}
\newbox\pippobox

\def\6{\partial}

\newcommand{\comments}[1]{}
\input Starburst.fd

\allowdisplaybreaks[3]

\usepackage{type1cm}         

\usepackage{makeidx}         
\usepackage{graphicx}        
\usepackage{multicol}        
\usepackage[bottom]{footmisc}

\usepackage{newtxtext}       %
\usepackage{newtxmath}       

\makeindex             

\begin{document}

\author{Matteo Baggioli}
\title{Applied holography\\
A practical mini-course}
\subtitle{-- SpringerBriefs in Physics --}
\maketitle

\frontmatter

%
%
\preface

\epigraph{Tra il dire e il fare c'\'e di mezzo il mare.}{Italian Proverb}
When I was asked by my friend and colleague Ayan Mukhopadhyay to give a series of lectures on Applied Holography, I wanted to create something that could be useful for students or in general researchers, who are willing to learn the fundamental techniques of the field and then apply them towards new directions. Being already approximately $20$ years since this tool has been first employed, there are already several, and I must say excellent, reviews and books available \cite{McGreevy:2016myw,McGreevy:2009xe,Nastase:2007kj,Ramallo:2013bua,Ammon:2015:GDF:2834415,Aharony:1999ti,CasalderreySolana:2011us,Zaffaroni_2000,Polchinski:2010hw,Natsuume:2014sfa,zaanen2015holographic,Hartnoll:2016apf,Hartnoll:2009sz}. They cover in great details the history, the derivation and the fundamental features of the AdS-CFT correspondence with a concrete view towards its applications to Condensed Matter, Heavy Ion physics, Hydrodynamics and strongly coupled field theory in general. Re-organizing and re-writing their content in my own words would have been totally useless. At the same time, there are several reviews regarding advanced numerical methods \cite{Krikun:2018ufr,Berti:2009kk,Dias:2015nua,Jansen:2017oag,Chesler:2013lia,Andrade:2017jmt,Zhang:2016coy}, with which I cannot (and do not want to) compete. I refer to them for the reader interested in digging into much more advanced problems which need those complex techniques.\\

I remember when, not long ago, I had to go through the various reviews by myself and what I felt at the time. The first thing that came to my mind is an Italian proverb saying \textit{''Tra il dire e il fare c'\'e di mezzo il mare''}. I am not sure how this should sound in proper English but Google suggests something like \textit{''It's easier said than done''}. In simpler words, reading through the lectures and books of the experts, everything seemed so obvious and obtainable in one line of computations, but then tons of hidden subtleties, caveats and derivations, taken for granted, were appearing everywhere. In this mini-course, I plan to go over them again with you, as if I were your school-mate. I selected a small and reasonable collection of benchmark computations and techniques which, in my personal opinion and experience, represent the necessary package to be able to float in this research landscape and to produce novel work within it. I tried to be the more pedagogical possible and to share with the reader all the tricks, warnings and shortcuts I have accumulated in these past years.\\

Do not take me wrong! I definitely recognize the pedagogical value of getting stuck with a problem and scratching your head with it. These lectures have not to be taken as a shortcut to avoid those fundamental moments of the learning curve. On the contrary, they have to be used at a second step to compare and integrate the possible resolution methods and results. They are not a substitute  car for your trip but rather a nice accessory which can make your trip more comfortable and entertaining like a good song driving on the highways. In this direction, I will also suggest exercises, which I will not solve, but which can be solved using the techniques I will present and whose results can be checked directly using the existing literature. I will try to be the most complete I can, and where it will not be possible I will indicate the concrete references where more details can be found.\\

The various examples I chose are just an excuse to learn the fundamental analytic and numerical techniques. As such, I will not focus too much on the physics of the problems and the interpretation of the results but rather on the practical steps of the problem-solving process. This is certainly a practice that I highly discourage during a creative process of research\footnote{Before jumping into whatever computations you have in mind, ask yourself why you are doing that! If the response is simply: ''Because I can'', you are definitely on the wrong path.} but that, in this case, for pedagogical purposes, might be efficient.  The different benchmarks are accompanied by available and open-source Mathematica \cite{Mathematica10} codes, which I used to obtain some of the results presented in the main text and most of the figures appearing in it\footnote{As an aside personal note, try to avoid becoming slaves of Mathematica. Mathematica is an amazing instrument which can speed up a lot of tedious computations but between Mathematica and you, you are the only one able to think (whatever the hyped machine-learning fans claim). When you realize (and it happened to me) that you use Mathematica to do $8*4$, then something is going wrong.}. The codes can be found at \url{https://members.ift.uam-csic.es/matteo.baggioli/lectures/}.
\newpage

The lectures are organized as follows:
\begin{enumerate}[label={\textbf{Section \arabic*.}},leftmargin=\widthof{[Stepeee-III]}+\labelsep]
    \item In this section, I will briefly introduce and motivate the tool we will be using and its origin. This part, even if already explained in several previous works, is necessary to convince the reader that the dictionary and the framework we use is not a collection of God-given laws as those which Moses received at Mount Sinai. Despite the counter-intuitive and not obvious nature of the playground, there are strong motivations behind its origin. At the same time, I will make an effort to motivate why such new tools are necessary to make advancements in certain open questions and how we should use them. If you already know the topic you can skip this section completely; if you do not care about these details or you are in a hurry, I suggest to read only subsection \ref{skipskip} for a brief advertising slogan of the duality.
    \item In this second part, I will go through the basic concepts of the duality and the dictionary. I will use the simplest example of a bulk scalar field to introduce the standard way of extracting physical information about the dual field theory from simple gravitational computations. This part can be a bit tedious and abstract but it is fundamental to understand what we will do later on, in the various applications. If you get bored along this section you can always jump ahead and come back to it at the moment you need it. Learning is not a linear process but rather an out-of-time order one.
    \item This section is the first one containing practical examples and fundamental techniques. More specifically I will explain the following points:
    \begin{itemize}
    \item[(i)] How to derive, with Mathematica, the Einstein equations for the background and for the perturbations (\textit{example: the shear mode}).
    \item[(i)] How to use the membrane paradigm and its power (\textit{example: the Kovtun-Starinets-Son (KSS) bound}).
    \item[(ii)] How to numerically obtain the Green function of a specific operator at finite frequency (\textit{example: the viscosity and elasticity of holographic massive gravity (HMG)}).
    \item[(iii)] How to obtain the low frequency expansion of a correlator via analytic methods in the bulk (\textit{example: the violation of KSS and the elastic modulus in HMG}).
    \item[(iv)] How to use near-horizon arguments to extract analytically the behaviour of certain physical quantities close to $T=0$ (\textit{example: the scaling of $\eta/s$ at zero temperature in HMG}).
    \end{itemize}
    \item This section contains the second block of concepts. In particular:
    \begin{itemize}
    \item[(i)] How to analytically study the stability of the solution by using the BF bound argument (\textit{example: the holographic superconductor}).
    \item[(ii)] How to numerically verify the instability of the background solution at finite temperature (\textit{example: the holographic superconductor}).
    \item[(iii)] How to numerically compute the Green function at finite frequency when the operators are mixing (\textit{example: the electric conductivity}).
    \item[(iv)] How to exploit certain properties of the bulk theory to derive the transport coefficients at zero frequency (\textit{example: the DC conductivities}).
    \item[(v)] How to extract and the quasinormal modes for a simple decoupled operator using numerical and perturbative techniques (\textit{example: the decoupled phonon mode in HMG}).
    \end{itemize}
    \item In this last section we will investigate how the holographic duality geometrizes the RG flows between two different fixed points. In particular we will consider:
    \begin{itemize}
    \item[(i)] The correspondence between the symmetries of the field theory and the isometries of the corresponding geometric gravitational picture (\textit{example: AdS and Lifshitz spacetimes}).
    \item[(i)] The construction of (zero temperature) holographic RG flows by using interpolating geometries (\textit{example: Lifshitz flows}).
    \end{itemize}
\end{enumerate}
Once you master with confidence all the techniques mentioned above, you could easily publish the same papers I did, and even more! At the same time, you will be able to learn fast the opportune generalizations of those tools to attack more complicated problems you might be interested in. The rest is on you! This is your driving license, but it does not mean you know how to drive, it just means you are ready to get on your car and learn by yourself and accumulate experience and knowledge.\\

I truly hope you will find these notes as useful as their writing has been to me. It gave me the opportunity of thinking about and re-derive by myself all those points that when you read a paper they are sold as obvious, and they are not! Things are simple once you did them. Now it is your turn to do them!\\

It is very likely that you will find through the next pages a lot of typos and even incorrect statements. I could say I did it on purpose to force you to spot them; of course that is not true, but your duty to spot them is still valid. Think of yourself not as a simple reader but more as the referee of this manuscript. Every comment, correction, suggestion and discussion will be useful to me, to you and especially for everyone that will use this book to learn. I therefore encourage you to contact me in any of these cases to help me improve the content of this manual. I like to think about it as a living creature (like Science), evolving and getting better in time, and this is possible only through your feedback.
\newpage
\subsection*{Tour guide}
Along the way you will find several colored boxes:
\begin{mdframed}[style=MyFrame4]
\textit{The yellow boxes contain technical and independent computations and/or proofs complementary to the main text. If you not are interested in these details, you can just skip them completely.}
\end{mdframed}
\begin{mdframed}[style=MyFrame]
\textit{The blue boxes contain tricks that I have learned during my years of research. These manoeuvres speed up computations and make them simpler.}
\end{mdframed}
\begin{mdframed}[style=MyFrame2]
\textit{The red boxes contain warnings about common mistakes or bad practices. They can save you in several situations.}
\end{mdframed}
\begin{mdframed}[style=MyFrame3]
\textit{The green boxes contain exercises for you. Most of them are unsolved in these notes but the results can be found in the existing literature which I will refer to. For some of them, I will provide directly Mathematica notebooks with the solutions and you are always welcome to contact me to discuss them.}
\end{mdframed}

%
%

\extrachap{Acknowledgements}

It is my pleasure to sincerely thank Ayan Mukhopadhyay and Indian Institute Of Technology Madras, Chennai to provide the spark that started all of this. You gave me the opportunity, the motivation and the courage to write this up and share all I have learned in these last years of research. I am grateful to all the people that contributed to my current understanding of the problem and to the knowledge I acquired in these last years. Supervisors, collaborators, colleagues, speakers, competitors, enemies, lecturers: you all have been of great help. I hope this could be of the same relevance for all the students interested in this topic. I am particularly grateful to my colleagues who wasted some of their time to read a preliminary version of this book. I finally acknowledge the generous support of the Spanish MINECO’s “Centro de Excelencia Severo Ochoa” Programme under grant SEV-2012-0249.

\tableofcontents

\mainmatter
%
%
%
\chapter{A Strings-less introduction to AdS-CFT}
\label{intro} 
\hspace{0.2cm} \includegraphics[width=0.5\textwidth]{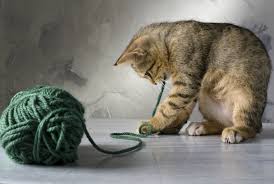}\\

\epigraph{Two is better than one.}{\textit{Boys Like Girls}}
In this first section we introduce the fundamental tool we will be using in the rest of the book. For different reasons, different people refer to it as:
\begin{itemize}
    \item \textbf{AdS-CFT correspondence}: This name is based on its historical derivation and it focuses on the first and most detailed example of the correspondence where one side is a Conformal Field Theory (CFT) and the other side is gravity in Anti De Sitter (AdS) spacetime. I feel this label is somehow outdated in the sense that, as you will see in the following, the correspondence can now be employed for situations which are much more general than a simple CFT and AdS spacetime. Almost $100\%$ of the cases we will consider do not fall in such class! 
    \item \textbf{Holography}: This label focuses on a single, but maybe the most surprising, aspect of the tool, which is the fact that the two sides of the correspondence do not live in the same number of dimensions. This noun is fancy but quite limited and it will fill up your email inbox with invitations to Optics conferences and Star Wars meet-ups.
    \item \textbf{Gauge-Gravity duality}: This last case is my favourite since it outlines the main and most understandable feature of the tool. It is a duality between theories of gravity and Gauge theories. It does not specify any feature of the two sides but it rather underlines its very general character.
\end{itemize}
Whatever you prefer to call it, we are going to discuss in detail the points which I believe are the fundamental motivations and features behind it. Given the applied aim of the course, we will completely ignore the following issues: (I) we will refer to String Theory and extended objects in extra-dimension as little as possible. We will not discuss the original motivation of the duality proposed by Maldacena in his very famous work \cite{Maldacena:1997re} (see also \cite{Witten:1998qj}). (II) We will always work in the so-called Bottom-up version of the duality. This means we will not care about any UV completion or string embedding of our theory. This is clearly a fair question and point of view; personally, I have always found quite an oxymoron the attitude of pretending to be using a complicated String Theory construction to shed light on some low energy question. (III) We will be using just ''simple'' classical and weakly coupled gravity construction to explore the physics of strongly coupled and many body (large $N$) field theories. (IV) On the same line, we will ignore any Swampland or naturalness discussion \cite{Giudice:2008bi,doi:10.1063/1.2735124,Palti:2019pca}. We will require the consistency of our theory only from simple and world-wide accepted criteria such as unitarity, absence of ghosts and superluminal modes and (linear) stability\footnote{This is not totally true. As we will explain later, instabilities of the gravitational theory could be totally physical and interesting rather than a pathology of the theory.}.
\section{If you want to skip this section, just read this}\label{skipskip}
\begin{figure}[h!]
    \centering
    \includegraphics[width=0.41\linewidth]{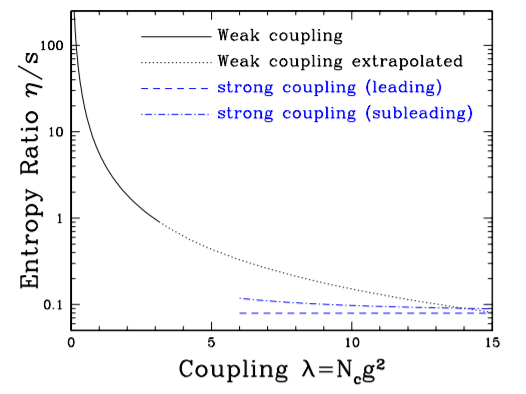}%
    \quad
    \includegraphics[width=0.535\linewidth]{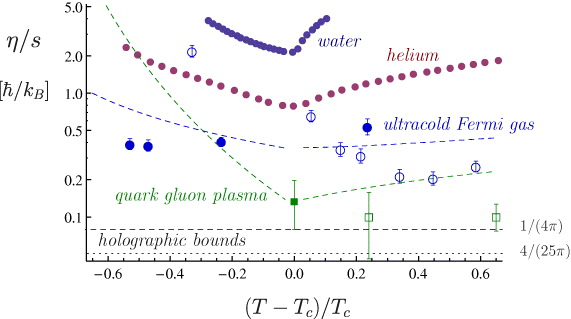}
    \caption{\textbf{Left: }Comparison between the weak coupling and strong coupling results for the $\eta/s$ ratio. \textbf{Right: }The experimental value for the QGP is around $\eta/s \sim 2.5/4\pi\,\sim0.19$. Figures taken from \cite{Huot:2006ys} and \cite{Adams_2012}.}
    \label{fig:strongweak}
\end{figure}

\begin{figure}
    \centering
    \includegraphics[width=0.3\linewidth]{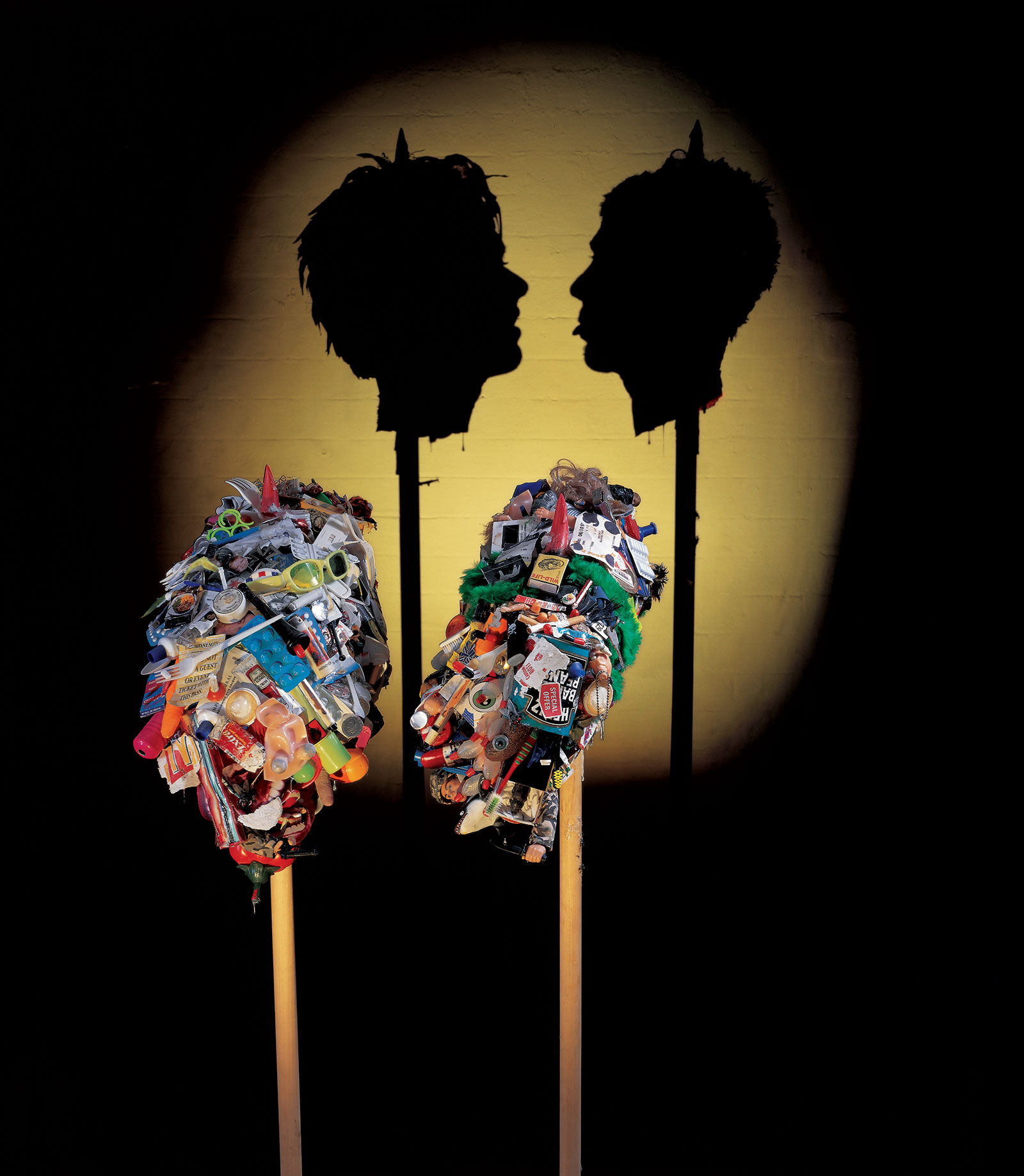}%
    \quad
    \includegraphics[width=0.65\linewidth]{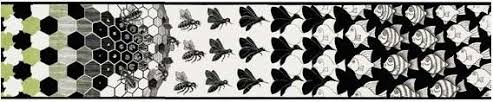}
    \caption{Artistic representations of the meaning of duality. The right panel is a famous painting from the very interesting collection of the Dutch artist M.C.Escher.}
    \label{fig:art}
\end{figure}

\begin{figure}[h!]
    \centering
    \includegraphics[width=0.6\linewidth]{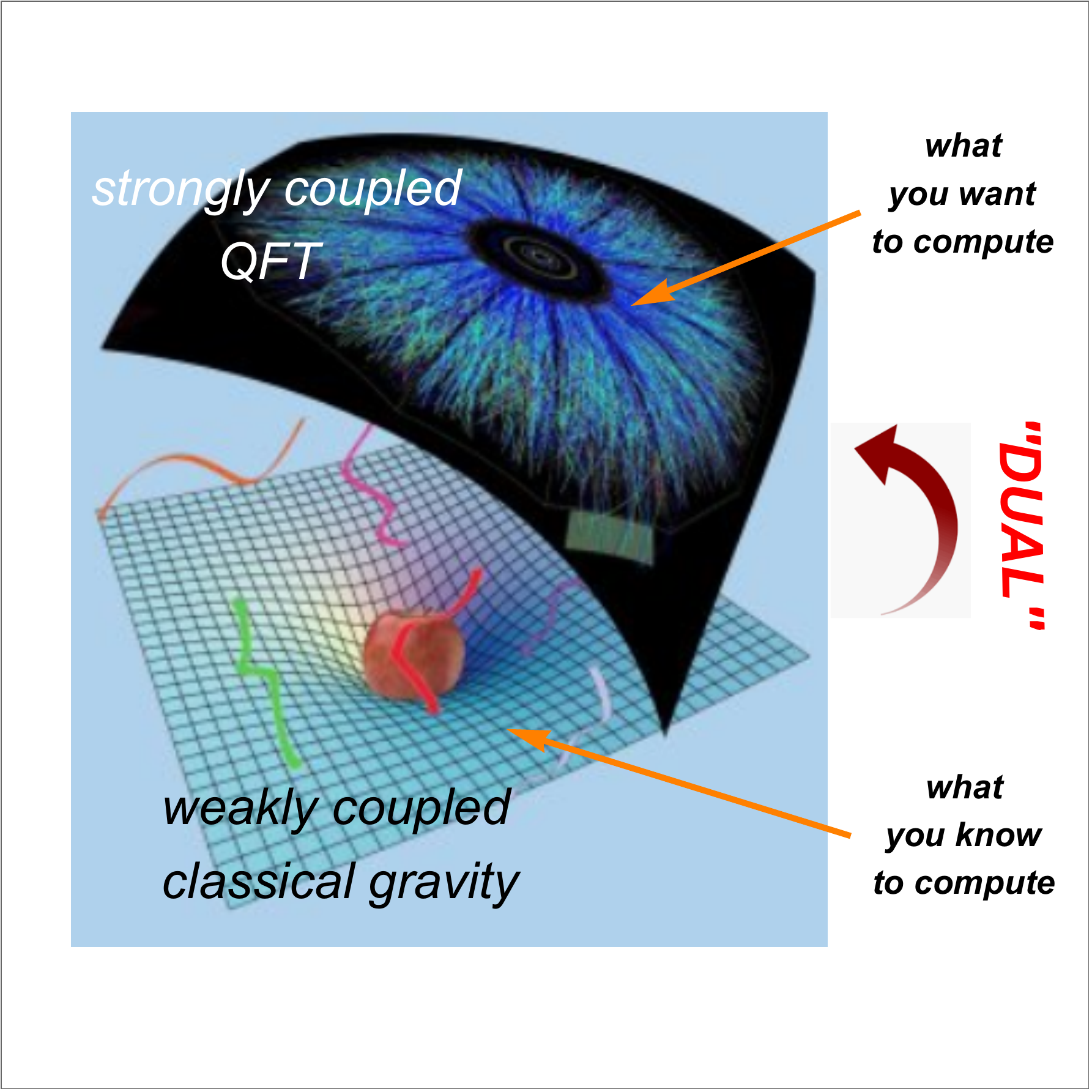}
    \caption{What do you need to know about AdS-CFT.}
    \label{fig:skip}
\end{figure}

\begin{figure}
    \centering
    \includegraphics[width=0.7\linewidth]{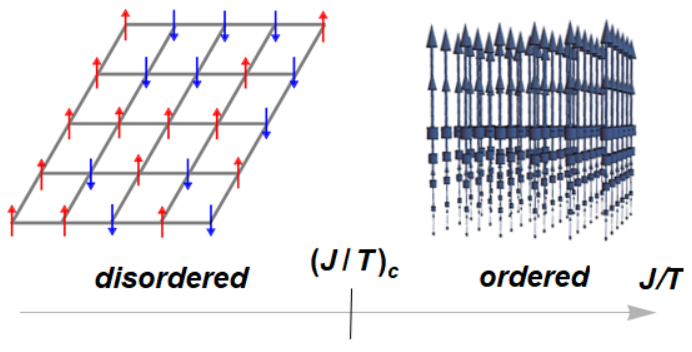}
    \caption{The transition between the ordered and disordered phases in a 2D Ising model.}
    \label{fig:ising}
\end{figure}

Perturbation theory is a super powerful tool, which is exploited in physics in any possible direction \cite{fradkin2013field,tsvelik2007quantum,fernandez2000introduction,scherer2003introduction,kato2012short}. To introduce the idea, let us consider a generic problem which we know how to solve explicitly and let us add a small deformation to it, parametrized by a coupling $\mathcal{C}$:
\begin{equation}
    \text{problem}\,=\,\underbrace{\text{problem}_0}_\text{\parbox{2cm}{\centering we know how\\[-4pt] to solve this}}\,+\,\mathcal{C}\,*\,\blacksquare
\end{equation}
where $\mathcal{C}\,\ll\,1$, namely it is ''small''\footnote{Small or big does not mean anything in physics. The way this is properly done is by making the coupling dimensionless or, in other words, to compare it with a characteristic quantity of the system with the same mass dimension.}.
Now, let us imagine we want to compute a physical observable $\mathbb{O}$. What we can do is to treat the deformation order by order and express the final result as:
\begin{equation}
    \mathbb{O}(\mathcal{C})\,=\,\mathbb{O}_0\,+\,\sum_{n=1}^{m}\,\mathcal{C}^n\,\mathcal{A}_n \label{ss}
\end{equation}
\textit{i.e.} a perturbative expansion in terms of this ''small'' coupling.\\
Because of the ''smallness'' of the coupling, each term in the sum will contribute less and less in the final result for the observable $\mathbb{O}$, meaning we will get a good result without the need of going towards very large $m$, but just truncating the series after few terms\footnote{We will not consider here the problem that most of the series as \eqref{ss} are asymptotic series, with finite or even zero radius of convergence. This implies that increasing the number of terms $m$ will not improve the final result after a certain step.}.\\
The perturbative procedure is formalized in quantum field theory by the use of Feynman diagrams which allow to express the full process in a very simple and graphic fashion. Just to show one example of what we were discussing, let us take the following quantum field theory know as $\lambda \phi^4$:
\begin{equation}
    \mathcal{L}(\phi)\,=\,\frac{1}{2}\,\left[\partial_\mu \phi \partial^\mu \phi\,-\,m^2\,\phi^2\right]\,-\,\frac{\lambda}{4!}\,\phi^4
\end{equation}
In the limit of small $\lambda$ the perturbative solution of this model can be found in all the QFT textbooks \cite{Peskin:1995ev,srednicki2007quantum}.\\
To show off the power of the pertubative method, I cannot avoid mentioning the results for the electron g-factor -- the dimensionless ratio of its magnetic moment to its angular momentum. From the Dirac equation governing the dynamics of the electron as a spin $1/2$ particle, we get that $g=2$ (this value is exactly twice the results obtained with classical mechanics) \cite{chalupsky2018classical}. Quantum effects, encoded in QED corrections, produce deviations from this value which can be both measured experimentally \cite{2008PhRvL.100l0801H} and computed analytically \cite{Aoyama:2017uqe}, with an incredible precision. The comparison between experiments and perturbation theory, according the data of \cite{PhysRevD.86.010001}, reads:
\begin{align*}
  & {a}_{exp}\,=\,11659208.9(5.4)(3.3) \,\times\, 10^{-10}\\
   & {a}_{theory}\,=\,116591802(2)(42)(26) \,\times\, 10^{-11}\\
   & {a}_{exp}\,-\,{a}_{theory}\,=\, 287(63)(49) \times 10^{-11}\,\,!!!
\end{align*}
where we define the anomalous magnetic moment $a \equiv (g-2)/2$. This result can be obtained using $12,672$ vertex	Feynman	diagrams at the $10^{th}$ order of the perturbative expansion which would look like more or less like this:
\begin{figure}[h!]
    \centering
    \includegraphics[width=0.36\linewidth]{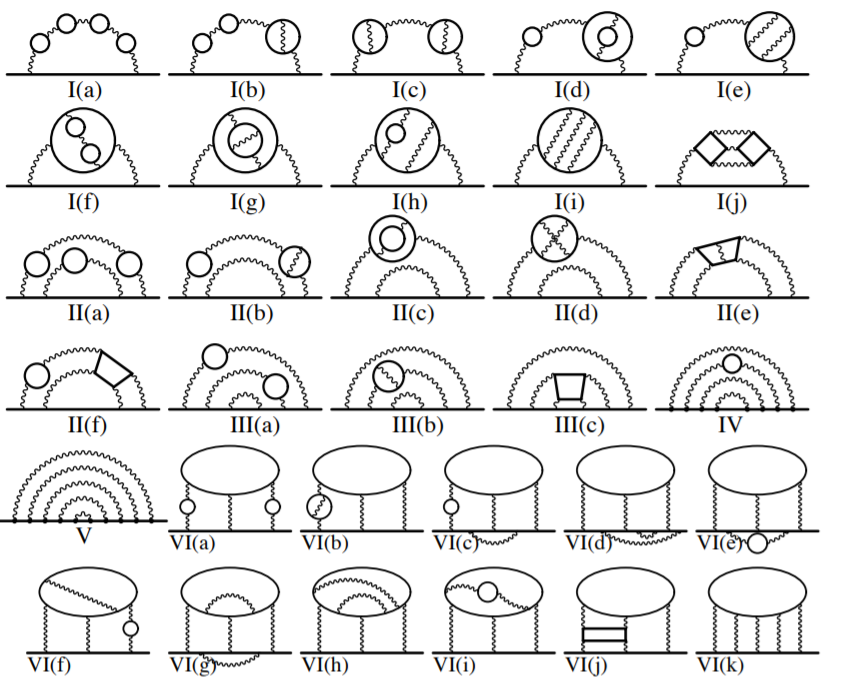}
\end{figure}\\
going on for approximately $50$ pages! Summing all these perturbative contributions, we get an agreement between theory and experiments with a precision of $10^{-11}$, is not that amazing?\\
Now, the perturbative methods work remarkably, but what if their assumption is not valid anymore? What if the coupling governing the interactions is not weak and small, but rather an $\mathcal{O}(1)$ number (in the opportune dimensions)? If we blindly follow the perturbative prescription, what we get would be something like this:
\begin{equation}
\bigcirc(\lambda)\,=\,\sum_{n=0}^{\infty} \lambda^n\,c_n\,=\,c_0\,+\,c_1\,\includegraphics[width=0.6cm]{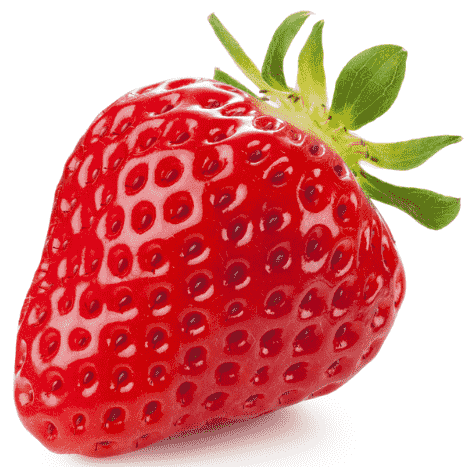}\,+\,c_2\includegraphics[width=0.6cm]{straw.png}+\,c_3\,\includegraphics[width=0.8cm]{straw.png}\,+\dots\,=\,\includegraphics[width=1.3cm]{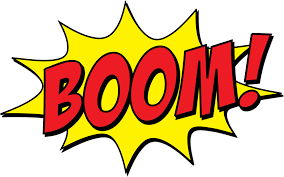}
\end{equation}
where our final result is completely meaningless and totally in disagreement with the experimental data. The simplest example of this failure is the computation of the viscosity-entropy ratio in the Quark Gluon Plasma. The experimental values \cite{Teaney:2003kp} indicates that this ratio is order:
\begin{equation}
    \frac{\eta}{s}\,\sim\,2.5\,\frac{1}{4\,\pi}\,\frac{\hbar}{k_B}
\end{equation}
From a different point of view, perturbation theory, in terms of the ’t Hooft coupling $\lambda$, gives the following result:
\begin{equation}
    \frac{\eta}{s}\,=\,\frac{A}{\lambda^2\,\log(B/\sqrt{\lambda})}
\end{equation}
where $A,B$ are non-universal numbers \cite{Arnold:2003zc,Huot:2006ys}. The weak-coupling computation gives a value of $\eta/s \gg 1$, in evident tension with the experimental data (see fig.\ref{fig:strongweak}). The QGP appeared to be a strongly coupled system, for which the perturbative methods are of no help. We will see later on how Holography has shown to be extremely useful in providing a solution to this conundrum \cite{Policastro:2001yc}.
 \\
What is Gauge-Gravity? It is a duality\footnote{For a nice discussion about the precise meaning of ''duality'' and some concrete examples see \url{http://users.physics.harvard.edu/~mwilliams/documents/phys143a_duality.pdf}.}, a very powerful one. The best way to understand the meaning of a duality is using art, as in fig.\ref{fig:art}.
\\
The main idea is that the same system can be analyzed from different points of view and described by different degrees of freedom. A system, that appears complicated and without a clear guidance principle from one perspective, can become elegant and simple from another, like the shades in the right panel of fig.\ref{fig:art}. The trick is to change the framework used to describe the system, move from the bees to the fishes; solve the system in the picture which results the simplest, and then translate back the result in the original variables.\\
This is exactly what Holography does (see fig.\ref{fig:skip}). It maps a hard, if not even impossible, problem into a simpler one which is described by completely different degrees of freedom in a completely different background. The two descriptions are conjectured\footnote{To be fair the duality has never been proven rigorously and mathematically. There are anyway several hints and indications that it might hold. Moreover, in the context of String Theory, there are several explicit examples \cite{Bobev:2013cja,Freedman:2013ryh,Beisert:2010jr}, where the physical observables can be computed exactly from both sides (the field theory and the gravitational theory) and they perfectly match.} to be dual and therefore the results obtained from the simple side (the gravitational one for us) can be mapped back in terms of the field theory language, to obtain the field theory observables desired.
\\
The idea of duality is of course much older that Holography and it has been widely employed in Physics and Mathematics \cite{at,Castellani:2018icz}. The most famous example is probably the wave-particle duality of light. Depending at which phenomenon dealing with light you are looking at, it is more convenient to think about light as a particle (the photon) or as a wave (EM wave).\\
In the following, we show explicitly the power of the idea of duality by applying it to the Ising model in two dimensions. The reader interested in more details can look at \cite{RevModPhys.52.453}. The Ising model is a simple model defined via discrete variables which can only take value $\pm 1$ (up or down). The variables are defined on a lattice and their dynamics is described by the following Hamiltonian:
\begin{equation}
    H\,=\,-\,J\,\sum_{\langle i j \rangle}\,S_i\,S_j\,,\,\quad \textit{with}\quad J\,>\,0 \quad \textit{and}\quad S_i\,=\,\pm 1
\end{equation}
where the spin variables interact between themselves via the coupling $J$ and only via nearest neighbours. For this concrete example, we choose the value of $J$ positive, which classifies the model as ferromagnetic. It is simple to realize that having all the spins pointing in the same direction does minimize the energy of the system. At low temperatures, the electrons will organize themselves to create indeed such configuration; this is called the ordered phase. At higher temperatures, there will be thermal fluctuations which will allow some electrons to overcome the energy cost associated with going against the others. Choosing the temperature to be sufficiently high, these thermal fluctuations dominate, and the spins will point in essentially random directions; this is called the disordered phase. See fig.\ref{fig:ising} for a cartoon of the situation. 
\\
What is the critical temperature at which this phase transition happens? In general, this is a very hard question to tackle analytically, but here comes the trick. For simplicity, we consider the system on a squared two-dimensional lattice. For practical purposes, let us perform the following re-definition:
\begin{equation}\label{rere}
    H\,=\,K\,\sum_{\langle i j \rangle}\,S_i\,S_j\,,\quad \textit{with}\quad K\equiv J/T
\end{equation}
where $K$ is now a dimensionless parameter, determining the strength of the $J$ interaction with respect to the thermal scale $T$.
Given this Hamiltonian, we can compute the partition function in a standard way:
\begin{equation}
    Z(K)\,=\,\sum_\alpha \,e^{-\,\beta\,E_\alpha(K)}\,, \quad \textit{with}\quad \beta=1/T
\end{equation}
After some not too tedious manipulations, we obtain that at high temperature the partition function reads:
\begin{equation}
    Z_{high}(K)\,=\,\left(\cosh K\right)^{N_l}\,2^N\,\left(1\,+\,N\,\left(\tanh K\right)^2\,+\,\dots\right)\label{highT}
\end{equation}
with $N_l$ the number of links and $N$ the number of sites of the 2D lattice.\\
In the oppositve regime, at low temperature, we get:
\begin{equation}
    Z_{low}(K)\,=\,2\,e^{K\,N_l}\,\left(1\,+\,N\,\left(e^{-2K}\right)^4\,+\,\dots\right)\label{lowT}
\end{equation}
Now, this does not seem very useful. Nevertheless, one can notice that, up to a multiplicative constant in front, the high and low temperature expansions \eqref{highT}, \eqref{lowT} can just be obtained by passing different arguments to the same series. In particular, by performing the substitution:
\begin{equation}
    e^{-\,2\,\tilde{K}}\,\Longleftrightarrow\,\tanh K \label{dualdual}
\end{equation}
one can pass from one limit to another. This is called the Kramers-Wannier duality \cite{PhysRev.60.252}. The two phases, high and low temperature, are dual under the identification above. This result is very powerful. At the critical point, the transformation above has to be singular, therefore:
\begin{equation}
    e^{-\,2\,\tilde{K_c}}\,\Longleftrightarrow\,\tanh K_c\quad \rightarrow T_c\,=\,\frac{2\,J}{\log \left (1\,+\,\sqrt{2}\right)} 
\end{equation}
which gives us immediately the value of the critical temperature analytically. This result is in perfect agreement with the numerical simulations.
This is a very famous and simple case where the system is self-dual, dual to itself. This means that the system and its degrees of freedom on the two sides ($T<T_c$ and $T>T_c$) are the same. Despite its simplicity, this example displays already a very interesting feature. Instead of dialing the temperature, we can think of changing the coupling $J$. The high temperature regime corresponds in this language to the weakly coupled regime and the low temperature one to the strongly coupled regime\footnote{This is evident using the redefinition in eq.\eqref{rere} $K\equiv J/T$.}. The duality transformation \eqref{dualdual} is relating a theory at strong coupling to a weakly coupled theory (which in this case, due to the self-dual nature, it is the same). So what is AdS-CFT? It is just a much more general version of the Ising duality.\\ In particular, AdS-CFT is an upgraded duality in the following senses:\\[0.3cm]
\noindent
\fcolorbox{black}{black!15}{\parbox{\dimexpr\textwidth-2\fboxsep-2\fboxrule}{%
  \color{red!50!black}%
\begin{tabular}{ r|c|c| }
\multicolumn{1}{r}{}
 &  \multicolumn{1}{c}{\textbf{Duality in 2D Ising model}}
 & \multicolumn{1}{c}{\textbf{AdS-CFT}} \\
\cline{2-3}
two sides of the duality & same theory & different theories \\
\cline{2-3}
d.o.f. in the two sides & ''same'' & different \\
\cline{2-3}
$\#$ dimensions & same & different\\
\cline{2-3}
duality transformation & local & highly non-local
\\
\cline{2-3}
coupling constant & strong-weak & strong-weak\\
\cline{2-3}
solvability & both sides & only one side\\
\cline{2-3}
\end{tabular}}}\\[0.4cm]
but the idea is analogous. Moreover, it is even more relevant because only one side of the duality can be solved\footnote{Technically, this is due to our limitations, not those of the duality.} and it therefore can shed light on the other, and strongly coupled, side, where almost no other tool is available\footnote{One could always do numerical simulations like lattice computations, Monte Carlo, etc \dots but as we will see later those techniques present very strong limitations.}.\\
In summary, the only thing you need to know about AdS-CFT \footnote{For the purposes of these lectures.} is that it is a very non trivial, but at the same time very powerful, duality which can be employed to solve strongly coupled theories, for which no other reliable method is available.
\section{The Holographic principle a.k.a. ''do we really need all these dimensions'' ?}
As already hinted in the previous discussion, one of the most intriguing, and at the same time disorientating, feature of the AdS-CFT correspondence is the fact that is holographic. This means that the dual theories do not live in the same number of spacetime dimensions but rather the gravitational theory is defined always in one dimension more than the field theory side. Somehow, it is like if one of the dimension of the gravitational theory is not necessary and one could describe the same system in terms of field theories getting rid of it. Is this crazy? No, it is actually known and it has to do a lot with Black Hole physics.\\
Black holes (BHs) are fascinating objects which are expected to exist in every galaxy of our universe. Recent observations have attracted a lot of attention also outside the physics community \cite{Akiyama:2019cqa,PhysRevLett.116.061102,PhysRevLett.116.241102}. BHs are thermal objects with a finite temperature and entropy \cite{Carlip:2014pma} and they follow the laws of standard thermodynamics.
The second law of thermodynamics requires that the entropy of a closed system will never decrease. Including the black hole entropy in the entropy ledger gives us what is known as the generalized second law of thermodynamics (GSL) (Bekenstein 1973 \cite{BekBH}):
\begin{equation}
\Delta S_0\,+\,\Delta S_{BH}\,=\,\Delta S_{total}\,\geq\,0\,.
\end{equation}
where $S_{BH}$ is the black hole entropy  and $S_0$ the entropy of the ''rest'' outside the black hole horizon.
This takes us immediately to the definition of the so-called Holographic Principle (see \cite{holprinrev} for a review and \cite{holprin1,holprin2} for the original works by t'Hooft and Susskind).\\
In an ordinary system, with no gravity, the number of degrees of freedom $\mathrm{N}$ is extensive and it scales with the volume of the system, as $\mathrm{N}\sim e^V$. This means that the entropy obtained from the Von Neumann relation $S\sim \log\, \mathrm{N}$ will be proportional to the volume itself $S\,\sim\,V$.
\begin{figure}[h!]
    \centering
    \includegraphics[width=0.6\linewidth]{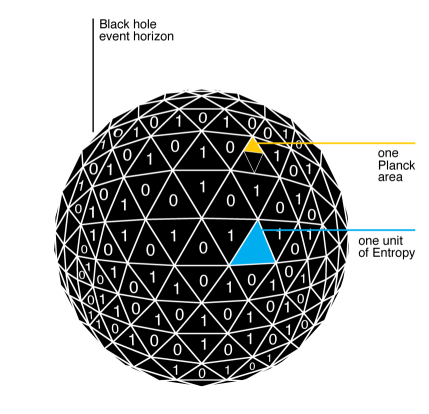}
    \caption{The information is all encoded on the surface of the black hole, making one of the spacetime dimensions completely ''useless''.}
    \label{fig:BH}
\end{figure}\\
When gravity kicks in, the story is different! The, already mentioned, \textit{Holographic principle} claims that in this case: \textit{the maximum entropy of a region of volume $V$ corresponds to the entropy of the biggest black hole that fits in it}.\\
This can be formulated as:
\begin{equation}
S_{max}\,=\,S_{BH}
\end{equation}
and here comes the novelty. The entropy of a BH object is not proportional to its volume $V$, but rather to the area of its event horizon. This is the famous discovery of Bekenstein and Hawking, which we can write as:
\begin{equation}
S_{BH}\,=\,\frac{\mathcal{A}}{4\,\pi\,G_N}
\end{equation}
where $\mathcal{A}$ is the area of the event horizon, the surface around the black hole inside which nothing, not even light, can escape.
Following the holographic principle, this means that in a theory with gravity the entropy of a region of volume $V$ is proportional to the area of the biggest ''ball'' surrounding it as shown in fig.\ref{fig:BH}. This also means that all the dynamics of that system confined in the volume $V$ is encoded in the degrees of freedom living in one dimension less, on its surface. In a sense, there is one redundant dimension that from the point of view of the amount of information is completely irrelevant. We can understand everything inside that volume, by just looking at what happens at its surface. This is the first hint that a theory of gravity in $d+1$ dimensions contains the same amount of information of a correspondent (we will explain in which sense) field theory in $d$ dimensions.
\begin{mdframed}[style=MyFrame4]
\begin{center}
    \textbf{Sketchy proof of the Holographic principle.}
\end{center}\vspace{0.15cm}
Suppose for a moment that the entropy of the system is bigger than the entropy of the largest BH fitting $S>S_{BH}$; now, let us start throwing matter inside the system (increasing both its entropy and its energy). At a certain point, the system will collapse into a black hole state with entropy $S_{BH}$. If we compute the total entropy variation for the process described, we would discover that:
\begin{equation}
\Delta S\,<\,0
\end{equation}
which violates the generalized second law of thermodynamics! Therefore, by contradiction, we proved the Holographic principle. For many more details see \cite{bousso2002holographic}.
\end{mdframed}
\begin{mdframed}[style=MyFrame4]
\begin{center}
    \textbf{What is Stephen Hawking famous for.}
\end{center}\vspace{0.15cm}
The Bekenstein-Hawking formula states that the entropy of a BH is proportional to the area of its horizon:
\begin{equation}
    S\,=\,\frac{\mathcal{A}}{4\,G_N}\label{fam}
\end{equation}
where $G_N$ is the Newton constant.\\
There is a more general result, found by Wald, which defines the entropy as a Noether charge \cite{Wald:1993nt,Visser:1993nu,Brustein:2007jj} and it is valid in more general situations, even beyond the Einstein-Hilbert action of General Relativity. Let us re-derive the famous result \eqref{fam} with this more powerful formula. Let us start from an action:
\begin{equation}\label{themodel}
\mathcal{S}\,=\,\frac{1}{16\,\pi\,G_N}\int\,d^4x\,\sqrt{-g}\,\left(\,R\,+\,\mathcal{L}_{matter}\,\right)
\end{equation}
We can compute the total entropy using the Wald formula \cite{Wald:1993nt,Visser:1993nu,Brustein:2007jj}:
\begin{equation}
S\,=\,-2\,\pi\,\int_{\Sigma}\,d^2x\,\sqrt{-h}\,\frac{\delta\,\mathcal{L}}{\delta\,R_{\mu\nu\rho\sigma}}\,\epsilon_{\mu\nu}\,\epsilon_{\rho\sigma}
\end{equation}
where  $\epsilon_{\mu\nu}$ is the binormal on the horizon $\Sigma$ and h the induced metric.
Let us assume a general ansatz for the metric:
\begin{align}
&ds^2\,=\,-\,h(r)\,dt^2\,+\,\frac{dr^2}{f(r)}\,+\,r^2\,dx^i\,dx^i
\end{align}
This implies that the binormal vector has only two non vanishing components:
\begin{equation}
\epsilon_{tr}\,=\,-\,\epsilon_{rt}\,=\,\sqrt{\frac{h(r)}{f(r)}}
\end{equation}
and therefore the Wald formula simplifies to:
\begin{equation}
S\,=\,-8\,\pi\,\int_{\Sigma}\,d^2x\,r_h^2\,\frac{h(r_h)}{f(r_h)}\,\frac{\delta\,\mathcal{L}}{\delta\,R_{rtrt}}
\end{equation}
Following our action \eqref{themodel} we obtain:
\begin{equation}
\frac{\delta\,\mathcal{L}}{\delta\,R_{\mu\nu\rho\sigma}}\,=\,\underbrace{\frac{1}{32\,\pi\,G_N}\,\left(g^{\mu\rho}g^{\nu\sigma}\,-\,g^{\mu\sigma}g^{\nu\rho}\right)}_{EH\,\,action}
\end{equation}
We are interested in the $rtrt$ component which simplifies to:
\begin{equation}
\frac{\delta\,\mathcal{L}}{\delta\,R_{rtrt}}\,=\,\frac{1}{32\,\pi\,G_N}\,g^{rr}g^{tt}\,\,=\,-\,\frac{1}{32\,\pi\,G_N}\frac{f(r_h)}{h(r_h)}
\end{equation}
We therefore conclude that:
\begin{equation}
S\,=\,\frac{1}{4\,G_N}\,\int_{\Sigma}\,d^2x\,r_h^2\,\,=\,\frac{\mathcal{A}_h}{4\,G_N}
\end{equation}
which is indeed the famous Hawking-Bekenstein formula.
\end{mdframed}
\begin{mdframed}[style=MyFrame3]
\begin{center}
    \textbf{Exercise \#1 : playing with Wald formula.}
\end{center}\vspace{0.15cm}
Consider an Horndeski gravity model \cite{Horndeski1974,Deffayet:2013lga}, whose action reads:
\begin{equation}
    \mathcal{S}\,=\,\int d^4x \sqrt{-g}\,\left[R\,-\,2\,\Lambda\,+\left(g^{\mu\nu}\,-\,G^{\mu\nu}\right)\,\sum_{I=1}^2\,\partial_\mu \phi^I \partial_\nu \phi^I \right]
\end{equation}
where $G^{\mu\nu}\equiv R^{\mu\nu}\,-\,\frac{1}{2}g^{\mu\nu}R$ is the Einstein tensor. Assume a simple ansatz:
\begin{equation}
ds^2\,=\,-\,h(r)\,dt^2\,+\,\frac{dr^2}{f(r)}\,+\,r^2\,dx^i\,dx^i\,,\quad \phi^I\,=\,k\,x^I
\end{equation}
Find the entropy density of the BH in this theory and show the Horndeski parameter $\gamma$ modifies the Bekenstein-Hawking formula \eqref{fam}. The complete results can be found in \cite{Baggioli:2017ojd}.\\
Try now with:
\begin{equation}
    \mathcal{L}\,=\,\frac{R}{2\,\kappa}\,-\,\Lambda\,+\,c_1\,R^2\,+\,c_2\,R_{\mu\nu}R^{\mu\nu}\,+\,c_3\,R_{\mu\nu\rho\sigma}R^{\mu\nu\rho\sigma}
\end{equation}
Do you agree with \cite{Kats:2007mq}? Do you see anything special for the specific combination of $c_1,c_2,c_3$ which is known as Gauss Bonnet gravity \cite{doi:10.1063/1.1665613}?
\end{mdframed}
\section{The extra-dimension and the RG flow}
\begin{figure}[h!]
    \centering
    \includegraphics[width=0.75\linewidth]{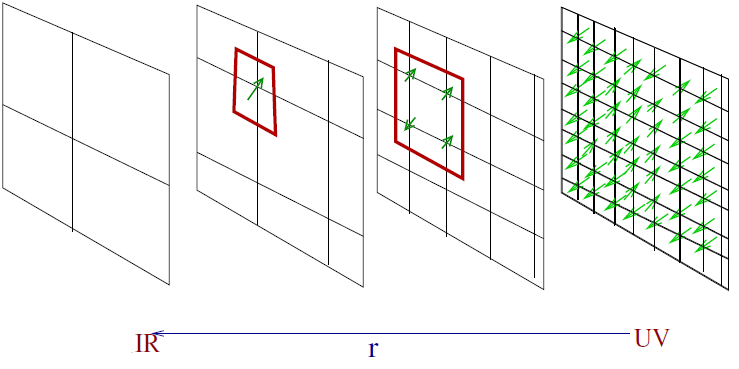}
    \caption{A pictorial representation about how the RG flow defines for us an extra (and warped) dimension whose geometric representation is very similar to AdS. Figure taken from \cite{adscftMcGreevy}.}
    \label{fig:RGads}
\end{figure}

In the previous section we discussed how this extra dimension entering in the duality can be understood from a gravitational point of view. In this section we ''flip the pan'' and we take the field theory point of view.\\
In a general local and well-defined quantum field theory, the observables depend on the energy scale we are looking at. From a more rigorous perspective, we can study the dynamics of the couplings of the theory under a change in the energy scale $\mu$, by looking at the so-called beta-function equation \cite{Peskin:1995ev,srednicki2007quantum} :
\begin{equation}
\mu\,\partial_\mu\,g(\mu)\,=\,\beta_g(g(\mu))\,.
\label{RGeqs}
\end{equation}
The important fact for us is that such equation is totally local in the energy scale and it can be thought of as a dynamical equation in an additional dimension $\mu$. In other words, we can now extend the system from depending on only the proper spacetime dimensions $(t,\vec{x})$ to a larger set of coordinates $(t,\vec{x},\mu)$. In this language, the beta-function equation \eqref{RGeqs} is simply the dynamical equation in the extra dimension $\mu$.  In order to understand this better, it is easier to think about the RG flow in the Wilsonian sense \cite{RevModPhys.47.773}. Changing the energy scale of the system simply corresponds to a coarse-graining process or in a way to use a different meter as shown in fig.\ref{fig:RGads}. Now we can take the images of our systems and put them aligned along the energy scale dimension. This procedure will create a $d+1$ spacetime which indeed looks like Anti de Sitter, where going from the boundary to the IR, the length scale dilatates as in the coarse-graining procedure.
\\
We can make this statement more precise, by counting  the number of degrees of freedom in the two sides and show that they match.
Let us start by considering a quantum field theory (QFT) in $d$ spacetime dimensions. For convenience, we introduce an IR and UV cutoff, represented by a lattice spacing $\epsilon$ and a finite spatial box of size $R$. The number of ''cells'' in the box is given by $\left(\frac{R}{\epsilon}\right)^{d-1}$. By defining the number of degrees of freedom for lattice spacing $c_{QFT}$, the total number of degrees of freedom in the QFT is:
\begin{equation}
N_{dof}^{QFT}\,=\,\left(\frac{R}{\epsilon}\right)^{d-1}\,c_{QFT}
\end{equation}
For a $SU(N)$ gauge theory, where the fields are $N \times N$ matrices, the counting in the large $N$ limit will give $C_{SU(N)}\sim N^2$.
Considering now the bulk gravitational theory in $d+1$ dimensions, the number of degrees of freedom is given by:
\begin{equation}
N_{dof}^{AdS}\,=\,\frac{A_\partial}{4\,G_N}
\end{equation}
where $A_\partial$ is the area of the region at the boundary as we explained in the previous section. Assuming the AdS metric, the area appearing in the expression above can be computed as:
\begin{equation}
A_\partial\,=\,\int_{\mathbb{R}^{d-1},z=\epsilon}\,d^{d+1}x\,\sqrt{-g}\,=\,\left(\frac{L}{\epsilon}\right)^{d-1}\,\int_{\mathbb{R}^{d-1}}\,d^{d-1}x
\end{equation}
where $\epsilon$ is the UV cutoff, the position of the AdS boundary and $L$ the size of the AdS length.
The last term in the expression above $\int_{\mathbb{R}^{d-1}}\,d^{d-1}x$ represents the volume of the AdS boundary $V_{\mathbb{R}^{d-1}}$, which is obviously infinite. If we request, as before, an infrared cutoff regulator $R$, such a volume becomes $
V_{\mathbb{R}^{d-1}}\,=\,R^{d-1}$. As a consequence, the area considered is now:
\begin{equation}
A_\partial\,=\,\left(\frac{R\,L}{\epsilon}\right)^{d-1}
\end{equation}
If we restore the the Planck length $G_N=(l_p)^{d-1}=\frac{1}{M_p^{d-1}}$, previously set to unit, then the total number of d.o.f. in the AdS spacetime can be written as:
\begin{equation}
N_{dof}^{AdS}\,=\,\frac{1}{4}\,\left(\frac{R}{\epsilon}\right)^{d-1}\,\left(\frac{L}{l_p}\right)^{d-1}
\end{equation}
We can finally compare the number of degrees of freedom in the two pictures  $N_{dof}^{QFT},N_{dof}^{AdS}$ and check that they match. The limit of classical gravity implies that:
\begin{equation}
\left(\frac{L}{l_p}\right)^{d-1}\,\gg\,1\,,
\end{equation}
\textit{i.e.} the curvature scale in Planck units is small.\\
Comparing the expressions, we can therefore conclude that a QFT has a classical gravity description  whenever $c_{QFT}$ is large, \textit{i.e.} in the large $N$ limit \footnote{To be more precise, this is just a necessary condition to have a well-defined gravity dual description. The question of determining which field theories have a gravity dual and which are all the necessary requirements to have it is largely to be determined. We thank Christopher Rosen for this clarification.}. Moreover, this classical description lives in one dimension more, which is representedin the QFT language by the RG flow scale. Several attempts to derive this point more rigorously have been made \cite{Lee:2013dln,Lee:2012xba,Lee:2010ub,Lee:2009ij,Mukhopadhyay:2017uik,Mukhopadhyay:2016fre,Behr:2015aat,Behr:2015yna}.
\begin{figure}
    \centering
    \includegraphics[width=0.8\linewidth]{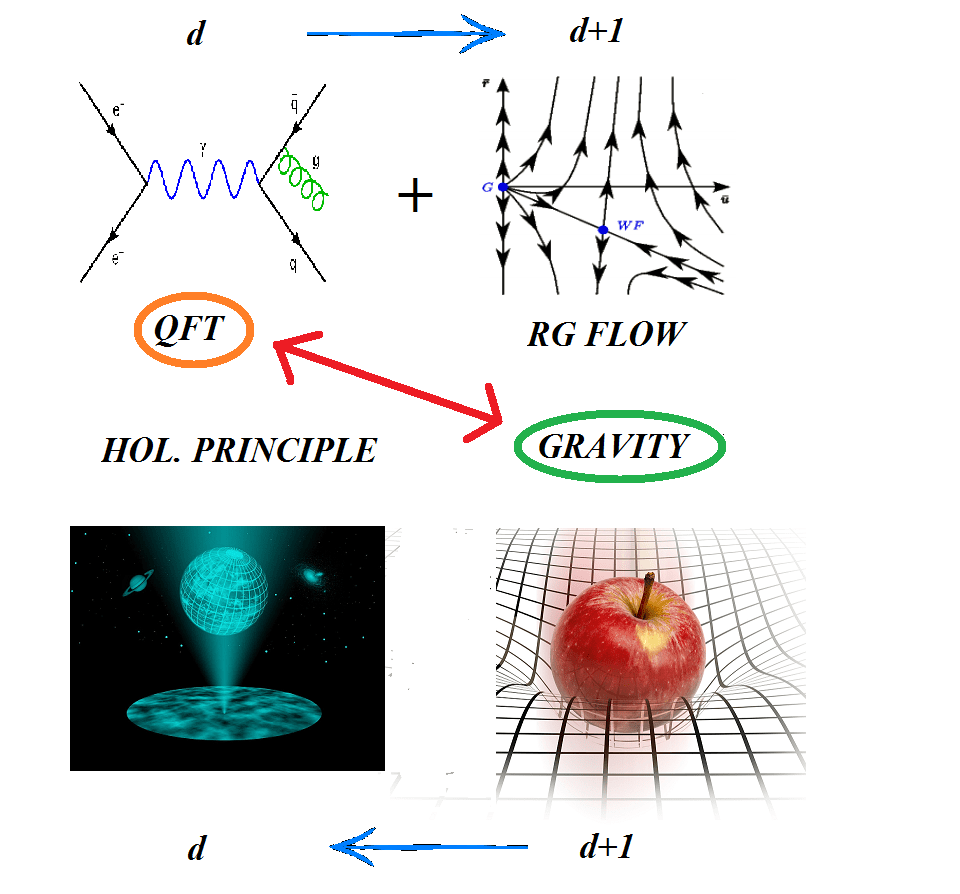}
    \caption{The net of relations between the $d+1$ dimensional gravity theory and the $d$ dimensional QFT. Figure taken from \cite{Baggioli:2016rdj}.}
    \label{fig:holo}
\end{figure}\\
In summary, we can understand the holographic nature of the duality from both the gravitational and QFT side by using the holographic principle and by thinking about the features of RG flows (see fig.\ref{fig:holo}).
\section{Large $N$ field theories and gravity}
\begin{figure}[h!]
\centering
\includegraphics[width=8.5cm]{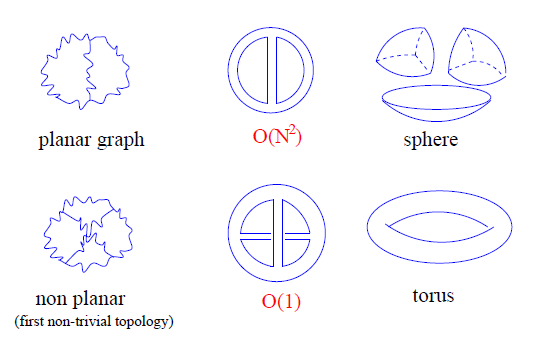}
\caption{Planar and non-planar graphs and their relation with Riemann surfaces. Figure taken from \cite{adscftZaffaroni}.}
\label{planar}
\end{figure}

Physicists are always in search for solvable models. Contrarily to what our intuition might suggest, quantum field theories become much simpler in the limit of very large number of degrees of freedom \cite{largeNQFT}.
G. t'Hooft was the first to realize it in 1974, in the concrete case of $U(N)$ gauge theories.  These models simplify extremely by taking  the rank of the gauge group $N$ to be large, $N\rightarrow \infty$ \cite{thooftN}. In this limit, several theories become solvable while the subleading effects can be taken into account in a perturbative expansion in $1/N$. This is for example a strong tool to try to solve QCD. The question whether the $N=3$ of QCD is large or not is subtle and will not be considered here.\\
The purpose of this section is to briefly show this idea and why it is relevant for us. 
To do that, we consider a $U(N)$ gauge theory defined by the following Lagrangian:
\begin{equation}
\mathcal{L}\,=\,Tr\,\left(F_{\mu\nu}^2\,+\,\mathcal{L}_{matter}\,\right)
\end{equation}
with $F_{\mu\nu}=\partial_\mu A_\nu-\partial_\nu A_\mu+i \,g_{YM}\,[A_\mu,A_\nu]$ the non-abelian field strength and $\mathcal{L}_{matter}$ the matter lagrangian, which generically includes fields in the fundamental and adjoint representations.\\
The main discovery by t'Hooft is that this theory substantially simplifies in the so-called \textit{t'Hooft limit}:
\begin{equation}
N \,\rightarrow\, \infty\,,\qquad g_{YM}\,\rightarrow\,0\,,\qquad \lambda\,=\,g_{YM}^2\, N\,=\,\text{fixed}
\end{equation}
where $\lambda$ is the \textit{t'Hooft parameter}. For many more details see \cite{Lucini:2012gg}. The \textit{t'Hooft limit} is a well defined limit and gives rise to a sensible perturbative framework.\\
For convenience, we re-define the fields in order to write the Lagrangian as :
\begin{equation}
\Longrightarrow\,\mathcal{L}\,=\,\frac{N}{\lambda}\,Tr\,\left(\,F_{\mu\nu}^2\,+\,\dots\,\right)
\end{equation}
Using this notation, the propagator brings a $\lambda/N$, vertices provide a factor $N/\lambda$ and loops a factor of $N$. At this stage, diagrams with different topology contribute with different powers of $N$. In particular, in the large $N$ limit only the \textit{planar} diagrams will survive. Have a look at figure \ref{planar}.
\\
The first diagram is planar; it can be drawn on a plane. The correspondent Riemann surface is a sphere, a topological object with genus $h=0$. The second graph is instead non planar. We cannot draw it on a plane, some of its lines will intersect in points which are not vertices. The corresponding surface is a Torus with genus $h=1$. At the same time, counting the propagators, vertices and lines we can convince ourselves that the first graph scales like $\sim N^2$, while the second one like $\sim N^0$. Therefore, in the large $N$ limit, the contribution from the second graph will be completely negligible with respect to the first one.\\
We can pursue a more systematical classification using topology, a fundamental branch of mathematics. More precisely given a Riemann surface with $F$ faces, $E$ edges and $V$ vertices, its Euler characteristic is given by:
\begin{equation}
2\,-2\,h\,=\,F\,-\,V\,+\,E
\end{equation}
At this point, we can just count the contributions for a given diagram:
\begin{equation}
    \underbrace{\left(\frac{\lambda}{N}\right)^V}_{vertices}\, \,\underbrace{N^F}_{loops}\,\, \underbrace{\left(\frac{N}{\lambda}\right)^E}_{propagators}\,=\,\mathcal{O}\left(N^{2\,-\,2\,h}\right)
\end{equation}
This means that the t'Hooft expansion organizes graphs accordingly to their topology. In particular, in the large $N$ limit the leading diagrams are those with $h=0$, \textit{i.e.} the planar -- all those which can projected onto a sphere. All in all, in this limit, we can write the free energy of the theory in a particularly simple form:
\begin{equation}
\mathcal{F}\,=\,\sum_{h\,=\,0}^\infty\,N^{2\,-\,2\-h}\,f_h(\lambda)
\label{largeNexpansion}
\end{equation}
\textit{i.e.} a double expansion in the rank $N$ and the coupling $\lambda$.\\
Why we are doing all of this? Because in String Theory perturbation theory is not done in powers of the coupling $g_s$ but indeed in the topology, in what is known as \textit{genus expansion}. See \cite{polchinski2001string,zwiebach2004first} for details about it. \\

The most striking result of this section is the incredible similarity between the perturbative string expansion\footnote{Which we did not show in details here. See for example the lecture by A.Uranga available at \url{https://members.ift.uam-csic.es/auranga/lect2.pdf}.} and the $1/N$ expansion of a Gauge theory. \eqref{largeNexpansion}. The two can be mapped into each other by the simple relation
\begin{equation}
N\,\sim\,g_s^{-1}\,,\qquad \alpha'\,\longleftrightarrow\,\lambda
\end{equation}
where $g_s$ is the string coupling and $\alpha'$ relates to the string tension.
This discussion brings us directly to the next section. After realizing that a quantum field theory can be surprisingly related to a theory of gravity with strings, we have to understand in which limits this duality becomes simple and exploitable.
\section{The assumptions: strengths and limits}
\begin{figure}[h!]
    \centering
    \includegraphics[width=0.47\linewidth]{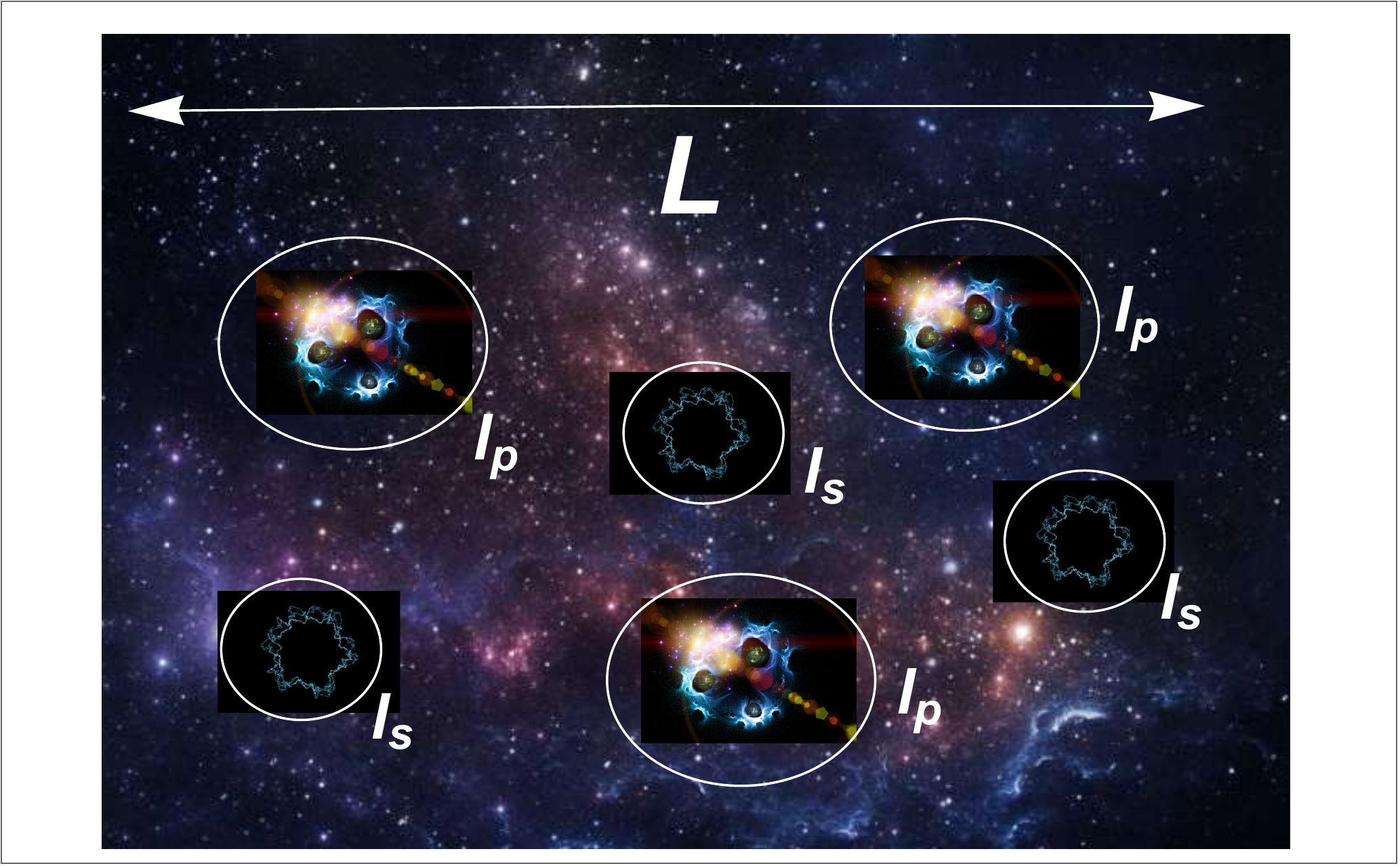}
    \includegraphics[width=0.47\linewidth]{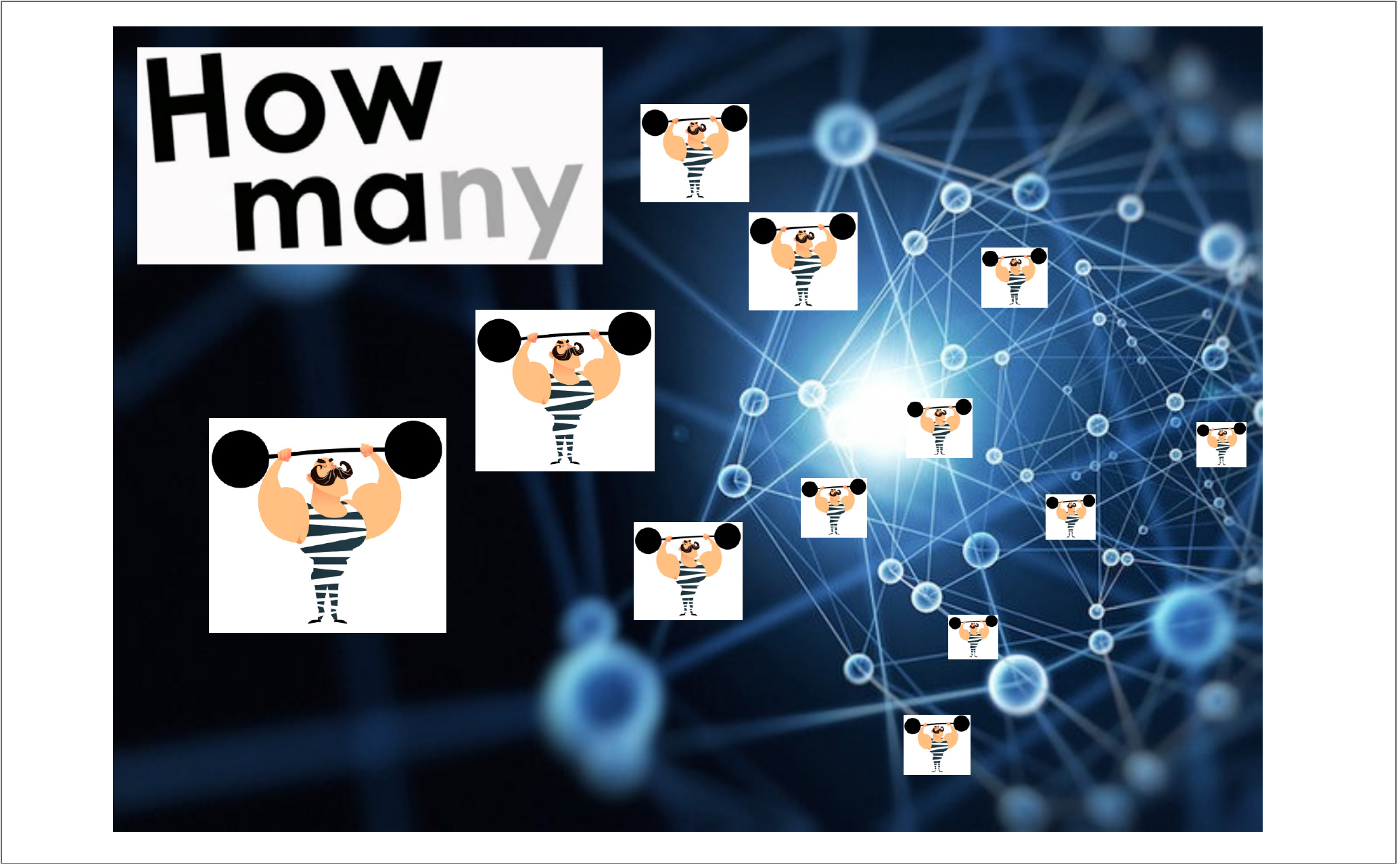}
    \caption{The limit of classical gravity with no extended objects and the dual field theory with many d.o.f.s strongly interacting between themselves.}
    \label{fig:limits}
\end{figure}

Let us start by the simple and honest statement that we do not know how to solve String Theory and we do not know how Quantum Gravity works (so far). As a matter of fact, we would like to work with classical gravity without strings vibrating around us. This is a problem that we know how to control. The quantum effects in gravity are governed by a length scale known as the Planck scale $l_p$, while the stringy effects are determined by the length of these stringy objects, $l_s$. The limit we would like to work in, what we denote as the \textit{classical gravity} limit, can be therefore defined by the two requirements:
\begin{equation}
    \frac{L}{l_p}\,\gg 1\,,\quad \frac{L}{l_s}\,\gg\,1
\end{equation}
where $L$ is the characteristic length of our theory. This is the limit displayed in fig.\ref{fig:limits}. On the other side, the field theory one, we have two independent parameters, the ''number of degrees of freedom'' $N$ and the coupling $\lambda$. Precise relations between the parameters on the gravity side and the field theory can be derived (we did derive one of them in the previous section) as follows:\\[0.15cm]
\noindent
\fcolorbox{black}{black!15}{\parbox{\dimexpr\textwidth-2\fboxsep-2\fboxrule}{%
  \color{red!50!black}%
\begin{center}
\begin{tabular}{ l|l }
\textbf{Field Theory} & \textbf{Gravity Theory} \\ \hline \\
  gauge symmetry rank $N$ & \,\, Planck Scale $l_p$ \\[0.2cm]
  t'Hooft coupling $\lambda\equiv N g^2$ & \,\,String Length $l_s$
\end{tabular}
\end{center}
\begin{equation}
    \left(\frac{L}{l_s}\right)^4\,=\,\lambda \quad \quad \left(\frac{l_p}{L}\right)^8\,=\frac{\pi^4}{2\,N^2}
\end{equation}}}\vspace{0.15cm}
Here comes the miracle. The simple limit of classical and string-less gravity corresponds in the field theory to consider the regime of large number of degrees of freedom $N\gg 1$ and strong coupling $\lambda \gg 1$. The second outcome is certainly more than welcome. As described in the introduction, field theory at strong coupling are very wild beasts that cannot be solved using perturbative methods. We can now use simple classical gravity to describe them. What about this large $N$ limit?
\\
If we think about $SU(N)$ QCD-like theories, it is very well defined what the large $N$ limit is. Then, it is a question of taste to judge if the corrections in real-world QCD:
\begin{equation}
    \frac{1}{N^2}\Big|_{N \rightarrow 3}\,=\,\frac{1}{9}
\end{equation}
are large or small, and if the large $N$ limit is satisfactory or not for QCD physics. That said, it is also fair to say that a great effort has been made to study the holographic results with finite $1/N$ corrections\footnote{Too many references here, sorry.}.\\
In Condensed Matter the situation is more subtle, because one could ask directly what is $N$ in those cases. In terms of vector $O(N)$ theories like spin models, the interpretation of $N$ is still solid. But what in general? I think the best way to answer this question is to remember that given a quantum field theory with a gauge symmetry of rank $N$ the central charge of the theory scales like $c \sim N^2$. The central charge counts directly the number of ''active'' degrees of freedom\footnote{An easy way to understand this is by noticing that:
\begin{equation}
    \langle T_{\mu\nu} T_{\rho\sigma} \rangle \,\sim\,c\,+\,\dots
\end{equation}
meaning that the central charge controls the two-point function of the stress tensor. The latter is indeed the quantity associated to the transport of energy and momentum in the field theory and it is therefore counting the numbers of those ''active'' degrees of freedom. For details have a look at \url{http://www.damtp.cam.ac.uk/user/tong/string/four.pdf}. We thank Amadeo Jimenez Alba for this comment.}. Putting these arguments together, we realize immediately that:
\begin{equation}
    N\,\rightarrow \infty\,\,\Longleftrightarrow\,\,\text{large number of d.o.f.}
\end{equation}
and that the large $N$ limit simply corresponds to considering a system with a very large number of degrees of freedom, a strongly interacting many-body system.
In this sense, this limit can be thought as a sort of mean-field theory approximation which selects the saddle point of the path integral.\\
The relevant question to ask is to which extent and in which terms the large $N$ approximation affects the physical results which we obtain. In this direction, I would like just to show you a simple example where large $N$ has a very important role.\\
The holographic superconductor \cite{Hartnoll:2008vx,Hartnoll:2008kx} is probably one of the most famous models in applied holography. At this level, we are not interested in the details of the model but the only thing we have to know is that it implement the spontaneous symmetry breaking of a $U(1)$ global symmetry\footnote{Yes, it is a superfluid in this sense. Unless you do some tricks \cite{Domenech:2010nf}.} in the dual field theory picture. The simplest setup is constructed in an asymptotic AdS$_4$ spacetime, which corresponds to considering the dual field theory in $D=2$ spatial dimensions. Why is this a problem? Because in statistical physics there is a very famous theorem, which goes under the name of \textit{Mermin-Wagner theorem}, which says that continuous symmetries cannot be spontaneously broken at finite temperature in dimensions $D \leq 2$. What is happening here? First, notice that this theorem is not violated by the Ising model example in section \ref{skipskip} because there the symmetry is discrete, \textit{i.e.} $\mathbb{Z}_2$. In this case, the symmetry is continuous and the theorem holds. Are holographic superconductors violating the theorem? Why?\\
In order to understand this point, we need to say few words about the theorem \cite{PhysRevLett.17.1133,Coleman1973,Merkl1994}. The main idea is that in $D \leq 2$ the quantum fluctuations of the Goldstone bosons are strong enough to destroy the ordered phase, and in particular the coherence of the condensate. Let us try to make this statement slightly more formal following \cite{PhysRevD.11.1701}. Consider a simple theory where a complex scalar operator $\Psi$ acquires dynamically a finite vacuum expectation value (VEV),  $v$. This can easily be achieved using the famous double well potential. The solution with the finite VEV induces the spontaneous symmetry breaking (SSB) of the $U(1)$ symmetry, under which the scalar is charged. Let us consider the fluctuations around that ground state by writing:
\begin{equation}
    \Psi(x)\,=\,\left(v\,+\,\sigma(x)\,\right)\,e^{i\,\theta(x)}
\end{equation}
where $\theta$ is the Goldstone mode associated to the SSB. Considering the fluctuations around the ground state, the VEV of field, which determines the order of the state, can be written as (see \cite{PhysRevD.11.1701} for details):
\begin{equation}
    \langle \Psi(x) \rangle \,\sim\,v\,e^{-\frac{1}{2}\langle \theta(0) \theta(0) \rangle}
\end{equation}
Moreover, the correlation function of the Goldstone bosons can be computed from the simple integral:
\begin{equation}
    \langle \theta(x) \theta(0) \rangle\,\sim\,\int \frac{d^Dk}{(2\pi)^D}\,\frac{e^{i\,k\,x}}{k^2}
\end{equation}
It is easy to prove, just by dimensional analysis, that this correlation function is IR divergent for $D \leq 2$. This means that in $D \leq 2$ :
\begin{equation}
    \lim_{x \rightarrow 0}  \langle \theta(x) \theta(0) \rangle\,=\,\infty\,\,\longrightarrow \,\, \langle \Psi(x) \rangle\,=\,0
\end{equation}
or in words, the quantum fluctuations of the Goldstone bosons destroy the order of the broken phase.\\
From this simple argument, we can immediately realize two facts. First, in the case of discrete symmetries this is not happening because there are no Goldstone bosons. Second, the SSB is destroyed because of quantum effects; nothing can be observed in the classical limit. Now, recall that in the large $N$ limit the quantum effects are suppressed by $1/N$ factors. This limit makes somehow the system classical. To be precise here we should write:
\begin{equation}
    \langle \Psi(x) \rangle \,\sim\,v\,e^{-\frac{1}{2}\frac{1}{N}\langle \theta(0) \theta(0) \rangle}
\end{equation}
such that in the $N \rightarrow \infty$ limit the quantum fluctuations are totally frozen. This is exactly how the Mermin-Wagner theorem fails in the large $N$ limit and it is exactly the reason why we have SSB in $D=2$ in the case of holographic superconductors.\\
This has been proven explicitly in \cite{Anninos:2010sq}, by considering $1/N$ effects in the holographic superconductor model. There, the authors computed the first loop correction to the large $N$ limit and they observed indeed that in $D=2$ the SSB pattern gets completely spoiled, in agreement with the Mermin-Wagner theorem.\\

In summary, large $N$ is a very useful limit but we need to be careful about the results that it spits out. It is certainly a concern and an open question to understand to which extent holography in the large $N$ limit could describe strongly correlated materials, for which the quantum effects are of primary relevance. Potential problems, such as the absence of disorder driven (Anderson) localization phenomena \cite{Grozdanov:2015qia}, have been already discovered.
\section{Why is it actually useful or needed?}
\begin{figure}[h!]
    \centering
    \includegraphics[width=0.7\linewidth]{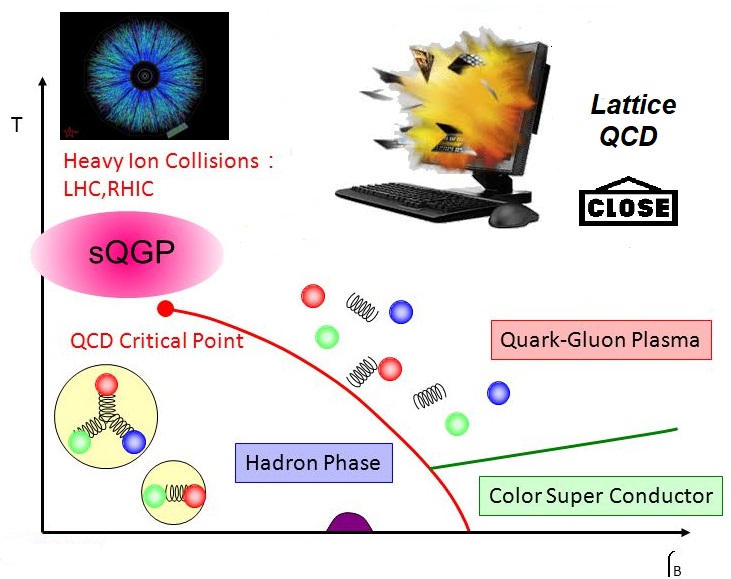}
    \caption{The phase diagram of QCD.}
    \label{fig:QCD}
\end{figure}

\begin{figure}
    \centering
    \includegraphics[width=0.8\linewidth]{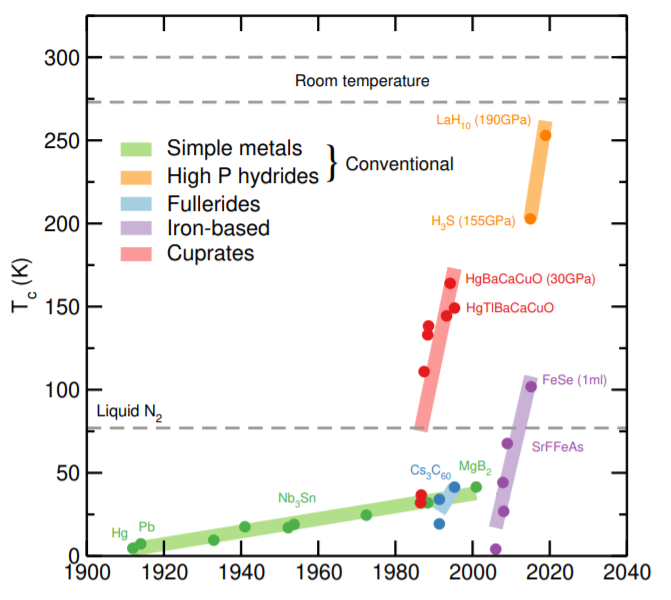}
    \caption{The High-Tc superconductors timeline. In more recent years, the record of $250K$ has been set \cite{Drozdov2019} in LaH$_{10}$ at a pressure of about $190$ Giga-Pascals! The figure is taken from \cite{Pickard_2019}.}
    \label{map}
\end{figure}

After outlining the features, the virtues and the flaws of the holographic tool set, let us briefly motivate a bit more specifically which type of situations or systems can be tackled with it. The two examples that you will find everywhere, and I really mean everywhere, are:
\begin{enumerate}
    \item QCD and Quark Gluon Plasma ;
    \item High-Tc superconductors, Strange Metals and the cuprates.
\end{enumerate}
\begin{figure}[h!]
    \centering
    \includegraphics[width=0.4\linewidth]{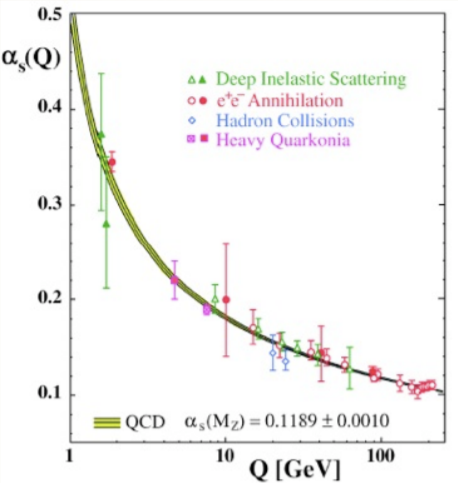}\quad
\includegraphics[width=0.545 \linewidth]{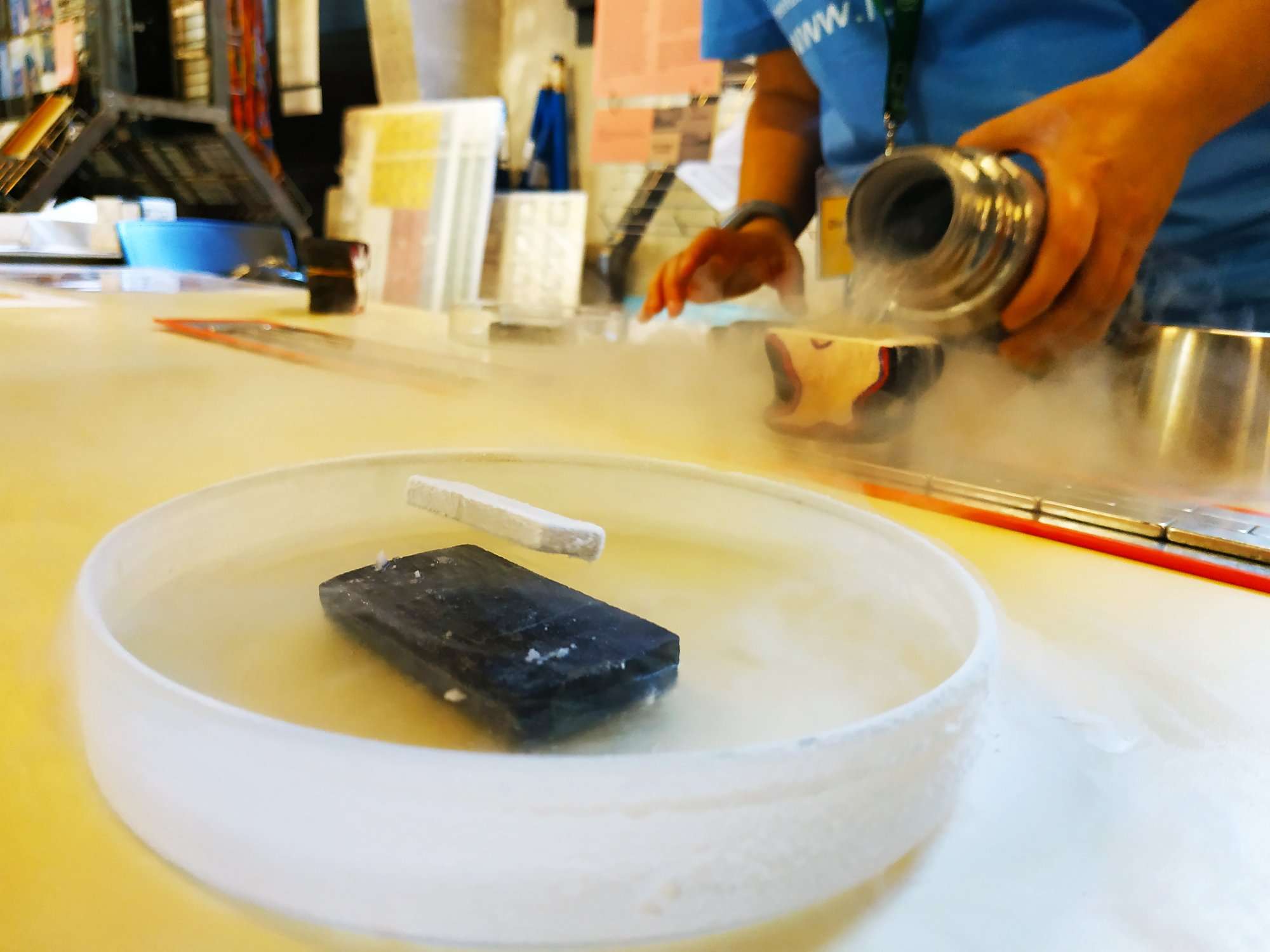}
    \caption{Two typical examples of physics at strong coupling: QCD and High-Temperature Superconductivity. \textbf{Left: }The running of the QCD coupling. \textbf{Right: }The first time I saw with my eyes a High-Tc superconductor at work. Picture taken at the Madrid Researchers' Night 2019.}
    \label{fig:strong}
\end{figure}
Let us try to explain a bit why these situations are very complicated and interesting, and why Holography could provide some help in understanding them.\\[0.2cm]
What is the Quark Gluon Plasma (QGP)? It is the state of matter that you find in QCD at very large temperature and/or density \cite{Sarkar:2010zza} (see fig.\ref{fig:QCD}). At these extremely large energies, the quarks are not anymore confined but they become free, together with the gluons, and they constitute a hot quantum soup. This new quantum state of matter can be observed experimentally in heavy-ion collisions, for example at RHIC (\url{https://www.bnl.gov/rhic/}) or at CERN in the ALICE experiment (\url{https://home.cern/science/experiments/alice}). Why is it so interesting? Because it is believe to be the state of matter of the first milliseconds of our Universe after the Big-Bang. Additionally, it is expected to be at work in very hot Neutron stars. There are several fundamental questions (elliptic flow, jet quenching, small viscosity, etc $\dots$) linked to the QGP; we refer to \cite{CasalderreySolana:2011us} for an extensive review. Why is the QGP so hard to tackle with standard techniques? \MB{HERE} At high energy the QCD coupling should be small and therefore the physics should be simple. Unfortunately, in the QGP regime the coupling is still order $\mathcal{O}(1)$ and perturbative methods are totally unreliable. The only tool left is Lattice QCD \cite{Gupta:1997nd}, \textit{i.e.} brute-force computer calculations. In principle, this is a reliable quantitative tool but: (I)  lattice-regularized calculations are CPU-expensive especially approaching the thermodynamic limit of very large lattice size; this means that all the features of the QGP which are related to long-wavelength physics are very hard to be explored with this method. For the same reason,
it is in practice very difficult to carry out lattice calculations using light quark masses which  yield realistically light pion masses. (II) Apart from this practical issues, there are two more fundamental and conceptual problems. The first one is the famous \textit{sign problem} which renders any Monte Carlo simulation impossible at finite baryon chemical potential. The action acquires an imaginary term and it can no longer be interpreted as a probability distribution in the simulation method. (III) Finally, lattice computations are not very suitable to considering real time properties such as calculating transport coefficients and answering questions about, say, far-from-equilibrium dynamics or jet quenching. Despite existing critiques \cite{Csaki:2008dt}, AdS-CFT represents a well-suitable tool to tackle this kind of system. We will mention in the next sections some of the major results in this direction.
\\[0.3cm]
Another, always mentioned, case is that of High-Tc superconductors. What are they? Standard superconductivity can be understood theoretically using BCS theory \cite{bogoliuubov1963theory} which led Bardeen, Cooper and Schrieffer to the Nobel Prize in Physics in $1972$. The main idea is that the electrons in a metal get paired via phonons interactions and they create bosonic bound states, known as Cooper pairs. Once these bosonic pairs are created, they can undergo condensation and produce a ground state with a macroscopic occupation number. All the experimentally observed properties of Superconductivity can be explained by this simple framework \cite{tinkham2004introduction}. Well, actually all the experimentally observed properties before $1986$ \cite{1986ZPhyB..64..189B}! In that year, the first High-Tc superconductor \cite{Leggett2006} was discovered in the IBM laboratories. Since then, a lot of very diverse materials displaying High-Tc superconductivity have been discovered; see fig.\ref{map} for an historical map.
\\
Why do these new materials constitute a challenge to the standard condensed matter wisdom?
\begin{enumerate}
    \item The BCS theory is expected to be reliable only for $T_c \,<\,30\,K$. This is certainly not the case for High-Tc superconductors, see fig.\ref{map}. The naive argument to understand this problem is realizing that the critical temperature in BCS theory is proportional to the electron-phonon coupling. Large $T_c$ means that such a coupling becomes strong and therefore the weakly coupled perturbative logic assumed not reliable \footnote{More recently, superconductivity at very high temperature ($\sim 250 K$) as been obtained for certain materials under extreme pressures \cite{Drozdov2019}. It is not yet clear if such phenomenon can be understood within BCS theory and from a standard electron-phonon interaction \cite{song2017phonon}. We thank A.Garcia Garcia for pointing this out.}.
    \item The BCS theory is built upon the superconducting instability of a normal metallic phase which follows Fermi Liquid theory. The normal states of the High-Tc superconductors are in contrast with Fermi liquid theory \cite{Cooper603}. They are usually referred to as \textit{Strange metals}\footnote{Be careful. There is another class of ''strange'' materials, which are defined \textit{Bad Metals}. The latter violate the Mott-Ioffe-Regel bound \cite{PhysRevLett.74.3253}. A priori, it is not clear if this is connected or not with the $T-$linear resistivity of Strange Metals.}, and they display a very discussed linear in $T$ resistivity\footnote{Fermi liquid theory predicts that in standard metals $\rho \sim T^2$, which comes from simple arguments related to the electron-electron scattering and the phase space in presence of a Fermi surface. There is no known weakly coupled logic or scattering mechanism which is known to give linear in $T$ resistivity $\rho \sim T$.}.
    \item BCS theory predicts a universal relation between the superconducting mass gap $\Delta$ and the critical temperature $T_c$, which is universal and tested in the lab. More precisely:
    \begin{equation}
        2\,\Delta(T=0)\,\sim\,3.54\,k_B\,T_c
    \end{equation}
    High-Tc superconductors give values around\footnote{Suprisingly enough, the holographic models \cite{Hartnoll:2008vx} are quite close to these values.}:
    \begin{equation}
        2\,\Delta(T=0)\,\sim\,9\,k_B\,T_c
    \end{equation}
    which is taken as another indication of the strong coupling physics behind them \footnote{One can modify BCS theory into the so-called Eliashberg theory \cite{hirsch2001electron}. The resulting theory is still a ''boring'' mean-field theory of superconductivity but it can describe materials such as Pb or Hg, for which the ratio $\Delta/T_c$ is higher than the BCS prediction.  Therefore, a non BCS value of the ratio only means that the superconductor is more strongly correlated but not that Fermi liquid theory breaks down as is the case in Cuprates. We thanks A.Garcia Garcia for this comment.}.
    \item The condensate in BCS theory is \textit{S-wave}, which means a spin-0 object; the Cooper pair is a spin-0 composite boson. There are strong indications that High-Tc superconductors are \textit{D-wave}, displaying a more complicated condensate which is indeed a spin-2 tensor.
\end{enumerate}
All in all, High-Tc superconductors cannot be explained by BCS theory and no well accepted substitutive theoretical framework exists so far. Moreover, there are several (experimental and theoretical) indications that strong coupling is playing an important role. Holography might be a viable tool to solve this condensed matter mistery.
\section{The Power: why I like to think about it as ''EFT 2.0''}
\textbf{Disclaimer:} This section contains mostly personal opinions built on the discussions and interactions with colleagues and experts in the topic.\\[0.2cm]
Bottom-up holography is certainly strongly reminiscent of the common effective field theory approach. It is not derived from any microscopic UV theory nor from a formal embedding in string theory or any other quantum gravity framework you can think of. The main ingredients are composite operators, fields whose microscopic structure remains unknown and which are labelled simply by their spin and their conformal dimension (or mass). Every time we discuss about a $U(1)$ current, we do not know if that current comes from some charged scalars:
\begin{equation}
    J^\mu\,=\,i\,q\,\left(\psi \partial^\mu \psi^*\,-\,\psi^* \partial^\mu \psi\right)
\end{equation}
or from some fermions (our ''wanna-be'' electrons):
\begin{equation}
    J^\mu\,=\,\bar{\Psi}\,\gamma^\mu\,\Psi\,;
\end{equation}
it is just a conserved current carrying $U(1)$ charge. This is totally analogous to say the way we formulate the Ginzburg Landau formalism. We write an action in terms of a complex order parameter field $\zeta$, but we do not know what is that made of. In BCS theory that would be a Cooper pair for example. The amazing outcome, in common to all the EFTs, is that we can already answer a lot of relevant questions without knowing it.\\
So, if we do not know any microscopic, how do we construct our theory? Symmetries! That is indeed the holy grail for effective field theories. Symmetries are very powerful constraints on the allowed interactions and on the emerging dynamics. If Holography is just an effective field theory like the others why we are even discussing it? Because there is much more.\\
Effective field theories are built using symmetries and a perturbative expansion in a small parameter. In particle physics it might be the mass dimension of an operator; in hydrodynamic the gradients of the thermodynamic quantities around equilibrium. Let us think about hydrodynamics and elasticity theory as benchmark models. Hydrodynamic is built in a gradient expansion. In issipative relativistic hydrodynamic \cite{Kovtun:2012rj} the stress tensor is expanded in a formal derivatives expansion which looks like:
\begin{equation}
   \underbrace{T^{\mu\nu}}_\text{\parbox{1cm}{\centering stress\\[-4pt] tensor}}\,=\,\underbrace{\epsilon}_\text{\parbox{1cm}{\centering energy\\[-4pt] density}}\,\underbrace{u^u}_{\text{velocity}}\,u^v\,+\,\underbrace{p}_{\text{pressure}}\,\Delta^{\mu\nu}\,-\,\underbrace{\eta}_\text{\parbox{1cm}{\centering shear\\[-4pt] viscosity}}\,\sigma^{\mu\nu}\,-\,\underbrace{\zeta}_\text{\parbox{1cm}{\centering bulk\\[-4pt] viscosity}}\,\Delta^{\mu\nu}\partial^\lambda u_{\lambda}\,+\,\dots
\end{equation}
Elasticity theory \cite{landau1986theory,chaikin2000principles} is constructed via a similar expansion, this time in the strain applied to a material:
\begin{equation}
    \underbrace{\sigma_{ij}}_{\text{stress}}\,=\,\underbrace{K}_\text{\parbox{1cm}{\centering bulk\\[-4pt] modulus}}\,\delta_{ij}\,\epsilon_{ii}\,+\,2\,\underbrace{G}_\text{\parbox{1cm}{\centering shear\\[-4pt] modulus}}\,\left(\underbrace{\epsilon_{ij}}_{\text{strain}}\,-\,\frac{1}{D}\,\delta_{ij}\,\epsilon_{ii}\right)\,+\,\dots
\end{equation}
All the effective field theories contain a  rapidly increasing number of undefined coefficients, which the EFT itself cannot determine. Hydrodynamic theory can tell you that you will observe a shear diffusion mode whose diffusion constant is proportional to the viscosity, but it will never tell you how much is the viscosity of water. Elasticity can provide you with the speeds of propagation of the transverse and longitudinal phonons in terms of the elastic moduli, but it will never give a numerical value for them. In other terms, the structure of the EFT is always of the type:
\begin{equation}
    \text{EFT}\,\equiv\,\sum_i\,c_i\,\mathcal{O}^i
\end{equation}
where $\mathcal{O}^i$ are specific operators with increasing dimension and $c_i$ unknown coefficients, known as \textit{Wilson coefficients}. EFTs are very predictive but without a microscopic description (or experiments of course) you will never get a value for those coefficients. Holography does it for you! Not only it tells you that Einstein-Hilbert gravity on the Schwarzschild background is a relativistic hydrodynamic system at finite temperature, but it tells you also that for such system $\zeta=0$ and $\eta/s=1/4\pi$ and much more \cite{Policastro:2002se,Policastro:2002tn}.\\
There is more. People have been (and still are) struggling for a long time about deriving hydrodynamics from an action principle \cite{Grozdanov:2013dba,Jensen:2018hse,Jensen:2017kzi,Glorioso:2018wxw,Haehl:2017zac,Crossley:2015evo,Haehl:2018lcu,Haehl:2015foa}, Holography gives you the action immediately\footnote{Here we keep a very optimistic view.}:
\begin{equation}
    \mathcal{S}\,=\,\int d^dx\,\sqrt{-g}\,\left[R\,-\,2\,\Lambda\right]\,;
\end{equation}
it is just not written in the usual hydrodynamic variables\footnote{To be fair, also all the recent constructions trying to formulate an action using Keldysh-Schwinger techniques do not write the action in transparent variables which could immediately identified from a physical point of view.}.\\
This last discussion brings us to another point. Effective field theory derived from an action principle (see for example the nice formalism of \cite{Baggioli:2019abx} or the simple EFT for phonons \cite{Leutwyler:1996er}) are by construction unitary. Unitarity makes the introduction of dissipation very hard if not even impossible (see for example \cite{Endlich:2012vt} for a perturbative attempt) \footnote{I have to mention that non-Hermitian models can be a fresh view on this problem \cite{bender2007making}.}. In Holography, this comes for free. Dissipation in the boundary theory is induced naturally because of the presence of a BH object with a dissipative horizon. From the membrane paradigm \cite{1986bhmp.book.....T,PhysRevD.18.3598}, we know that the horizon dynamics is equivalent to that of a dissipative fluid. And surprisingly enough, the description is perfectly unitary (but not local).\\
Moreover, there is another issue with EFT. Effective field theories rely on the existence of a separation of scales, a mass gap, between some light modes, which will constitute the relevant degrees of freedom, and some heavier modes, which can be integrated out because decoupled from the low-energy dynamics \cite{Cohen:2019wxr}. What happens in theories which are scale invariant? Building an EFT for a scale invariant system sounds like an oxymoron, given the fact in such system the spectrum is continuous and there is no separation of scales. Again, holography knows it better and it can provide this effective framework also for situations which are scale invariant. This is crucial to investigate systems at quantum phase transitions or in the quantum critical region, where scale invariance is a well-known and observed feature.\\[0.2cm]
Finally, let us try to compare holography with other standard condensed matter tools. The situation can be summarized by considering Landau Fermi liquid theory and Boltzmann theory. The efficacy and success of Landau Fermi liquid theory is the idea that electrons can be considered free, at the cost of considering them as quasiparticles with renormalized parameters. Quasiparticles can be emergent and collective objects which are very different from the elementary electrons, phonons etc but they have the huge advantage that they can considered free. Their presence can be revealed via scattering experiments, as a well-defined and coherent peak in the spectraul function. What happens if there are no quasiparticles? Are there systems with no quasiparticles? Apparently this is the case and they are very interesting \cite{Hartnoll2014}. The absence of quasiparticles is signaled by the violation of the Mott-Ioffe-Regel bound \cite{hussey2004universality} and it is observed in strongly correlated systems \cite{allen2002quasi}. There are several standard condensed matter frameworks which are immediately in trouble because of this fact. It is sufficient to think about the Landau Fermi Liquid theory or the Boltzmann machinery to compute transport \cite{harris2012introduction}. Transport without quasiparticles is certainly something where Holography can be of incredible help \cite{Hartnoll2014}.\\[0.2cm]
All in all, given this short and personal discussion, I like to promote the role of the holographic tool as EFT 2.0! Its construction and its nature has certainly a lot in common with the standard effective field theory methods, but, as explained in this section, its power and outputs go much beyond it. Most of the flaws of standard EFT techniques are overtaken in an elegant way by Holography. Nevertheless, it is very important to keep in mind its EFT origin when using it.
    \begin{figure}
        \centering
        \includegraphics[width=0.8\linewidth]{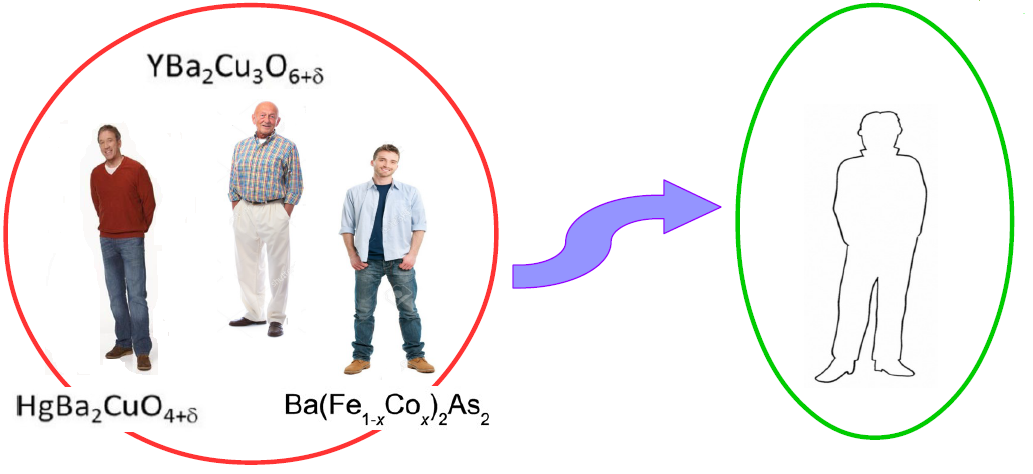}
        \caption{The EFT nature of Bottom-Up Holography. The details of the microscopic are completely washed out and what remains are the universal low energy features. There holography can be your best friend!}
        \label{fig:eft}
    \end{figure}

%
%
%
\chapter{A Practical Understanding of the Dictionary}
\label{intro2} 
\hspace{0.2cm} \includegraphics[width=0.5\textwidth]{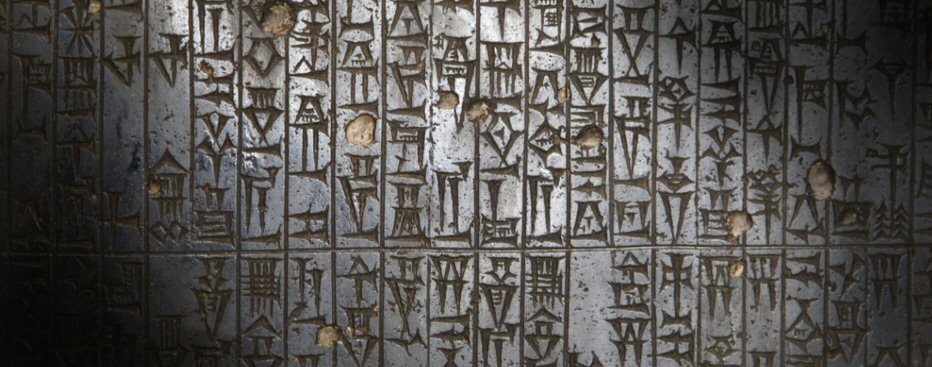}\\
\epigraph{Religion is a culture of faith; science is a culture of doubt.}{\textit{Richard Feynman}}
As already mentioned, holography is a duality between a $d-$dimensional field theory, defined by a series of operators $\mathcal{O}_i$, and a $d+1$ gravitational theory, described by a collection of dynamical fields $\phi^I$ living in a $d+1$ dimensional bulk. The first ingredient we have to learn is how to connect these two sides. This section contains more technical details regarding this bridge, which is usually referred to as \textit{the Dictionary}. The material presented in this section builds on the several reviews cited in the introduction and on some chapters of my own PhD thesis, that you can find here \cite{Baggioli:2016rdj}.
\begin{figure}[h!]
    \centering
    \includegraphics[width=0.7\linewidth]{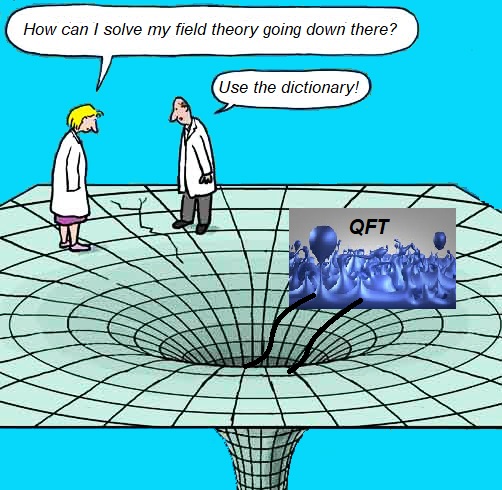}
    \label{fig:dic}
\end{figure}
\section{The bulk and the boundary: how to bring them together}
The dictionary is mostly determined by the map between the operators $\mathcal{O}_i$ of the (not necessarily) conformal field theory and the bulk fields $\Phi_i$.\\
The gravitational theory is defined by a $d+1$ dimensional bulk action:
\begin{equation}
\mathcal{S}_{bulk}^{(d+1)}\,\left(\Phi_i:\,g_{\mu\nu}\,,\,A_{\mu}\,,\,\phi\,,\dots\right)\,=,\int d^{d+1}x\,\sqrt{-g}\,\left[R\,-\,\frac{1}{4}\,F^2\,-\,\frac{1}{2}\,\partial_\mu \phi \partial^\mu \phi\,+\,\dots\,\right]
\end{equation}
where the bulk fields $\Phi^I$ can have arbitrary mass terms $m^I$, spin $J^I$ and interaction terms. Moreover, the fields $\Phi^I$ might be charged under various bulk symmetries.\\
In the weakly coupled classical limit\footnote{As already explained, this coincides with the large $N$ and strong coupling limit of the dual field theory. For the rest of these lecture,s only this case will be considered.}, this is simply the theory of classical fields on a dynamical curved spacetime. All the dynamics will involve only the classical equations of motion for these fields and the metric $g_{\mu\nu}$, and all the stringy and quantum effects will be totally neglected. In simple words, the only knowledge you need to attack these computations is classical General Relativity.\\
For simplicity, we will present the dictionary in the simplest and original case of a conformal field theory. This is not a necessary assumption and the dictionary can be generalized for very general and even non-relativistic field theories (see for example \cite{Chemissany:2014xsa}).
Under these assumptions, the (conformal) field theory\footnote{See \cite{blumenhagen2009introduction,gaberdiel2000introduction} for much more information about conformal field theories.} side is defined by a collections of \textit{operators}
\begin{equation}
\mathcal{L}_{CFT}\,\equiv\,\sum_i\,c_i\,\mathcal{O}_i
\end{equation}
The important point is to understand how to relate the bulk fields and the field theory operators:
\begin{equation}
    \underbrace{\phi^I(r,t,x^i)}_{d+1\,-\,\text{bulk}}\quad\longleftrightarrow\quad \underbrace{\mathcal{O}^I(t,x^i)}_{d\,-\,\text{field theory}} 
\end{equation}
The first step to understand this point is to write down a deformation on the CFT side, which is represented by an operator $\mathcal{O}$ and its correspondent source $\phi_0$:
\begin{equation}
\mathcal{L}\quad \rightarrow \quad \mathcal{L}_{CFT}\,+\,\int\,d^dx\,\phi_0\,\mathcal{O}
\end{equation}
Once we introduce sources, we are in the position of computing the following object:
\begin{equation}
e^{W(\phi_0)}\,=\,\langle\,e^{\int\,\phi_0\,\mathcal{O}}\,\rangle_{QFT}
\end{equation}
where $W(\phi_0)$ represents the generator functional for all the correlation functions of the operator $\mathcal{O}$.\\
More specifically, the $n-$points function of the operator $\mathcal{O}$ can be obtained by acting $n$ times with the functional derivatives of $W(\phi_0)$ with respect to the source $\phi_0$:
\begin{equation}
\langle\,\underbrace{\mathcal{O}\,\dots\,\mathcal{O}}_{1\,,\,\dots\,,\,n}\,\rangle\,=\,\frac{\delta^n\,W}{\delta\,\phi_0^n}\Big|_{\phi_0=0} \label{stan}
\end{equation}
and then setting the source to zero. Using the recipe of \eqref{stan}, we can derive all the desired physical observables of our field theory. We just need to learn how to compute \eqref{stan} from the gravity point of view.\\
Here comes the most fundamental concept of the dictionary, which is known as the GPKW (Gubser, Polyakov, Klebanov, Witten) master rule \cite{Witten1,GPKW1}. The functional generator of the field theory can be simply obtained by computing the on-shell $d+1$ dimensional gravitational action. More precisely, the following identification holds:
\begin{equation}
\boxed{e^{W(\phi_0(x))}\,=\,\Big\langle\,e^{\int\,\phi_0(x)\,\mathcal{O}}\,\Big\rangle_{QFT}\,\textbf{=}\,e^{\mathcal{S}_{\text{bulk}}[\phi(x,r)]}\,=\,\mathcal{Z}_{gravity}\left[\phi_0(x)\equiv \phi(x,r)_{\text{boundary}}\,\,\right]}
\end{equation}
The source for the operator $\mathcal{O}$ is identified with the boundary value\footnote{We will be more precise in the following.} of the dual bulk field $\phi$. The on-shell action is going to be written in terms of such boundary value and then the functional derivative prescription is straightforward\footnote{Not really! There are several things one has to be careful about it. Coming soon!}. Let us proceed step by step. The first thing to learn is how to select the couple of dual objects $\{\phi_0,\mathcal{O}\}$. The answer is in the symmetries! In more detail, the operator of the field theory and its dual bulk field need to have the same quantum numbers according to the $O(2,d-1)$ group. For example, we can write down
\begin{equation}
\mathcal{L}_{CFT}\,+\,\int\,d^dx\,\sqrt{g}\,\left(g_{\mu\nu}\,T^{\mu\nu}\,+\,A_\mu\,J^\mu\,+\,\phi\,\mathcal{O}\,+\dots\right)
\end{equation}
where the dual pairs are\\[0.15cm]
\noindent
\fcolorbox{black}{black!15}{\parbox{\dimexpr\textwidth-2\fboxsep-2\fboxrule}{%
  \color{red!50!black}%
\begin{equation}
\centering
\begin{tabular}{|c|c|}
\hline 
\textbf{Field theory operator} & \textbf{Bulk field} \\
\hline
\text{Stress tensor} $T^{\mu\nu}$ & bulk graviton $g^{\mu\nu}$ \\
\text{U(1) current} $J^{\mu}$ & bulk gauge field $A^{\mu}$ \\
\text{scalar operator} $\mathcal{O}$ & bulk scalar $\phi$ \\
\text{Antisymmetric two-form current} $J^{\mu\nu}$ & bulk two forms $B^{\mu\nu}$ \\
\dots & \dots \\
\hline
\end{tabular}
\end{equation}}}\vspace{0.15cm}
Obviously, this is just the step zero of the dictionary. We do not intend to spend more time on this. Most of the properties of the dictionary will be encountered in the next sections by looking at concrete examples. For more details, we always refer to more comprehensive review about the AdS-CFT correspondence.
\section{$\Lambda$CFT, a scalar in AdS: renormalization and correlators (duty first)}
Let us explain the most important features of the AdS-CFT dictionary by considering the simplest scenario possible: a scalar bulk field in AdS.\\
We consider the following background metric:
\begin{equation}
ds^2\,=\,\frac{L^2}{z^2}\,\left(\,dz^2\,-\,dt^2\,+\,d\vec{x}^2\,\right)\label{pp}
\end{equation}
which parametrizes the AdS $d+1$ spacetime with holographic radial coordinate $z$. Importantly, the AdS boundary is located at $z=0$ \footnote{To be precise that is a conformal boundary. Fixing $z=0$ in \eqref{pp}, we obtain a metric which is only conformal to Minkowski $ds^2\,=\,e^{2\,\sigma(z)}\,ds^2_{Mink}$.}.\\
On top of this geometry, we consider the action for a massive scalar field given by:
\begin{equation}
\mathcal{S}\,=\,-\,\frac{1}{2}\,\int\,d^{d+1}x\,\sqrt{-g}\,\left[\,g^{MN}\,\partial_M\,\phi\,\partial_N\,\phi\,+\,m^2\,\phi^2\right]\label{actionex}
\end{equation}
It is a simple exercise to obtain the corresponding equations of motion
\begin{equation}
\frac{1}{\sqrt{-g}}\,\partial_M\,\left(\sqrt{-g}\,g^{MN}\,\partial_M\,\phi\,\right)\,-\,m^2\,\phi\,=\,0\,.
\end{equation}
and write them explicitly on the geometric background defined in \eqref{pp}, obtaining
\begin{equation}
z^{d+1}\,\partial_z\,\left(\,z^{1-d}\,\partial_z\,\phi\,\right)\,+\,z^2\,\delta^{\mu\nu}\,\partial_\mu\,\partial_\nu\,\phi\,-\,m^2\,L^2\,\phi\,=\,0\,.\label{close}
\end{equation}
It is now convenient to Fourier transform the scalar field with respect to the boundary coordinates $x^\mu=(t,\vec{x})$ :
\begin{equation}
\phi(z,x^\mu)\,=\,\int\,\frac{d^dk}{(2\,\pi)^d}\,e^{i\,k\,\cdot\,x}\,f_k(z)
\end{equation}
We can solve the equation \eqref{close} asymptotically, close to the UV boundary $z=0$ , by using a power law ansatz $f_k(z)\sim z^\beta$ for the scalar field. Doing that, we obtain the simple indicial equation:
\begin{equation}
\beta\,(\beta\,-\,d)\,-\,m^2\,L^2\,=\,0\,.
\end{equation}
which fixes the asymptotic powers in terms of the bulk mass of the scalar field as:
\begin{equation}
\beta_{\pm}\,=\,\frac{d}{2}\,\pm\,\sqrt{\frac{d^2}{4}\,+\,m^2\,L^2}
\end{equation}
Collecting the various steps, we obtain the generic solution for the scalar field close to the boundary, which reads:
\begin{equation}
\phi(x,z)\,\sim\,A(x)\,z^{d\,-\,\Delta}\,\left(1\,+\,\dots\right)+\,B(x)\,z^\Delta\,\left(1\,+\,\dots\right)
\label{UVscalarexp}
\end{equation}
where we have defined:
\begin{equation}
\Delta\,=\,\beta_+\,=\,\frac{d}{2}\,+\,\nu\,,\qquad
\nu\,=\,\sqrt{\frac{d^2}{4}\,+\,m^2\,L^2}
\end{equation}
The expansion of any bulk field is always of this type, more precisely:
\begin{equation}
    \Phi(z,t,\vec{x})\,=\,\textit{leading}(t,\vec{x})\,z^{\Delta_L}\,\left(1\,+\,\dots\right)\,+\,\textit{subleading}(t,\vec{x})\,z^{\Delta_S}\,\left(1\,+\,\dots\right)\label{cc}
\end{equation}
where the powers $\Delta_{L,S}$ are obtained from the equation of motion for the bulk field solved close to the AdS boundary. Assuming $\Delta_L<\Delta_S$, the term we defined (not surprisingly) \textit{leading} is the dominant one close to the UV boundary (in this case $z=0$), while the other term is usually called the \textit{subleading}. Notice that between these two powers there might be other terms in the expansion. Those terms are not independent and can be written down in a recursive way in terms of the undetermined coefficients $\textit{leading},\textit{subleading}$. These two coefficients are the two independent integration constants, which follow from the 2$^{nd}$ order differential equation in the bulk.\\
The holographic dictionary\footnote{In this course we will only consider the standard quantization procedure. In the case both the leading and subleading terms in the expansion \eqref{cc} correspond to ''normalizable'' contributions, an alternative quantization scheme can be used (see \cite{Marolf:2006nd} for details about it). This alternative scheme is connected to a double-trace deformation of the boundary theory \cite{Witten:2001ua,Witten:2003ya,Yee:2004ju} and it has nice physical interpretations \cite{Andrade:2013wsa,Jokela:2013hta,Aronsson:2017dgf,Baggioli:2019aqf,Armas:2019sbe}.} tells us\footnote{We will see in a while why this is the case. We do not have to believe it. It is a dictionary, not a bible!} that the leading term in such expansion corresponds to the source of the dual operator $\mathcal{O}$. In the literature, the source for the dual operator is usually indicated with $\phi_0(x)$ and it is formally defined:
\begin{equation}
\phi_0(x)\,\equiv\,\lim_{z\rightarrow 0}\,z^{\Delta-d}\,\phi(z,x)
\end{equation}
where $\Delta$ is the dimension of the operator $\mathcal{O}$.\\
Now, let us consider the original action \eqref{actionex}, together with the asymptotic expansion of our bulk field \eqref{UVscalarexp}. After some algebra, and opportune renormalization maneuvers \cite{Skenderis1}, the on-shell action\footnote{Corresponding to assume the equations of motion for the scalar.} reduces to a boundary term of the form:
\begin{equation}
\mathcal{S}_{bdy}\,\sim\,\int\,d^dx\,\sqrt{-\gamma}\,A(x)\,B(x)\,=\,\int\,d^dx\,\sqrt{-\gamma}\,\phi_0(x)\,\mathcal{O}(x)\label{fin}
\end{equation}
where $\gamma$ is the induced metric on the boundary, \textit{i.e.} the fixed geometrical background where the field theory lives in. This proves explicitely that the boundary action becomes of the type:
\begin{equation}
    \mathcal{S}_{bdy}\,\sim\,\int\,d^dx\,\sqrt{-\gamma}\,\,\text{source}\,*\,\text{VEV}
\end{equation}
in which the term $\phi_0\equiv A(x)$ is the source, and $B(x)$ the expectation value $\langle \mathcal{O}\rangle$ of the CFT operator $\mathcal{O}$.\\
At this stage, we can just use the AdS/CFT prescription for the generating functional:
\begin{equation}
\mathcal{Z}_{QFT}\,=\,\Big\langle\,exp\,\left[\int\,\phi_0\,\mathcal{O}\right]\Big\rangle_{QFT}\,=\,\mathcal{Z}_{gravity}[\phi \rightarrow \phi_0]
\end{equation}
where $\mathcal{Z}_{gravity}[\phi \rightarrow \phi_0]$ is the partition function (i.e. the path integral) of the gravitational theory evaluated over all bulk functions which have $\phi_0$ as the asymptotic value at the boundary of AdS:
\begin{equation}
\mathcal{Z}_{gravity}[\phi \rightarrow \phi_0]\,=\,\sum_{\{\phi\rightarrow\phi_0\}}\,e^{\mathcal{S}_{gravity}}
\end{equation}
This looks quite hard, especially because we do not know how to treat quantum gravity. Nevertheless, the limit we took makes gravity classical. This is to say that we can proceed using the saddle-point approximation:
\begin{equation}
\mathcal{Z}_{QFT}\,\approx\,e^{\mathcal{S}_{\text{gravity}}^{\text{on-shell}}[\phi\rightarrow\phi_0]}
\end{equation}
Modulo renormalization problems \cite{Skenderis1}, this is a much simpler task. The generating functional becomes:
\begin{equation}
log\,\mathcal{Z}_{QFT}\,=\,\mathcal{S}_{\text{gravity}}^{\text{ren}}[\phi\rightarrow\phi_0]
\end{equation}
where the label ''ren'' indicates the opportunely normalized boundary action. In the simple case we are considering, this is the final result in eq.\eqref{fin}.
This is the end of the story. All the $n-$point functions of the operator $\mathcal{O}$ can be obtained by computing derivatives with respect to the source $\phi_0$ as:
\begin{equation}
\langle\,\mathcal{O}(x_1)\,\dots\,\mathcal{O}(x_n)\,\rangle\,=\,\frac{\delta^{(n)}\,\mathcal{S}_{\text{gravity}}^{\text{ren}}[\phi]}{\delta\phi_0(x_1)\,\dots\,\delta\phi_0(x_n)}\,\Big|_{\phi_0=0}
\end{equation}
For completeness, let us discuss the case of $2-$points functions in more detail, using \textit{linear response theory} (see for example \cite{Linear1} for more details).\\
In this picture, we can define the $2-$points function (sometimes called also the \textit{correlator}, the \textit{Green function} or in particle physics the \textit{propagator}) as:
\begin{equation}
\langle\,\mathcal{O}(x)\,\rangle_{\phi_0}\,=\,\int\,d^dy\,\mathcal{G}(x\,-\,y)\,\phi_0(y)
\end{equation}
where:
\begin{equation}
    \mathcal{G}(x)\,\equiv\,\langle\,\mathcal{O}(x)\,\mathcal{O}(0)\,\rangle
\end{equation}
This simply means that the (linear) response in the system upon introducing an external source $\phi_0$ is just the convolution of the $2-$points function with the source itself. Later on, we will encounter plenty of concrete physical examples.\\
Now, things simplify consistently if we work in momentum space where the convolution integral becomes a simple product
\begin{equation}
\langle\,\mathcal{O}(k)\,\rangle_{\phi_0}\,=\,\mathcal{G}(k)\,\phi_0(k)
\end{equation}
All in all, we obtain our final expression for the Green Function, which reads
\begin{equation}
\mathcal{G}(k)\,=\,\frac{\langle\,\mathcal{O}(k)\,\rangle_{\phi_0}}{\phi_0(k)}
\end{equation}
and which will be used a lot in the following sections. We can already anticipate that for our scalar example, it follows from eq.\eqref{fin} that the green function is simply:
\begin{equation}
    \mathcal{G}_{\phi}(k)\,=\,\frac{B(k)}{A(k)}
\end{equation}
In the framework of AdS/CFT, computing a 2-points function reduces to:
\begin{equation}
\langle\,\mathcal{O}(k)\,\mathcal{O}(0)\,\rangle\,=\,\mathcal{G}_E(k)\,=\,\lim_{z\rightarrow 0}\,z^{d\,-\,\Delta}\,\frac{\Pi^{ren}(z,k)}{\phi_0(z,k)}
\end{equation}
Before continuing, let us explain another point that so far was mentioned only as a recipe. We said before that $\Delta$ in the expansion \eqref{UVscalarexp} is the conformal dimension of the dual operator $\mathcal{O}$. Do you believe it? No! Let us prove it.\\

Let us introduce a UV cutoff $z=\epsilon$ and let us compute the action on this shifted boundary. Using simple algebra (try please!), we get the following boundary action
\begin{equation}
\mathcal{S}_{bdy}\,\sim\,L^d\,\int\,d^dx\,\phi_0(x)\,\epsilon^{-\Delta}\,\mathcal{O}(\epsilon,x)
\end{equation}
In order to show that $\Delta$ is the conformal dimension of the dual operator, we have to remember how it is defined.  The conformal dimension is simply the eigenvalue of the dilatation operator. Consider a scaling transformation:
\begin{equation}
t'\,=\,\lambda\,t\,,\qquad x^i\,'\,=\,\lambda\,x^i\,,\qquad z'\,=\,\frac{z}{\lambda}\,.\label{trasf}
\end{equation}
in terms of the parameter $\lambda$.
In order to preserve the boundary expansion, the source $\phi_0$ has to transform according to:
\begin{equation}
\phi_0'(x')\,=\,\lambda^{d-\Delta}\,\phi_0(x)
\end{equation}
Using this ingredient the transformed action becomes:
\begin{equation}
\int\,d^dx'\,\sqrt{-\gamma}\,\phi_0'(x')\,\mathcal{O}'(x')\,=\,\int\,d^dx\,\sqrt{-\gamma}\,\phi_0(x)\,\lambda^{\Delta}\,\mathcal{O}'(x')
\end{equation}
At this point, since the boundary theory is conformal (or at least scale invariant), the boundary action has to be invariant under the scaling transformation in eq.\ref{trasf}.
This immediately implies that the operator $\mathcal{O}$ has to transform as:
\begin{equation}
\lambda^{\Delta}\,\mathcal{O}'(x')\,=\,\mathcal{O}(x)
\end{equation}
which defines its conformal dimension to be exactly $\left[\mathcal{O}\right]\,=\,\Delta$ !\\
\begin{mdframed}[style=MyFrame3]
\begin{center}
    \textbf{Exercise \#2 : Conformal dimension and Operators.}
\end{center}
Operators in quantum field theory are classified according to their conformal dimensions. They can be irrelevant, relevant or marginal: $\Delta>d,\,\Delta=d,\,\Delta<d$.\\
Consider our bulk scalar field with mass $m^2$. Classify the dual operator in function of the mass of the bulk field $m^2$. For which mass is the operator relevant? When does the dimension of the operator become complex? By answering this question, you will obtain the so-called BF bound \cite{BFpaper}, which for a scalar states reads:
\begin{equation}
m^2\,\geq\,-\,\left(\frac{d}{2\,L}\right)^2
\end{equation}
Surprisingly, a scalar with negative mass in AdS spacetime is not unstable!\\
Finally, a good criterium to understand the choice of the expectation value of an operator is its finiteness. $\langle \mathcal{O}\rangle$ has to be a \textit{normalizable} mode, on which we can perform the path integral of our CFT. The requirement of having such a scalar field mode normalizable (see \cite{wald1} for details about how to define such norm in curved spacetime) fixes its power $\sim z^\beta$ to satisfy:
\begin{equation}
\beta\,>\,\frac{d\,-\,2}{2}
\end{equation}
Try to get this inequality; \cite{Marolf:2006nd} can possibly be helpful.
In this case, this lower bound coincides with the so called \textit{unitarity bound}, which the conformal dimension of the scalar has to satisfy in order for the CFT to retain unitarity (see \cite{unbounds}).
\end{mdframed}\vspace{0.3cm}

\textbf{Time to get our hands dirty.}\\[0.15cm]
It is time to pass from the theory to the practice and see how this all shebang works.\\
Let us consider again the bulk action for our massive scalar field:
\begin{equation}
\mathcal{S}\,=\,-\,\frac{\eta}{2}\,\int\,dz\,d^{d}x\,\sqrt{-g}\,\left[\,g^{MN}\,\partial_M\,\phi\,\partial_N\,\phi\,+\,m^2\,\phi^2\right]
\end{equation}
where $\eta$ is just a normalization constant.\\
Using the equations of motions, we can derive the on-shell action as:
\begin{equation}
\mathcal{S}^{on-shell}\,=\,-\,\frac{\eta}{2}\,\int\,x\,d^dx\,\partial_M\,\left[\sqrt{-g}\,\phi\,g^{MN}\,\partial_N\,\phi\right]
\end{equation}
This action might potentially contain divergences. Therefore, let us consider a UV cutoff $z=\epsilon$ and re-write the on-shell action in the form:
\begin{equation}
\mathcal{S}^{on-shell}\,=\,\frac{\eta}{2}\,\int\,d^dx\,\left(\sqrt{-g}\,\phi\,g^{zz}\,\partial_z\,\phi\,\right)_{z=\epsilon}
\end{equation}
By defining the conjugate momentum $\Pi$:
\begin{equation}
\Pi\,=\,-\,\frac{\partial \mathcal{L}}{\partial\,(\partial_z\phi)}\,=\,\eta\,\sqrt{-g}\,g^{zz}\,\partial_z\,\phi
\end{equation}
the on-shell action takes a very simple form:
\begin{equation}
\mathcal{S}^{on-shell}\,=\,\frac{1}{2}\,\int_{z=\epsilon}\,d^dx\,\Pi(z,x)\,\phi(z,x)
\end{equation}
For simplicity, we Fourier transform the field $\phi$ and its conjugate momentum
\begin{equation}
\Pi(z,x^\mu)\,=\,\int\,\frac{d^dk}{(2\,\pi)^d}\,e^{i\,k\,\cdot\,x}\,\Pi_k(z)\,,\qquad\phi(z,x^\mu)\,=\,\int\,\frac{d^dk}{(2\,\pi)^d}\,e^{i\,k\,\cdot\,x}\,f_k(z)
\end{equation}
and we obtain:
\begin{equation}
\mathcal{S}^{on-shell}\,=\,\frac{1}{2}\,\int\,\frac{d^dk}{(2\pi)^d}\,\Pi_{-k}(z=\epsilon)\,f_k(z=\epsilon)
\end{equation}
Considering the UV expansion defined in eq.\eqref{UVscalarexp}, the behaviour of the conjugate quantity close to the AdS boundary is just:
\begin{equation}
\Pi_{k}\,(z)\,=\,\eta\,L^{d\,-\,1}\,\left[(d\,-\,\Delta)\,A(k)\,z^{-\Delta}\,+\,\Delta\,B(k)\,z^{\Delta-d}\,\right]
\end{equation}
We are now in the position to write down our on-shell action at the cutoff position $z=\epsilon$, for which we obtain:
\begin{equation}
\mathcal{S}^{on-shell}\,=\,\frac{\eta}{2}\,L^{d-1}\,\int\,\frac{d^dk}{(2\pi)^d}\,\left[\epsilon^{-2\nu}\,(d-\Delta)\,A(-k)\,A(k)\,+\,d\,A(-k)\,B(k)\,\right]
\end{equation}
As anticipated, this action has a divergent piece which scales like $\sim \epsilon^{-2\nu}$.\\
In order to renormalize the theory, we consider a local counterterm of the form:
\begin{equation}
\mathcal{S}_{ct}\,=\,-\,\frac{\eta}{2}\,\frac{d\,-\,\Delta}{L}\,\int_{z=\epsilon}\,d^dx\,\sqrt{-\gamma}\,\phi^2
\end{equation}
or, equivalently, in momentum space:
\begin{equation}
\mathcal{S}_{ct}\,=\,-\frac{\eta}{2}\,(d\,-\,\Delta)\,L^{d-1}\,\int\,\frac{d^dk}{(2\pi)^d}\,\left[\,\epsilon^{-2\nu}\,A(-k)\,A(k)\,+\,2\,A(-k)\,B(k)\,\right]
\end{equation}
where $\gamma$ is the induced metric at the boundary of AdS spacetime.\\
The full renormalized action, obtained by summing up the on-shell action with the counterterm above, is given by:
\begin{equation}
\mathcal{S}^{ren}\,=\,\frac{\eta}{2}\,L^{d-1}\,(2\,\Delta\,-\,d)\,\int\,\frac{d^dk}{(2\pi)^d}\,A(-k)\,B(k)
\end{equation}
Using what we have learned before, we can directly compute the $1-$point function as:
\begin{equation}
\langle\,\mathcal{O}(k)\,\rangle_{\phi_0}\,=\,(2\pi)^d\,\frac{\delta \mathcal{S}^{ren}}{\delta\phi_0(-k)}\,=\,2\,\nu\,\eta\,L^{d-1}\,B(k)
\end{equation}
where we have used that $2\Delta\,-\,d\,=2\,\nu$. Additionally, it follows immediately that the $2-$point function, or Green Function, for the same operator can be obtained as:
\begin{equation}
\mathcal{G}_E(k)\,=\,2\,\nu\,\eta\,L^{d-1}\,\frac{B(k)}{A(k)}
\end{equation}
In other words, the ratio between the subleading and the leading contributions of the bulk field provides the Green Function for the correspondent CFT operator dual to such a field! Keep this in mind, we will use it in every sauce in the following.\\

After checking explicitely some of the statements of the dictionary, how can be sure that this procedure gives us the correct results? For example, how can we know that the Green function of the scalar we obtain from the AdS computation matches the known results for a scalar operator in a CFT? Simple, we compute it!\\
Consider a redefinition of the original function $f_k$:
\begin{equation}
f_k(z)\,=\,z^{d/2}\,g_k(z)
\end{equation}
This new function satisfies the following equation:
\begin{equation}
z^2\,\partial_z^2\,g_k\,+\,z\,\partial_z\,g_k\,-\,\left(\,\nu^2\,+\,k^2\,z^2\,\right)\,g_k\,=\,0\,.
\end{equation}
which is a known-mathematical object, known as the modified Bessel equation. The two independent solutions of this equation can be taken to be $g_k=I_{\pm \nu}(kz)$, where $I_{\pm \nu}$ are the modified Bessel functions.
For small argument (\textit{i.e.} close to the AdS boundary), the modified Bessel functions behave as:
\begin{equation}
I_{\pm \nu}(z)\,\sim\,\frac{1}{\Gamma(1\,\pm\,\nu)}\,\left(\frac{z}{2}\right)^{\pm \nu}
\end{equation}
Assuming the generic solution to be a superposition of these two Bessel functions:
\begin{equation}
    f_k(z)\,=\,z^{d/2}\,g_k(z)\,=\,z^{d/2}\,\left(A \,I_{-\nu}(k z)\,+\,B\, I_{\nu}(k z)\right)
\end{equation}
we can expand it close to the UV boundary $z=0$ and obtain:
\begin{equation}
    f_k(z)\,\sim\,\frac{A \,2^{\nu }\, k^{-\nu } \,z^{\frac{1}{2} (d-2 \nu )}}{\Gamma (1-\nu )}+\frac{B\, 2^{-\nu }\, k^{\nu }\, z^{\frac{d}{2}+\nu }}{\Gamma (\nu +1)}\,+\,\dots
\end{equation}
Using that $\nu\equiv \Delta-\frac{d}{2}$ we can immediately verify that the two asymptotic powers are those described in the previous discussion:
\begin{equation}
    \underbrace{d-\Delta}_{\text{leading}}\quad , \quad \underbrace{\Delta}_{\text{subleading}}
\end{equation}
From the asymptotic analysis, we can therefore pin the meaning of the constants $A,B$ as the source $\phi_0$ and the VEV $\langle \mathcal{O} \rangle$ of the dual operator $\mathcal{O}$:
\begin{equation}
    \langle \mathcal{O} \rangle\,\equiv\,\frac{B\, 2^{-\nu }\, k^{\nu }}{\Gamma (\nu +1)}\quad,\quad \phi_0\,\equiv\,\frac{A \,2^{\nu }\, k^{-\nu }}{\Gamma (1-\nu )}
\end{equation}
In terms of the source and the vev of the dual field theory operator $\mathcal{O}$, we can now re-write the generic solution as:
\begin{equation}
f_k(z)=z^{d/2}\left[\phi_0\,\Gamma(1\,-\,\nu)\,\left(\frac{2\,k}{\nu}\right)^\nu\,z^{d/2}\,I_{-\nu}(kz)\,+\,\langle \mathcal{O} \rangle \,\Gamma(1\,+\,\nu)\,\left(\frac{2\,k}{\nu}\right)^{-\nu}\,I_{\nu}(kz)\,\right] \label{uniuni}
\end{equation}

The solution has to be taken to be regular at the horizon, $z=\infty$\footnote{This is a common practice that we will encounter often. The equations of motions in the bulk are $2^{nd}$ order differential equations, which need two boundary conditions to guarantee a unique solution. One of the boundary condition is always imposed at the boundary and it is usually the regularity of the bulk fields (we will see later how this condition is modified to compute retarded Green functions at finite temperature). A second boundary condition is fixed at the UV boundary and it corresponds to the choice of quantization in the dual field theory.}. For large argument, the Bessel functions have the following expansion:
\begin{equation}
I_{\pm\nu}\,\approx\,\frac{e^z}{\sqrt{2\,\pi\,z}}
\end{equation}
This means that the condition of having the complete solution $f_k(z)$ in eq.\eqref{uniuni} finite at the horizon $z=\infty$ fixes uniquely the ratio $\langle \mathcal{O}\rangle/\phi_0$ to be:
\begin{equation}
\frac{\langle \mathcal{O}\rangle}{\phi_0}(k)\,=\,-\,\frac{\Gamma(1\,-\,\nu)}{\Gamma(1\,+\,\nu)}\,2^{-2\,\nu}\left(\frac{k}{2}\right)^{2\,\nu}\,=\,\frac{\Gamma(-\,\nu)}{\Gamma(\nu)}\,\left(\frac{k}{2}\right)^{2\,\nu}
\end{equation}
In other words, we have just obtained the Green Function for our scalar operator from a purely bulk computation.
We can rewrite this result as:
\begin{equation}
\mathcal{G}(k)\,=\,2\,\nu\,\eta\,L^{d-1}\,\frac{\Gamma(-\,\nu)}{\Gamma(\nu)}\,\left(\frac{k}{2}\right)^{2\,\nu}
\end{equation}
and, in position space \footnote{We make use of the formula:
\begin{equation}
\int\,\frac{d^dk}{(2\pi)^d}\,e^{i\,k\,x}\,k^n\,=\,\frac{2^n}{\pi^{d/2}}\,\frac{\Gamma\left(\frac{d+n}{2}\right)}{\Gamma\left(-\,\frac{n}{2}\right)}\,\frac{1}{|x|^{d+n}}
\end{equation}}, as:
\begin{equation}
\langle\,\mathcal{O}(x)\,\mathcal{O}(0)\,\rangle\,=\,\frac{2\,\nu\,\eta\,L^{d-1}}{\pi^{d/2}}\,\frac{\Gamma\left(\frac{d}{2}\,+\,\nu\right)}{\Gamma(-\nu)}\,\frac{1}{|x|^{2\,\Delta}}
\end{equation}
which is indeed the well-known result for for a primary operator of dimension $\Delta$ in a conformal field theory (see for example \cite{blumenhagen2009introduction}!
\newpage
\begin{mdframed}[style=MyFrame3]
\begin{center}
    \textbf{Exercise \#4 : A vector in AdS.}
\end{center}\vspace{0.15cm}
For higher order p-forms  $A_{\mu_1,\dots,\mu_p}$ the indicial equation in  AdS reads:
\begin{equation}
(\Delta\,-\,p)\,(\Delta\,+\,p\,-\,d)\,=\,m^2\,L^2
\end{equation}
fixing the solutions to be:
\begin{equation}
\Delta_{\pm}\,=\,\frac{d}{2}\,\pm\,\sqrt{\left(\frac{d\,-\,2\,p}{2}\right)^2\,+\,m^2\,L^2}
\end{equation}
A known example is:
\begin{enumerate}
\item the gauge field $A_\mu$ ($p=1$) for which:
\begin{equation}
\Delta_{\pm}\,=\,\frac{d}{2}\,\pm\,\sqrt{\left(\frac{d\,-\,2}{2}\right)^2\,+\,m^2\,L^2}
\end{equation}
Note that in the case of $m=0$, when the U(1) gauge symmetry is unbroken, the conformal dimension of the current operator $J^\mu$ is equal to $\Delta=d-1$, which is indeed the conformal dimension for a conserved current.
\end{enumerate}
One can also consider symmetric fields, like the spin-2 metric field $g_{\mu\nu}$ for which:
\begin{equation}
\Delta\,(\Delta\,-\,d)\,=\,m^2\,L^2
\end{equation}
such that in the massless case, enjoying diffeomorphism invariance, the dimension of the dual stress tensor operator is $\Delta=d$, which implies indeed a conserved stress tensor $T^{\mu\nu}$.\\
Try to obtain these expressions. You might want to start with a vector in AdS:
\begin{equation}
    S\,=\,\int d^{d+1}x\,\sqrt{-g}\,\left[-\frac{1}{4}\,F_{\mu\nu}F^{\mu\nu}\,-\,\frac{m^2}{2}\,A_\mu A^\mu\right]
\end{equation}
and in particular from its EOM:
\begin{equation}
    D^\mu F_{\mu\nu}\,=\,m^2\,A_\nu
\end{equation}
Can you reproduce the results above?
\end{mdframed}

%
%
%
\chapter{The first big success: $\eta/s$ and Hydrodynamics}
\label{intro3} 
\hspace{0.2cm} \includegraphics[width=0.5\textwidth]{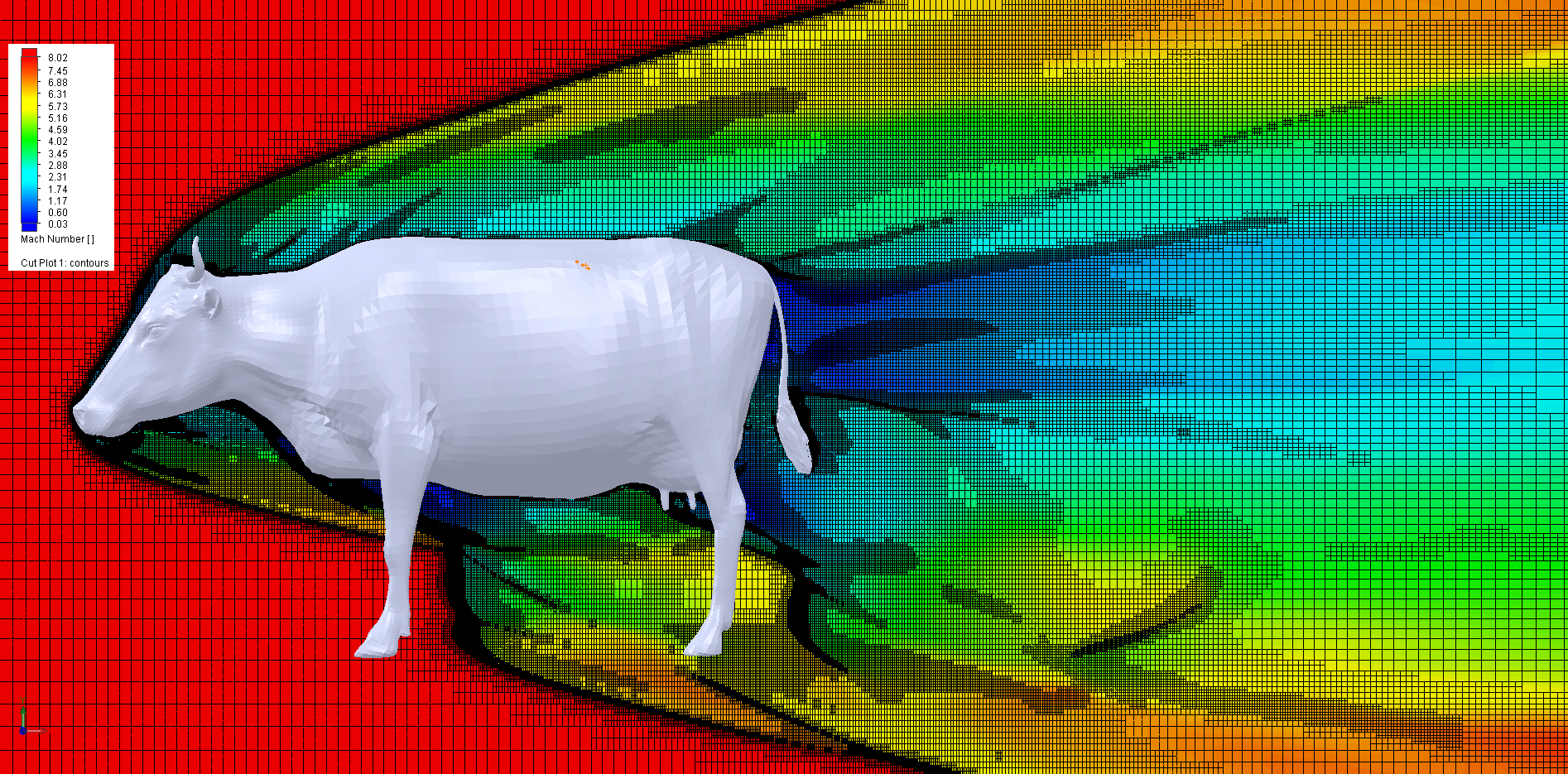}\\
\epigraph{It is life, I think, to watch the water. A man can learn so many things.}{\textit{Nicholas Sparks}}
\section{Droplets of Relativistic Hydrodynamics and retarded Green functions}
\begin{figure}[h!]
    \centering
    \includegraphics[width=0.99\linewidth]{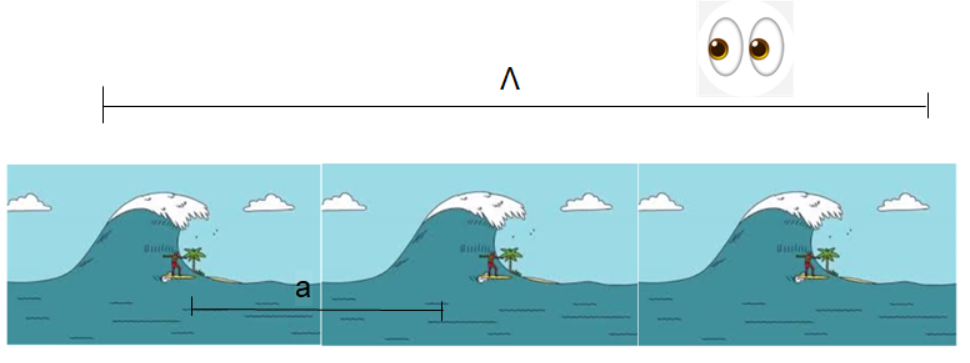}
    \caption{The long wavelength limit of the effective hydrodynamic description. We are looking at scales (in this case spatial, $\Lambda$) which are much larger than the characteristic length of the system $a$. The microscopic details are neglected as in every EFT and the resulting theory is a low energy approximation.}
    \label{fig:fluid}
\end{figure}
What is Hydrodynamics? It is just the effective field theory description of a system close to equilibrium, valid at sufficiently long times and sufficiently large distances \cite{landau2013fluid}. Notice that, despite the misleading name, it can be applied to very generic systems, and not necessarily only to fluids \cite{PhysRevA.6.2401}! Hydrodynamics is not the theory of fluids but it is much more general and it is simply the low energy effective description of a specific system. As such, it is constructed in a perturbative fashion as an expansion in the (time and spatial) gradients around equilibrium. In the convenient Fourier language, hydrodynamics stands as a expansion in powers of the frequency $\omega$ and the momentum $k$. At late times and large distances, the dynamics is governed, or rather dominated, by the longest living modes, which are usually referred to as \textit{hydrodynamic modes}\footnote{This term is used in several different connotations. A strong definition of ''hydrodynamic mode'' is linked to the requirement:
\begin{equation}
    \,\lim_{k \rightarrow 0}\omega_{\text{hydro}}(k)\,=\,0
\end{equation}
Here, we take a more relaxed definition according to which we will call hydrodynamic mode every excitation living in the hydrodynamic window:
\begin{equation}
    \omega/T,\,k/T\,\ll 1
\end{equation}}.
Formally, it is constructed by using conservation equations together with constitutive relations. The conservation equations are simply the consequence of the symmetries of the system and they are our starting point\footnote{Again, hydrodynamic is much powerful than that and it can be applied or generalized in the case of softly broken symmetries. See for example \cite{forster1975hydrodynamic,boon1991molecular,Grozdanov:2018fic}.}. The simplest possible scenario, the case of an uncharged fluid, involves only the conservation of the stress-energy tensor
\begin{equation}
    \partial_\mu\,T^{\mu\nu}\,=\,0 \label{conon}
\end{equation}
which follows by the translational invariance of the system. For simplicity, we will present the material using the relativistic framework\footnote{See \cite{Kovtun:2012rj,Policastro:2002se,Policastro:2002tn} for review about relativistic hydrodynamics and connections with holography.}.
The conservation equation \eqref{conon} does not bring us very far if it is not accompanied by a constitutive relation for the stress-energy tensor. Here, it is exactly where the gradient expansion and the low energy nature of hydrodynamics enter. Given the symmetries of the system, we can build the tensorial structure of the considered conserved current\footnote{In this case the stress-tensor but for example in presence of a $U(1)$ symmetry we will have also the corresponding $U(1)$ conserved current $J^\mu$.}. This is performed order by order in the variations around equilibrium. As an example, the constitutive relation for the stress tensor reads: \begin{equation}
    T^{\mu\nu}\,=\,\underbrace{\epsilon\,u^\mu\,u^\nu\,+\,p\,\Delta^{\mu\nu}}_{0^{st}\,order}\,\underbrace{-\,\eta\,\sigma^{\mu\nu}\,-\,\zeta\,\Delta^{\mu\nu}\,\partial_\lambda u^\lambda}_{1^{st}\,order}\,+\,\dots
\end{equation}
where we defined the objects:
\begin{equation}
    \Delta^{\mu\nu}\,\equiv \eta^{\mu\nu}\,+\,u^\mu u^\nu
\,,\quad \sigma^{\mu\nu}\,\equiv\,\Delta^{\mu\alpha}\Delta^{\nu\beta}\left(\partial_\alpha u_\beta\,+\,\partial_\beta u_\alpha\,-\,\frac{2}{d}\,\eta_{\alpha \beta}\,\partial_\mu u^\mu\right)
\label{cc2}\end{equation}
which are first order in derivatives, \textit{i.e.} the variations around equilibrium. Again, this expansion has simply to be thought as:
\begin{equation}
    T^{\mu\nu}\,=\,(\textit{slow})\,+\,(\textit{fast})\,+\,(\textit{faster})\,+\,(\textit{much faster})\,+\,\dots
\end{equation}
and in this sense it is a low energy effective theory, since every term carries more and more energy\footnote{Remember that ''faster'', or if you want higher frequency, implies higher energy.}.\\
\begin{figure}[h]
    \centering
    \includegraphics[width=0.85\linewidth]{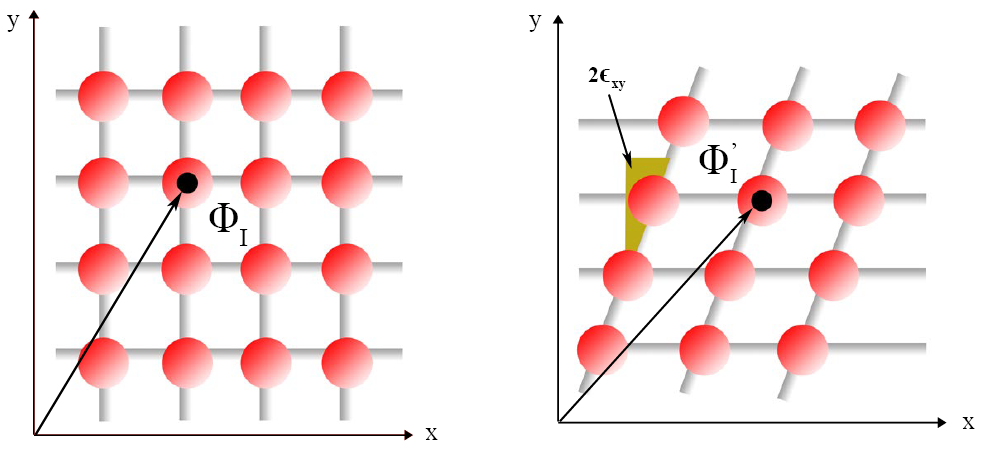}
    \caption{The geometric interpretation of the external shear $\varepsilon_{xy}$ in a $2-D$ object. The displacements $\phi^I$ are simply the shift in the position of the individual constituents. Picture taken from \cite{soon}.}
    \label{figshear}
\end{figure}\\
Let us pause for two important or at least interesting comments. (I) The hydrodynamic expansion (as most of the perturbative expansions in physics \cite{PhysRev.85.631}) is an asymptotic expansion and its radius of convergence is still under debate \cite{Withers:2018srf,Grozdanov:2019kge,Grozdanov:2019uhi} \footnote{To be precise, recent works point out that hydrodynamics is at most a divergent series in position space. In momentum space, and expanded around the $\omega=k=0$ point is a legitimate series with finite radius of convergence in the sense of the Puiseux series. I thank Saso Grozdanov for this clarification.}. This means that going at higher order does not necessarily constitute an improvement of the results after a certain point where non-perturbative (instanton-like) effects appear. Novel resummation mathematical techniques are employed to reach a more robust control on the expansion \cite{Heller:2016rtz,Baggioli:2018bfa,Buchel:2016cbj,Aniceto:2018uik,Aniceto:2015mto,Heller:2015dha}. (II) Hydrodynamics can be recasted in the standard language of effective field theory \cite{Dubovsky:2011sj}, but it is still a deep and open question how to formally derive it from an action principle \cite{Glorioso:2018wxw}.\\

\begin{figure}[h!]
    \centering
    \includegraphics[width=0.7\linewidth]{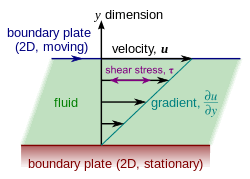}
    \caption{The standard picture of the shear viscosity coefficient in terms of an external shear rate. A plate moves at a constant speed $\vec{u}$ and a normal force is produced. Figure taken from \url{https://en.wikipedia.org/wiki/Viscosity}.}
    \label{figvisc}
\end{figure}

Let us now focus on a concrete example. Consider an uncharged relativistic ''fluid'' and more specifically its shear sector. Consider the linear response of the system as a reaction to an external shear strain. This response is generically encoded in the form of the (helicity $2$ part\footnote{$=$ traceless and transverse part.} of the) stress-tensor $T^{\mu\nu}$, which we build, as prescribed, in a perturbative expansion in frequency (and momentum). For simplicity, let us assume zero momentum, $k=0$, which corresponds to only spatially homogeneous response. In this case, the lowest orders of the expansion read:
\begin{equation}
    T_{\mu\nu}\,=\,G\,\varepsilon_{\mu\nu}\,-\,i\,\omega\,\eta\,\varepsilon_{\mu\nu}\,+\,\mathcal{O}(\omega^2) \label{gn}
\end{equation}
where $\varepsilon_{\mu\nu}$ is a shear strain, a mechanical deformation of the system which does not change the volume of the system itself. Technically, this tensorial object can be constructed via the symmetric derivative of the displacement $\phi^I$ and it has the simple geometric interpretation as the angle deficit between the original and deformed sample (see fig.\ref{figshear}). Notice that, in this language, the tensor $\sigma^{\mu\nu}$ appearing in the relativistic stress tensor in eq.\eqref{cc2} is simple the rate of variation of the external strain, namely:
\begin{equation}
    \sigma^{\mu\nu}\,=\,\partial_t\,\varepsilon^{\mu\nu}
\end{equation}
In other terms, the shear viscosity is simply the transport coefficient determining the response under an external shear rate, as in the classical picture of \ref{figvisc}.
\\

Now, given the expansion in eq.\eqref{gn}, we can define the two lowest order transport coefficients:
\begin{align}
  &  G\,=\,Re\left[\mathcal{G}^R_{xyxy}\left(\omega=k=0\right)\,\right]\,\,,\\
  &  \eta=\,-\,\lim_{\omega\, \rightarrow 0}\left\{\frac{1}{\omega}\,Im\left[\mathcal{G}^R_{xyxy}\left(\omega,k=0\right)\,\right]\,\right\}\, ,
\end{align}
where we indicated with $\mathcal{G}^R_{xyxy}$ the Green function of the stress tensor. These coefficients are known as the shear elastic modulus $G$ and the shear viscosity $\eta$. In linear response, it just follows from eq.\eqref{gn} that:
\begin{equation}
    \mathcal{G}^R_{xyxy}\left(\omega,k=0\right)\,\equiv\,\frac{T_{xy}\left(\omega,k=0\right)}{\varepsilon_{xy}}
\end{equation}
which is just an application of the general recipe:
\begin{equation}
  \colorboxed{blue}{ \text{response} (\omega,k)\,=\,\text{GreenFunction}(\omega,k)\,\,\text{source}(\omega,k)}
\end{equation}
As explained in the previous section, such an object can be derived holographically by using the procedure explained in the previous section and summarized as:
\begin{align}
\centering
   &\colorboxed{blue}{\text{GreenFunction}(\omega,k)\,\sim\,\frac{\text{subleading}(\omega,k)}{\text{leading}(\omega,k)}}\nonumber\\
    & \text{field}(u)\,=\,\text{leading}(\omega,k)\,u^{\Delta_l}\,\left(1\,+\,\dots\right)\,+\,\text{subleading}(\omega,k)\,u^{\Delta_s}\,\left(1\,+\,\dots \right)\nonumber\\
    & \text{close to the boundary}\,u\,=\,0\,\quad \text{with}\,\Delta_l<\Delta_s \label{rout}
\end{align}
which is exactly what we will do.
\section{How to deal with Einstein equations}
\begin{figure}[h!]
    \centering
    \includegraphics[width=0.9\linewidth]{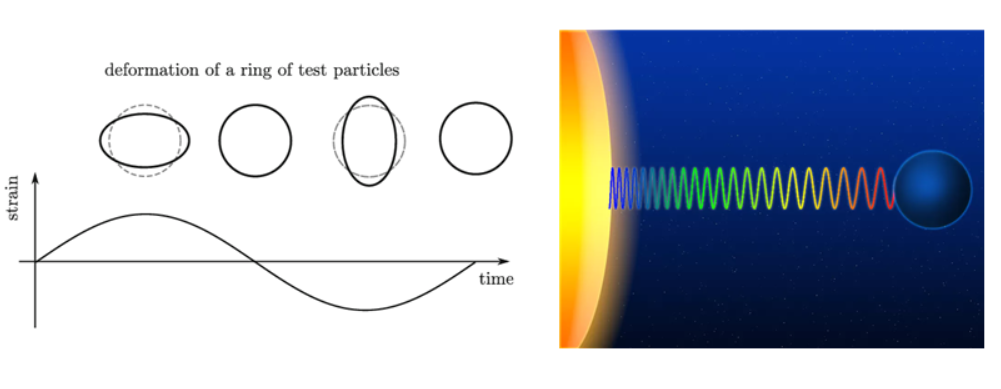}
    \caption{The external strain deforms a test particle exactly the wave a gravitational wave deforms the geometry. }
    \label{figg}
\end{figure}

Now, let us get serious. The bulk field dual to the stress tensor $T^{\mu\nu}$ is the bulk metric itself $g^{\mu\nu}$. This means that, in order to compute the linear response in eq.\eqref{gn}, we need to introduce metric perturbations which are transverse and traceless. What are they? Those are simply gravitational waves! We are shaking our background geometry, if you like our black hole, by throwing at it gravitational waves. And what we are interested in it is simply how the black hole will react. Not surprisingly, it has been proven that the holographic computation of the shear viscosity corresponds exactly to determining the absorption of gravitons at the horizon of the black hole \cite{Policastro:2001yc} (see fig.\ref{figg}).
\\
Let us start from the simplest gravitational action one can imagine: Einstein-Hilbert with a finite cosmological constant. The action is given by:
\begin{equation}
    S\,=\,\int d^4x\,\sqrt{-g}\,\left[R\,-\,2\,\Lambda \right]
\end{equation}
and the Einstein equations, derived using the standard variational principle (try if you never did it), are:
\begin{equation}
    R_{\mu\nu}\,-\,\frac{1}{2}\,R\,g_{\mu\nu}\,+\,\Lambda\,g_{\mu\nu}\,=\,0
\end{equation}
Let us derive the Scharzchild Black Hole solution using the ansatz:
\begin{equation}
    ds^2\,=\,\frac{1}{u^2}\,\left[\,-f(u)\,dt^2\,+\,\frac{du^2}{f(u)}\,+\,dx^2\,+\,dy^2\right]
\end{equation}
in which $u=0$ is the AdS boundary. You can convince yourself that the temperature and the entropy of the background read:
\begin{equation}
    T\,=\,-\,\frac{f'(u)}{4\,\pi}\,,\quad s\,=\,\frac{4\,\pi}{u_h^2}
\end{equation}
and the full solution is completed by:
\begin{equation}
    f(u)\,=\,1\,-\,\left(\frac{u}{u_h}\right)^3
\end{equation}
which is called the blackening factor. Notice that at $u=u_h$ the function $f$ displays a single zero. This property determines the position of the even horizon, at which $g_{tt}\rightarrow 0$ -- nothing can escape from there (see \cite{wald2010general} for details).\\
This is our solution, it is called the Schwarzschild Black Hole and it was obtained by Karl Schwarzschild while he was serving in the German army during World War I\footnote{This is a good thought when you feel you are working under stress. To cope with it with some laughs I suggest a visit to \url{http://www.thegrumpyscientist.com/}.}. Let us define our shear perturbation (the gravitational wave disturbance) as $\delta g_{xy}= e^{-i \omega t} h(u)$. The radial dependent perturbation obeys the simple equation:
\begin{equation}
    \left(\frac{f'}{f}+\frac{2}{u}\right) h'+h \left(\frac{2 \,f'}{u
   \,f}+\frac{\omega^2}{f^2}-\frac{2}{u^2}\right)+h''\,=\,0 \label{initial}
\end{equation}
which serves as the prototype of the typical equation you will find (and you will have to solve) in holography. See the attached Mathematica notebook for the derivation of all these equations.
\begin{mdframed}[style=MyFrame2]
\begin{center}
    \textbf{Warning \#1 : Using the variational principle in Mathematica is dangerous}
\end{center}\vspace{0.15cm}
One simple, but quite dangerous, way of obtaining the equations of motion with Mathematica is to use the implemented variational function.\\
This can be simply written as:
\begin{itemize}
    \item Define your action in terms of your bulk fields (functions of the bulk coordinates):
    \begin{equation}
        \text{action}\,=\,\mathcal{F}\,\left[\text{field}_1(x^\mu),\,\text{field}_2(x^\mu),\,\dots,\,\text{field}_n(x^\mu)\right]
    \end{equation}
    \item And then you can define directly the various equations of motion by doing:
    \begin{equation}
        \text{eq}_n\, =\,\text{
   EulerEquations}[\mathcal{F}\,, \text{field}_n(x^\mu), {x^1,x^2,\dots,x^n}][[1]]
    \end{equation}
\end{itemize}
This command will give you directly all the equations of motion but it can be dangerous. In particular, with this method you can not check directly if your ansatz is complete and if you switched on all the relevant perturbations. It might be that your ansatz is not enough and one equation of motion cannot be satisfied by your ansatz but you do not see that just because you are missing that equation.\\
The second and most robust way of proceeding is using a package for tensorial computations and implement the tensorial Einstein equations directly. Everybody (apart me) uses the \textit{diffgeo.m} package which can be found here \url{http://people.brandeis.edu/~headrick/Mathematica/}. I prefer another one which you can find in the Notebooks attached to this lectures.
\end{mdframed}
\begin{mdframed}[style=MyFrame3]
\begin{center}
    \textbf{Exercise \#5 : The temperature of a Black Hole}
\end{center}\vspace{0.15cm}
For a simple BH geometry of the type:
\begin{equation}
    ds^2\,=\,\frac{1}{u^2}\,\left(-\,f(u)\,dt^2\,+\,\frac{du^2}{f(u)}\,+\,d \vec{x}^2\right)
\end{equation}
the corresponding temperature reads:
\begin{equation}
    T\,=\,-\,\frac{f'(u_h)}{4\,\pi}
\end{equation}
This is a very well-known result which can be obtained in several ways (the most common is going to Euclidean time, compactify the time direction and require the absence of any deficit angle) and it is often used without questioning it.\\
What if the BH is more complicated? For instance a diagonal metric of the type:
\begin{equation}
    ds^2\,=\,-\,g_{tt}(u)\,dt^2\,+\,g_{uu}(u)\,du^2\,+\,g_{xx}\,d\vec{x}^2
\end{equation}
How can we derive the temperature?\\
One of the most generic way is to connect the temperature of the BH with the surface gravity at its horizon. The surface gravity is just the gravitational acceleration at the horizon of the black hole (see \cite{weinberg1972gravitation}). Surprisingly enough, the bigger the mass of the black hole, the lower the surface gravity at the horizon! We can define the temperature of the BH from the relation (see for example \cite{wald2010general}) :
\begin{equation}
    T\,=\,\frac{\kappa_h}{2\,\pi}\,=\,\frac{1}{2\,\pi}\,\sqrt{-\,\frac{n^{\mu;\nu}n_{\mu;\nu}}{2}}\Big|_{\text{horizon}}
\end{equation}
where $n^\mu$ is the Killing field whose null surface is the horizon (in this case, $\partial t$).\\
Using the ansatz above, you should get as a final result something like:
\begin{equation}
    T\,=\,-\,\frac{1}{4\,\pi}\,\sqrt{{g_{tt}}'\,{g^{uu}}'}\Big|_{\text{horizon}}
\end{equation}
Can you compute also the entropy? Remember is simply related to the area of the horizon $\mathcal{A}\,=\,\int g_{xx}^D(u_h)$, with $D$ the spatial directions. Notice that for flat horizons, the area is clearly infinite so what we refer to is the entropy density, which is always finite.
\end{mdframed}
\section{How to derive the KSS bound analytically}\label{secK}
\begin{figure}[h!]
    \centering
    \includegraphics[width=0.75\linewidth]{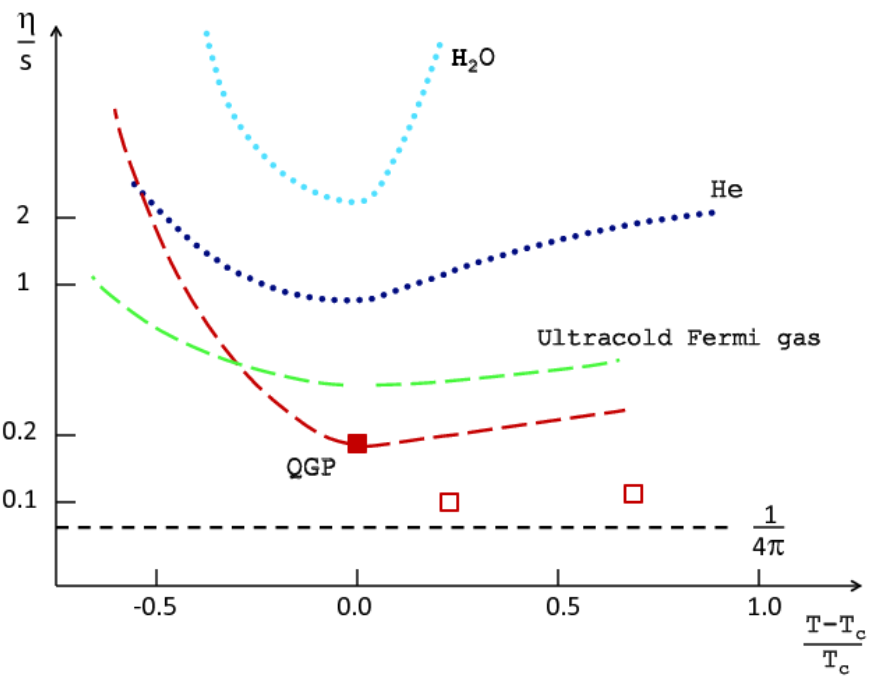}
    \caption{The $\eta/s$ ratio in common (and less common) fluids compared to the KSS bound $\eta/s >1/4\pi$ conjectured from holography. The figure is taken from \cite{Cremonini:2012ny}.}
    \label{fig:etas}
\end{figure}

The first exercise we are going to do with our brand new black hole is to derive the famous result of \cite{Policastro:2001yc}, which goes under the name of the Kovtun-Son-Starinets (KSS) bound. The statement is that the ratio between the viscosity $\eta$ and the entropy density is bounded from below by:
\begin{equation}
    \frac{\eta}{s}\,\geq\,\frac{1}{4\,\pi}\,\frac{\hbar}{k_B}
\end{equation}
This bound is supposed to be universal and induced by the saturation of the minimum quantum relaxation time scale $\tau \sim \hbar/k_B T$ which is known as Planckian time \cite{Zaanen:2018edk}. This bound has been tested experimentally in various systems (see fig.\ref{fig:etas}) and the lower the value of this ratio the more strongly coupled the corresponding system\footnote{Remember that the viscosity of a system is proportional to the mean free path $l_{mfp}$. The smaller the mean free path the strongest the interactions.}. For a recent and comprehensive review about this bound see \cite{Cremonini:2011iq}.
\\
\begin{mdframed}[style=MyFrame]
\begin{center}
    \textbf{Trick \#1 : Eddington–Finkelstein coordinates}
\end{center}\vspace{0.15cm}
What are they? They are coordinates adapted to radial null geodesics.
Why they are useful? Because they implement directly the ingoing boundary conditions at the horizon. More precisely, the ingoing boundary conditions translate into simple regularity at the horizon.\\
Given the BH geometry:
\begin{equation}
    ds^2\,=\,\frac{1}{u^2}\,\left[\,-f(u)\,dt^2\,+\,\frac{du^2}{u^2}\,+\,d\vec{x}^2\right]\label{awq}
\end{equation}
and a probe bulk field $\phi(u)$ on top of it, the solution close to the horizon are always of the type:
\begin{equation}
    \phi(u)\,=\,e^{\pm\,i\,\omega\,\int \frac{1}{f(z)}dz}\,\left(\phi_0\,+\,\phi_1\,(u-u_h)\,+\,\dots\right)
\end{equation}
where the $\pm$ corresponds to the ingoing or outgoing boundary conditions (this arbitrariness corresponds to the different types of Green functions one can define in the dual field theory: retarded, advanced, etc... See \cite{Son:2002sd}).\\
As explained in detail in \cite{Son:2002sd}, in order to obtain the retarded correlators (and eventually the quasinormal modes of the system) we need to impose ingoing b.c.s. at the horizon. That is why EF coordinates are very welcome. In EF coordinates the geometry in eq.\ref{awq} becomes:
\begin{equation}
    ds^2\,=\,\frac{1}{z^2}\,\left[-f(z)\,d\tau^2\,-\,2\,d\tau\,dz\,+\,d\vec{x}^2\right]
\end{equation}
Find the transformation that brings you from the Poincare coordinates to the EFT ones. In these new coordinates, the ingoing b.c.s. correspond simply to regularity at the horizon:
\begin{equation}
    \phi(z)\,=\,\phi_0\,+\,\phi_1\,(z-z_h)\,+\,\dots
\end{equation}
[A trick to see that you are doing right: in the eoms no term $\sim\omega^2$ has to appear anymore, only terms linear in the frequency!]\\
Another trick is to obtain the equations in the original Poincare coordinates but then remove the ingoing factor from the field variables by using the redefinition:
\begin{equation}
    \tilde{\phi}(u)\,=\,\phi(u)\,e^{i\,\omega\,\int \frac{1}{f(z)}dz}
\end{equation}
You should get exactly the same result, try!
\end{mdframed}\vspace{0.2cm}
What we want to do now is to prove that the Schwarzschild solution in Einstein-Hilbert theory saturates the KSS bound $\eta/s=1/4\pi$.\\
In order to do that, we will use the so-called \textit{membrane paradigm}, discussed at length in \cite{Iqbal:2008by}. Before proceeding, two observations are in order: (I) the equation for the shear perturbation \eqref{initial} has exactly the same form of that of a massless scalar on the same geometry; (II) at zero frequency, eq.\eqref{initial} implies the radial conservation of a specific quantity which will turn out to be very useful.\\
Consider the previously derived equation \eqref{initial} and perform the following rescaling  $h(u)\,=\,H(u)/u^2$\footnote{I thank Aravindh Swaminathan Shankar for correcting a previous typo in this transformation.}. The resulting equation is simply:
\begin{equation}
    H''\,+\,H'\,\left(\frac{f'}{f}\,-\,\frac{2}{u}\right)\,+\,\frac{\omega^2}{f^2}\,H\,=\,0 \label{eqi}
\end{equation}
and it is our starting point.
Eq.\eqref{eqi} follows from a simple gravitational  action for a massless scalar of the type
\begin{equation}
    S\,=\,-\,\frac{1}{2}\,\int\,d^{d+1}x\,\sqrt{-g}\,\frac{1}{q(u)}\,\left(\nabla H\right)^2
\end{equation}
where in our specific case $q(u)=cost.$ (more precisely, $q(u)=1$). Following the methods discussed in the previous section, we can obtain the boundary action as
\begin{equation}
    S_{boundary}\,=\,\int_\Sigma\,d^dx\,\Pi_H(\vec{x},t)\,H_0(\vec{x},t) \label{ciccia}
\end{equation}
where:
\begin{equation}
    \Pi_H\,=\,-\,\frac{g^{uu}\,\partial_u H}{q}\,,\quad H_0(\vec{x},t)\,=\,H|_{\text{boundary}}
\end{equation}
Two important facts follow
\begin{itemize}
    \item The equation for the shear mode can be re-written:
    \begin{equation}
        \partial_u\,\Pi_H(u)\,=\,0\,+\,\mathcal{O}(\omega^2)
    \end{equation}
    which means that $\Pi_H(u)$ is a radially conserved quantity at zero frequency. The first corrections, breaking such conservation, enter at order $\mathcal{O}(\omega^2)$.
    \item From the form of the boundary action \eqref{ciccia}, and using the standard dictionary, we obtain:
    \begin{equation}
        \langle \mathcal{O}_H (\vec{x},t) \rangle\,=\,\lim_{u \rightarrow 0}\,\Pi_H(u,\vec{x},t)
    \end{equation}
    where $\mathcal{O}_H (\vec{x},t)$ is the operator dual to the bulk field $H(u)$, which in our case is the stress tensor $\mathcal{O}_H (\vec{x},t) \equiv T_{xy}$.
\end{itemize}
Given the previous points, the Green function for the stress tensor operator can be consequently defined as
\begin{equation}
    \langle T_{xy}T_{xy}\left(\vec{x},t \right) \rangle \,=\,-\,\frac{\Pi_H\left(\vec{x},t \right)}{H_0\left(\vec{x},t \right)}
\end{equation}
where $H_0$ represents the source for the dual field theory operator, and $\Pi_H$ the corresponding expectation value.\\
Now comes the trick! The quantity $\Pi_H$ is radially conserved, implying that $\Pi_H|_\text{boundary}=\Pi_H|_\text{everywhere}$. It is convenient to compute its value at the horizon $u=u_h$. There, the form of the bulk field is constrained by the ingoing boundary conditions to be:
\begin{equation}
    H(u)\,=\,e^{-i \omega \int \frac{1}{f(y)} dy}\,\left(H_0\,+\,\dots\right)
\end{equation}
Using this expression, we obtain
\begin{equation}
    \Pi_H\,=\,\frac{1}{q_h\,u_h^2}\,i\,\omega\,H_0 \label{re}
\end{equation}
which gives:
\begin{equation}
    \langle T_{xy}T_{xy}\left(\vec{x},t \right) \rangle \,=\,-\,i\,\omega\,\frac{1}{q_h\,u_h^{2}}\,+\,\mathcal{O}(\omega^2) \label{ok}
\end{equation}
Notice that the conservation of the conjugate momentum holds only at $\omega=0$, so the result in eq.\eqref{re} contains corrections of order $\mathcal{O}(\omega^2)$. Nevertheless, those corrections are not affecting the computation of the shear viscosity, but only high order coefficients. That said, expression \eqref{ok} gives us the final answer:
\begin{equation}
    \eta\,=\,\frac{1}{q_h\,u_h^{2}}\,\,\quad \underbrace{\longrightarrow}_{q=1}\,\quad \,\,\eta\,=\,\frac{1}{u_h^2}
\end{equation}
Using the definition for the entropy density $s=4\pi/u_h^2$, we finally obtain:
\begin{equation}
    \frac{\eta}{s}\,=\,\frac{1}{4\,\pi}
\end{equation}
which is indeed the saturation of the KSS bound \cite{Kovtun:2004de}.\\
Notice that there is another thing which we have learned:
\begin{equation}
    G\,=\,0
\end{equation}
which means that our holographic system has no static elastic response. This tells us that the dual field theory represents an holographic fluid with zero shear elastic modulus. The presence of a finite shear modulus, together with the absence of propagating shear waves, is what usually define the difference between solids and fluids\footnote{This criterium is now under debate as a consequence of the experimental results of \cite{Noirez_2012,yang2017emergence} and the theoretical discussions in \cite{trachenko2015collective,Baggioli:2019jcm}, which seem to be in agreement with what Holography suggests \cite{Baggioli:2018nnp,Baggioli:2018vfc}.}. In the next section we will see how to generalize the model to account for viscoelastic and solid systems.
\begin{mdframed}[style=MyFrame3]
\begin{center}
    \textbf{Exercise \#5 : The violation of KSS in Gauss Bonnet.}
\end{center}\vspace{0.15cm}
Consider a more complicated action containing higher-derivative corrections:
\begin{equation}
    S\,=\,\int d^5x\,\sqrt{g}\,\left[R\,-\,2\,\Lambda\,+\,\frac{\lambda_{GB}}{2}\,\left(R^2\,-\,4\,R_{\mu\nu}R^{\mu\nu}\,+\,R_{\mu\nu\rho\lambda}R^{\mu\nu\rho\lambda}\right)\right]
\end{equation}
and try to compute $\eta/s$ again for this theory which is known as Gauss-Bonnet (GB) theory. You will find a very famous violation of the KSS bound \cite{Brigante:2007nu,Brigante:2008gz}. You will also realize that the only difference with the computations in the main text is that:
\begin{equation}
    q(u)\,=\,1\,-\,4\,\lambda_{GB}
\end{equation}
and can be interpreted simply by an effective Newton constant \cite{Brustein:2007jj}.\\
Try it! You will also learn that GB is a very specific case of higher derivative theories, where the formula for the entropy density is still given by the area law. Try with the most generic quadratic theory:
\begin{equation}
    S\,=\,\int d^5x\,\sqrt{g}\,\left[R\,-\,2\,\Lambda\,+\,\alpha\,R^2\,+\,\beta\,R_{\mu\nu}R^{\mu\nu}\,+\,\gamma\,R_{\mu\nu\rho\lambda}R^{\mu\nu\rho\lambda}\right]
\end{equation}
which is discussed in \cite{Kats:2007mq}. Can you see why the GB combination is so special?
\end{mdframed}
\section{How to make it less universal (and more elastic)}
Now, let us slightly modify the theory. In particular, let us consider a gravitational theory where the graviton is massive. The graviton mass modifies the propagation of the gravitational waves at large distances and it was introduced long time ago as a possible explanation for the acceleration of the universe which does not require any dark energy. Here we make a big (and for the moment not so physically motivated) jump, whose meaning will be clearer later.\\
Assuming isotropy and homogeneity, one can see that the former equation for the shear perturbation generally acquires a radially dependent mass given by:
\begin{equation}
    \mathcal{M}^2\,=\,g^{xx}\,T_{xx}\,-\,\frac{\delta T_{xy}}{\delta g_{xy}}
\end{equation}
which was derived in \cite{Hartnoll:2016tri} and it is the subject of one of our exercises.
\begin{mdframed}[style=MyFrame3]
\begin{center}
    \textbf{Exercise \# 7: The shear mass.}\end{center}\vspace{0.15cm}
    Derive that for an isotropic solution, the mass of the shear perturbation is:
    \begin{equation}
m^2(r)\,=\,g^{xx}\,T_{xx}\,-\,\frac{\delta\,T_{xy}}{\delta g_{xy}}
\end{equation}
Use the following definitions:
\begin{align}
   & ds^2\,=\,-\,g_{tt}(r)\,dt^2\,+\,g_{rr}(r)\,dr^2\,+\,g_{xx}(r)\,dx^I\,dx^I\\
   &T_{\mu\nu}\,=\,\text{diag}\,\left(T_{tt}(r),\,T_{rr}(r),\,T_{xx}(r),\,\dots,\,T_{xx}(r)\right)
\end{align}
\end{mdframed}
\begin{figure}[h!]
    \centering
    \includegraphics[width=0.45\linewidth]{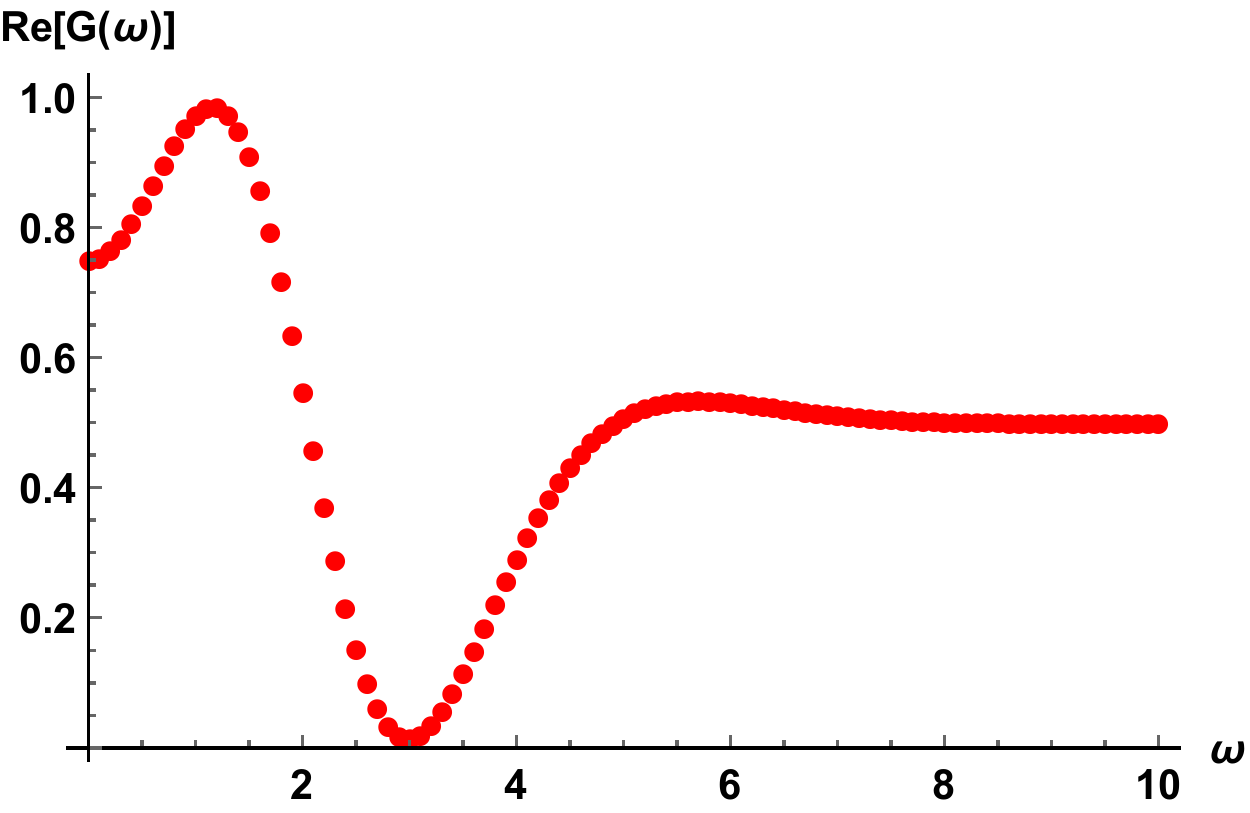}\quad 
    \includegraphics[width=0.45\linewidth]{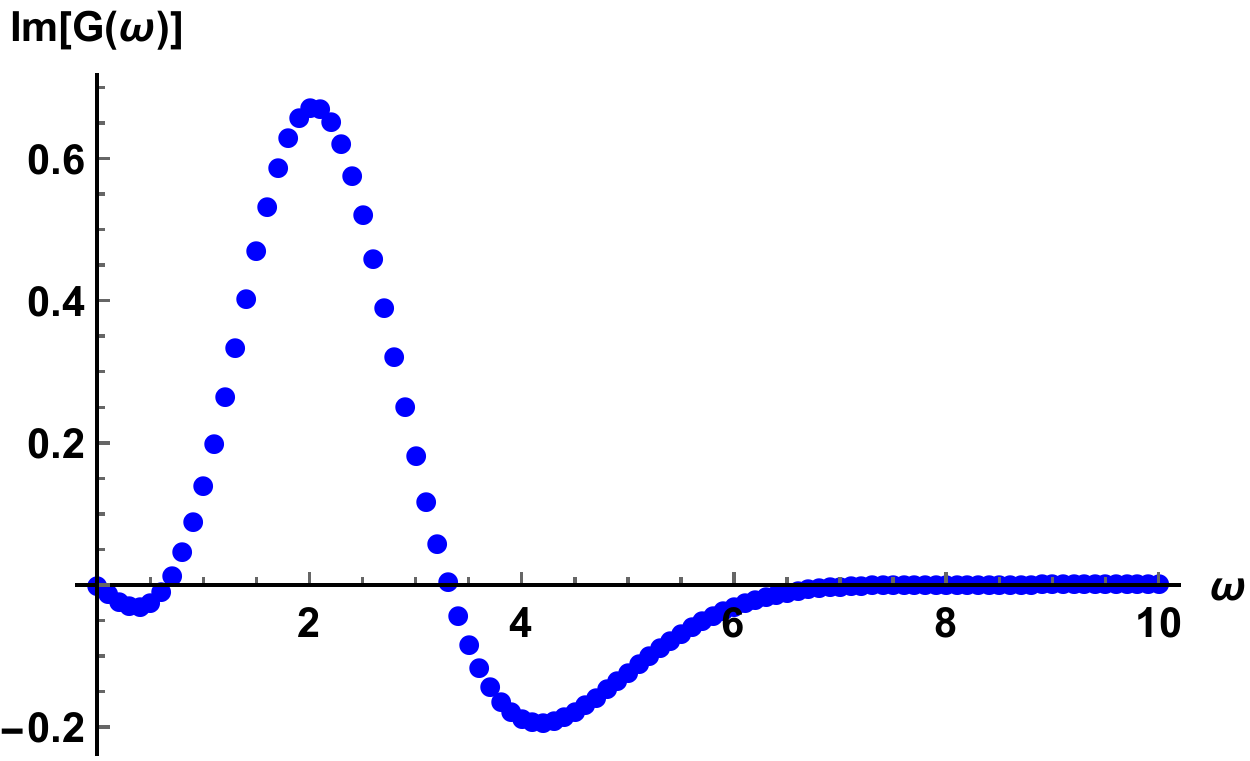}
    \caption{The frequency dependent behaviour of the correlator \eqref{green} extracted numerically from equation \eqref{EQ} for a random value of $m/T$ and for the potential $V(X)=X^3$. The numerical procedure, involving a simple matching technique, can be found in one of the notebooks available with this course.}
    \label{fig:sheargreen}
\end{figure}
For concreteness, let us focus on a specific model where the graviton is massive and which is discussed in \cite{Alberte:2016xja}. For more details about the origin of the model and its physics see \cite{Baggioli:2014roa,Alberte:2015isw,Baggioli:2019abx}. Here we jump directly to the equation for the shear mode, which in this case is:
\begin{equation}
h_{xy}(u) \left(-\frac{2 m^2 V_X\left(u^2,u^4\right)}{f(u)}-\frac{2 i \omega}{u f(u)}\right)+h_{xy}'(u) \left(\frac{f'(u)}{f(u)}+\frac{2 i \omega}{f(u)}-\frac{2}{u}\right)+h_{xy}''(u)=0\label{EQ}
\end{equation}
using EF coordinate. The UV asymptotic behaviour of the $h_{xy}$ field is:
\begin{equation}
h_{xy}\,=h_{xy\,(l)}(\omega)\,(1\,+\,\dots)\,+\,h_{xy\,(s)}(\omega)\,u^{3}\,(1\,+\,\dots)
\end{equation}
The AdS/CFT dictionary allows us to express the Green's function of the stress tensor as
\begin{equation}
 \mathcal{G}^{\textrm{(R)}}_{T_{xy}T_{xy}}(\omega)\,=\,\frac{2\,\Delta-d}{2}\,\frac{h_{xy\,(s)}(\omega)}{h_{xy\,(l)}(\omega)}\,=\,\frac{3}{2}\frac{h_{xy\,(s)} (\omega) }{h_{xy\,(l)} (\omega)}\label{green}
\end{equation}
The first task we want to achieve is to determine numerically the frequency dependence of the Green function in eq.\eqref{green}. This is done with a simple matching method in one of the notebooks available with this course. An example of the results is shown in fig.\ref{fig:sheargreen}.
\begin{mdframed}[style=MyFrame3]
\begin{center}
    \textbf{Exercise \# 7: A numerical Green function.}
    \end{center}\vspace{0.15cm}
    
    Use equation \eqref{EQ} to obtain numerically the green function \eqref{green} for various values of $m/T$ and various potentials $V(X)=X^n$.\\
    Try to implement the numerics with a matching method (as shown in the available notebook) and also with a simple shooting method from the horizon. Using this second method, what you have to do is:
    \begin{enumerate}
        \item Solve the equation close to the horizon, imposing regularity therein.
        \item Use the NDSolve routine of mathematica (\url{https://reference.wolfram.com/language/ref/NDSolve.html}) to solve numerically the differential equation from the horizon to the boundary, by imposing at the horizon the b.c.s. you found in the previous point.
        \item Read numerically the leading and subleading term of the field expansion close to the boundary and use equation \eqref{rout}.
        \item Plot it!
    \end{enumerate}
    Do you get the same results compared to the available notebook which uses the matching procedure?
    Do you notice any difference for $n<3/2$ and $n>3/2$?\\
    Try to find out what is the value of:
    \begin{equation}
        \lim_{\omega \rightarrow \infty}\, Re[\mathcal{G}^R_{xyxy}(\omega,k=0)]
    \end{equation} 
    for various $n$ and $m$. Can you correlate it with any thermodynamic property of the background? Tip: see \cite{Andrade:2019zey}.
\end{mdframed}

\vspace{0.2cm}

Now, after we learned the numerics, we want to do two things: (I) get the analytic result for this correlator at small graviton mass, \textit{i.e.} $m/T \ll 1$ and (II) understand analytically the behaviour at small temperatures, and link it with some geometric property of the solution. Notice that, in this case, because of the presence of a mass, the equation for the shear mode does not anymore correspond to a radial conservation equation and therefore the classical membrane paradigm cannot be used. Nevertheless, even without solving this equation, a lot can be learned. For instance, one can derive that, as far as the mass squared is positive, the KSS bound will be violated $\eta/s < 1/4\pi$ (see \cite{Hartnoll:2016tri}).\\

In order to give you all the flavours of the fundamental computations in holography, this time, we will use Poincare coordinates. More specifically, we consider equation:
\begin{equation}
\label{spin2}
    \left[f\partial_r^2 +\left(f'-2\frac{f}{r}\right)\partial_r + \left(\frac{\omega^2}{f} -4\,m^2 M^2(r)\frac{r^2}{L^2}\right)\right]h=0 \end{equation}
with $M^2(r) \equiv \frac1{2r^2} \hat V_X(r)$, which is nothing else than eq.\eqref{EQ} in the new coordinates.\\
Using the already mentioned techniques, the boundary action for this theory becomes, as always, of the type:
\begin{equation}
S_{\text{bdy}}=\int d^2x\int\frac{d\omega}{2\pi}\,h_0\,\mathcal F(\omega,r)h_0\Big|_{r=\epsilon}\;\label{bubul}
\end{equation}
where $\epsilon$ is the UV cutoff put by hands and $h_0$ is the value of the bulk field at the boundary $h_0 \equiv h(r=0)$.\\
According to the prescription given in \cite{Son:2002sd}, which is equivalent to what we discussed in the previous sections, the retarded Green's function ${\cal G}^R_{T_{ij}T_{ij}}$ can then be extracted from the on-shell boundary action as 
\begin{equation}\label{green2}
{\cal G}^R_{T_{ij}T_{ij}}=-\lim_{\epsilon\to \,0}2\,\mathcal F(\omega,r)\Big|_{r=\epsilon}\;,
\end{equation}
which follows simply by taking two functional derivative of the above boundary action \eqref{bubul} with respect to the source $h_0$.
To solve equation \eqref{spin2}, we assume the following ansatz
\begin{equation}
h(r)\,=\,h_0\,e^{-\frac{i\,\omega}{4\,\pi\,T}\,\log{f}}\,\left(\Phi_0(r)\,+\,\frac{i\,\omega}{4\,\pi\,T}\,\Phi_1(r)\,+\,\mathcal{O}(\omega)^2\right)\;
\label{anse}
\end{equation}
which: (I) implement the infalling boundary conditions with the appropriate exponential prefactor, (II) represents the leading order in the frequency expansion. The idea is to solve perturbatively equation \eqref{spin2} using this ansatz. Two remarks are important. First, the functions $\Phi_0$ and $\Phi_1$ are required to be regular at the horizon, and this will fix for us some of the integration constants. Second, near the boundary, we demand that 
\begin{equation}
\Phi_0(0)\,=\,1\,,\hspace{1cm}\Phi_1(0)\,=\,0\,.
\end{equation}
which is equivalent to impose a unitary source for the dual operator, $h_0=1$, in the language of \eqref{bubul}.
The equations for $\Phi_0$ and $\Phi_1$ can be obtained by solving eq.\eqref{spin2} with the ansatz \eqref{anse} order by order in the frequency $\omega$ and they are:
\begin{align}\label{Phi0}
&\frac{z^2}{f}\left(\frac{f}{z^2}\Phi_0'(z)\right)'-\frac{4 \,m^2\, r_h^4\, z^2\, M^2(z)}{L^2f(z)}\,\Phi_0(z)\,=\,0\,,\\\label{Phi1}
&\frac{z^2}{f}\left(\frac{f}{z^2}\Phi_1'(z)\right)'-\frac{4\, m^2\, r_h^4\,z^2\, M^2(z)}{L^2f(z)}\,\Phi_1(z)-\frac{2 \,f'}{f}\Phi_0'(z)+\Phi_0(z) \left(\frac{2 \,f'}{z
   \,f}-\frac{f''}{f}\right)\,=\,0\,,
\end{align}
To make the computations slightly more compact, we have re-defined the radial coordinate as $z=r/r_h$\footnote{This is equivalent of fixing $r_h=1$.}, and the primes denote derivatives with respect to $z$. Also, since we are interested only in the small graviton mass limit, we can safely neglect the corrections of the mass to the blackening factor $f(z)$. We will use the simple Schwarzschild background $f(z)=1-z^3$; this simplifies the calculations.\\
To proceed, we assume the potential:
\begin{equation}
    V(X)\,=\,X^n\,\,\quad \quad 2\,n\,=\,4\,+\,\nu
\end{equation}
which leads to a mass function of the form
\begin{equation} \label{mass}
    M^2(z)=\frac{z^\nu\,(r_h)^{\nu}}{2\,L^{2+\nu}}
\end{equation}
Using these notations, the graviton mass is now parameterized by the dimensionful parameter $m^2$. Since we are interested in the small mass limit, $m/T \ll 1$, we additionally expand the ansatz \eqref{anse} in a perturbative series in $m^2$ :
\begin{align}\label{Phisol}
\Phi_0=\sum_{n=0}^\infty m^{2n}\phi_n\;,\qquad \Phi_1=\sum_{n=0}^\infty m^{2n}\psi_n
\end{align}
with $\phi_0=1$ and $\psi_0=0$.\\
Collecting all our assumptions, the final equations we need to solve are:
\begin{align}\label{eq_phi}
\left(\frac{f}{z^2}\phi_1'\right)'&=4\frac{r_h^4}{L^2}M^2(z)=2\left(\frac{r_h}{L}\right)^{4+\nu}\,z^\nu\,,\\\label{eq_psi}
\left(\frac{f}{z^2}\psi_1'\right)'&=2\frac{f'}{z^2}\phi_1'\,
\end{align}
The quadratic on-shell action, computed at the boundary $z=\epsilon$, is given by
\be\label{act_bdy_0}
S_{\text{bdy}}\,=-\,\frac{L^2}{4r_h^3}\int d^3x\,\left(\frac{f}{z^2}\,h(z)\,h'(z)\right)\Bigg|_{z=\epsilon}\;
\ee
which expanded at first order in the graviton mass $m^2$ and the frequency $\omega$ reads
\begin{equation*}
S_{\text{bdy}}=\int d^3x\,\frac{L^2h_0^2}{4\,r_h^3}\left[m^2\left(\frac{f}{\epsilon^2}\phi_1'\right)+\frac{i\omega}{4\pi T}\left(3+m^2\left(\frac{f}{\epsilon^2}\psi_1'\right)-2\,m^2\log f\left(\frac{f}{\epsilon^2}\phi_1'\right)\right)\,+\,\dots\right]\,
\end{equation*}
where the ellipsis indicates terms $\mathcal{O}(\omega^2, \mathcal{O}(m^4)$.\\
At this order of approximation, the Green's function \eqref{green2} can be computed exactly. Its real and imaginary parts are determined by:
\begin{align}
&G=\text{Re} \,{\cal G}=-\frac{L^2}{2r_h^3}\,\lim_{\epsilon\to\, 0}\,m^2\left(\frac{f}{\epsilon^2}\phi_1'\right)\;\\
&\eta\,=\,\lim_{\omega \rightarrow 0} \left[-\,\frac{1}{\omega}\text {Im}\,{\cal G}(\omega)\right]\,=\,\frac{L^2}{2r_h^3}\,\lim_{\epsilon\to0}\,\frac{1}{4\pi T}\left[3+m^2\left(\frac{f}{\epsilon^2}\psi_1'\right)\right]
\end{align}
Interestingly, for any given $M^2(z)$, the equation of motion \eqref{eq_phi} can be integrated directly and it gives us a very generic expression for the shear elastic modulus
\be\label{eq_phi02}
G= 2 \,{m^2 r_h}\lim_{\epsilon\to\, 0}  \int_{\epsilon}^{1} M^2(z)\, dz\;.
\ee
which does not depend on the explicit form of the mass function. This result immediately tells us that, as soon as the graviton mass is not zero, a finite elastic modulus appears. The graviton mass makes somehow the black hole rigid, like a solid.
For the mass function \eqref{mass}, the equation for $\phi_1$ is solved by
\be\label{eq_phi2}
\frac{f}{z^2}\phi_1'=c_\nu\left(-z^{\nu+1}+1\right)\,,\qquad\nu\neq -1\,
\ee
where the integration constant $c_\nu$ is fixed by requiring regularity at the horizon $z=1$:
\be\label{cnu}
c_\nu=\frac{-2}{\nu+1}\left(\frac{r_h}{L}\right)^{4+\nu}\,.
\ee
Putting all together, we obtain the shear elastic modulus of our theory at leading order in $m^2$, which is given by the generic formula
\be\label{real}
G\,=\,-\,\frac{L^2}{2\,r_h^3}\,m^2\,c_\nu\,+\,\mathcal{O}(m^4),\qquad\nu\neq -1\,.
\ee
Notice that the shear modulus is positive, $G>0$, only for $\nu>-1$, which coincides with $n>3/2$. This has to do with the dual field theory interpretation of this model. Without entering the details, this model displays the spontaneous symmetry breaking of translations only for certain values of $n$ \cite{Alberte:2017cch}. Only in that limit, the shear modulus is well defined and accompanied by the presence of the corresponding Goldstone bosons, the phonons. Not surprisingly the speed of the phonons is determined by the elastic modulus that we are discussing \cite{Alberte:2017oqx,Ammon:2019apj}.
\begin{figure}
    \centering
    \includegraphics[width=0.3\linewidth]{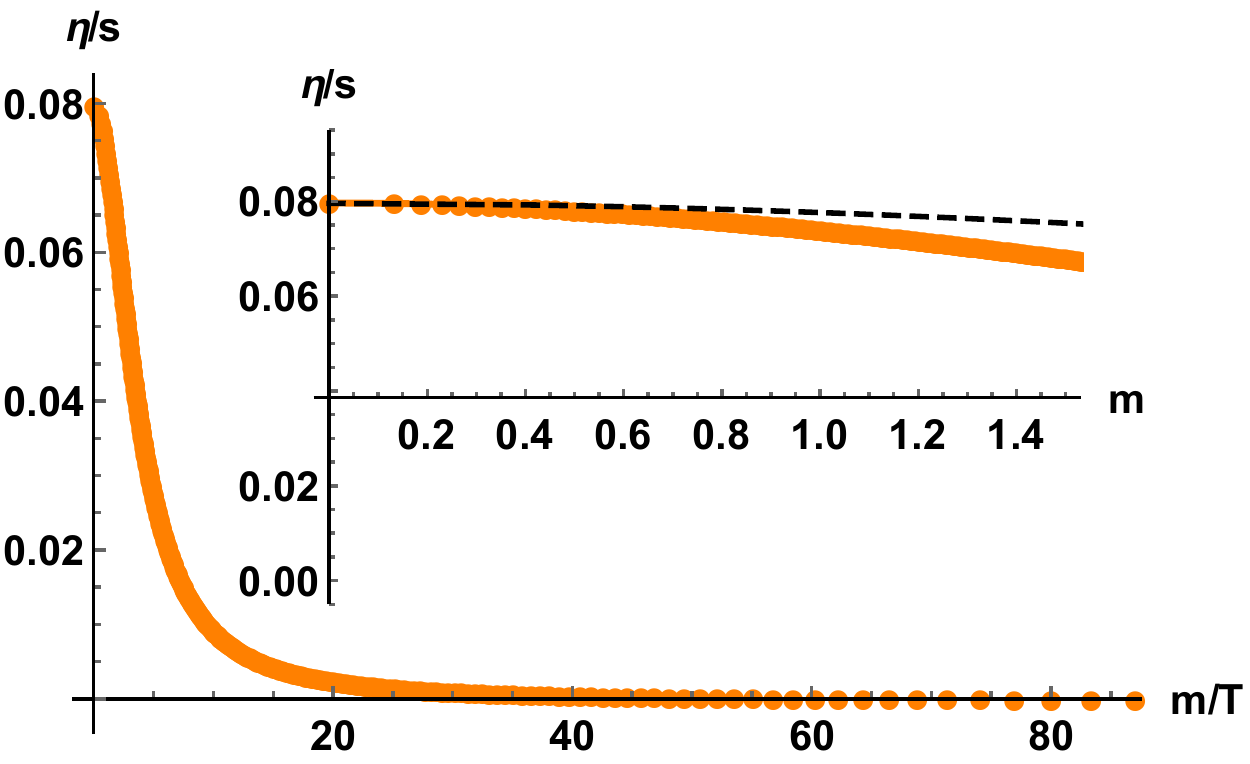}\quad 
    \includegraphics[width=0.3\linewidth]{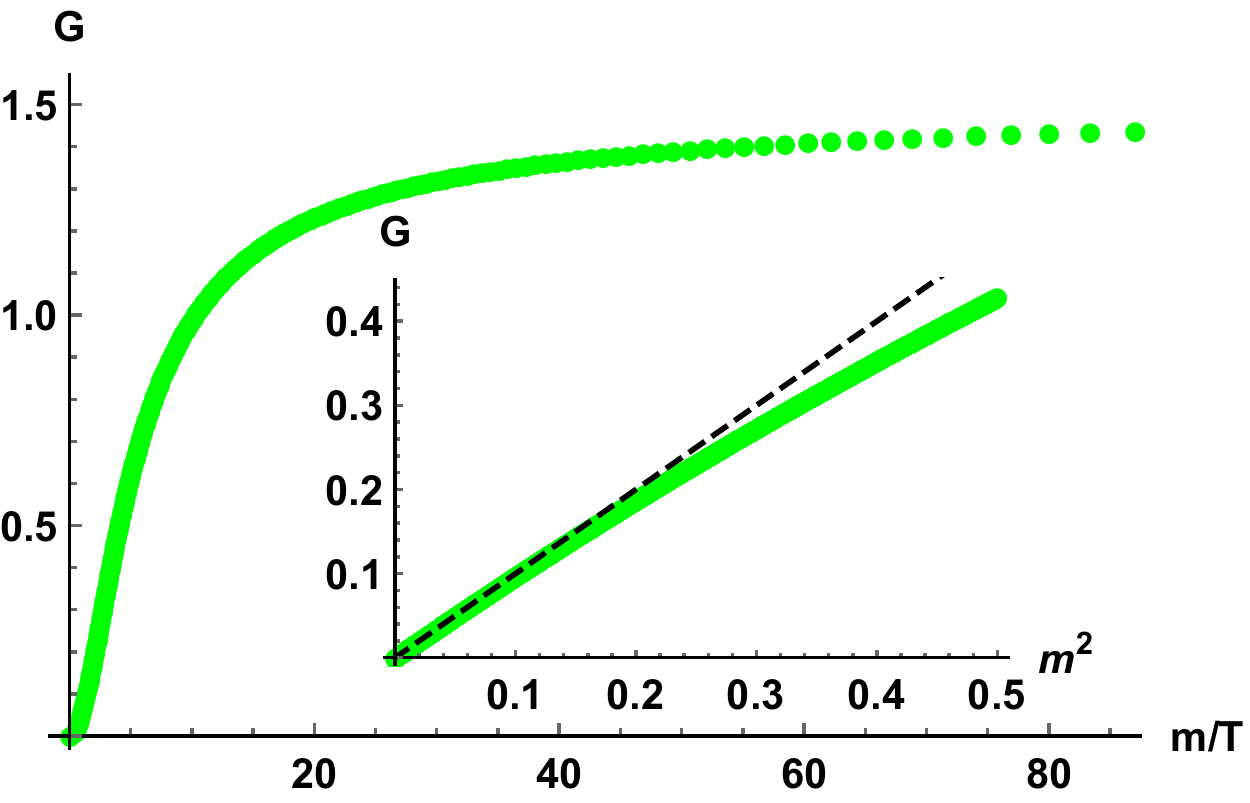}
    \quad 
    \includegraphics[width=0.3\linewidth]{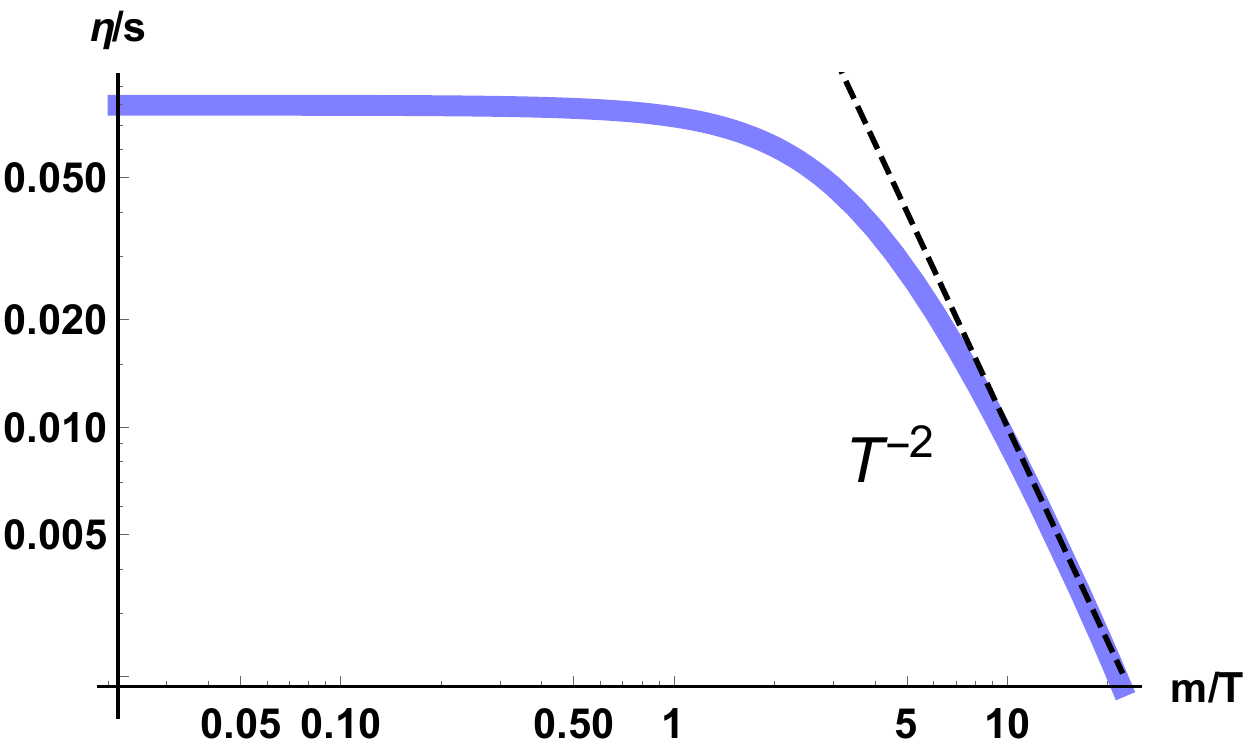}
    \caption{Comparison between the numerical results and the analytic perturbative formulas for the massive gravity model of \cite{Alberte:2016xja}. These plots are built using the notebooks available together with this course.}
    \label{fig:viscoelastic}
\end{figure}\\
We can perform the analogous steps for the imaginary part of the Green's function:
\be
\text {Im}\,{\cal G}=-\,\frac{L^2}{2r_h^3}\,\lim_{\epsilon\to0}\,\frac{i\omega}{4\pi T}\left[3+m^2\left(\frac{f}{\epsilon^2}\psi_1'\right)\right]
\ee
where this time
\begin{equation}
\frac{f}{z^2}\psi_1'\,=\,2\,c_\nu\,\int_{1}^z \frac{f'}{f}\,\left(1-x^{\nu+1}\right)\,dx\;
\end{equation}
By using this result, we are now able to find the viscosity to entropy density ratio
\begin{equation}
\frac{\eta}{s}\,=\,\frac{1}{4\,\pi}\left(1+\frac{2}{3}\,c_\nu\,m^2\,H_{\frac{1}{3}(\nu+1)}\,+\,\mathcal{O}(m^4)\right)\label{imi}
\end{equation}
where we use the mathematical definition of the Harmonic numbers
\be
\mathcal H_{\frac{1}{3}(\nu+1)}\equiv\int_{1}^0 \frac{f'}{f}\,\left(1-x^{\nu+1}\right)\;.
\ee
and we have used the fact that the temperature of the Schwarzschild-AdS solution with the emblackening factor $f(z)=1-z^3$ is simply $T=3/(4\pi r_h)$, and the entropy density is given by $s=2\pi(L^2/r_h^2)$.\\ For similar perturbative computations see also \cite{Hartnoll:2016tri,Burikham:2016roo}.
What we found is a violation of the KSS bound due to the introduction of a graviton mass. The interpretation of this result is still under debate.\\
In fig.\ref{fig:viscoelastic} we display the agreement at small $m/T$ of the analytic results with the numerical data.\\
\begin{mdframed}[style=MyFrame3]
\begin{center}
    \textbf{Exercise \# 8: Perturbative methods.}
\end{center}\vspace{0.15cm}
Try to obtain the next order in the analytic formulae \eqref{real},\eqref{imi} and see if now the agreement with the numerics shown in fig.\ref{fig:viscoelastic} improves.
\end{mdframed}
Before jumping to another topic, we want to address an additional question. We proved perturbatively that the $\eta/s$ ratio decreases upon increasing the dimensionless parameter $m/T$. How does this ratio behave at very small values of $T$? Does it go to zero? How?\\
In order to do that, we have to understand how the background geometry looks like at $T=0$. As we explained in the first section, moving along the radial extra-dimension in the bulk corresponds to moving in energy in the dual field theory. This means that the dynamics at high temperature ($=$ high energy) is determined by the dynamics of the bulk fields close to the boundary of the spacetime, whether the dynamics at low energy (which can be low temperature, low frequency, etc...) is governed only by what happens close to the horizon of the black hole geometry, in what is called the \textit{near-horizon geometry}.\\
In our model, and for all the charged black holes, the geometry of the black holes at zero temperature interpolates between:
\begin{equation}
   \underbrace{\text{AdS}_{d+1}}_{\text{UV region}}\quad \rightsquigarrow \quad \underbrace{\text{AdS}_2\,\times\,\mathbb{R}^{d-1}}_{\text{IR region}}
\end{equation}
What does that mean?\\
Consider the following BH metric:
\begin{equation}
    ds^2\,=\,\frac{1}{u^2}\,\left[-f(u)\,dt^2\,+\,\frac{du^2}{f(u)}\,+dx^2\,+\,dy^2\,\right]
\end{equation}
The function $f(u)$ can be expanded close to the horizon as:
\begin{equation}
    f(u)\,=\,-\,4\,\pi\,T\,(u-u_h)\,+\,\frac{1}{2}\,f''(u_h)\,(u-u_h)^2\,+\,\dots
\end{equation}
which means that at $T=0$ the function has a double zero. Consider indeed the case of zero temperature and perform the following coordinate transformation:
\begin{equation}
    \Upsilon\,=\,\frac{\varrho}{u\,-\,u_h}
\end{equation}
with $\varrho$ an arbitrary coefficient.
Now, the original geometry is given by
\begin{equation}
ds^2\,=\,-\,\frac{f_2\,\varrho^2}{2\,u_h^2}\,\frac{dt^2}{\Upsilon^2}\,+\,\frac{2}{u_h^2\,f_2}\,\frac{d \Upsilon^2}{\Upsilon^2}\,+\,\frac{1}{u_h^2}d \vec{x}^2
\end{equation}
where we have used $f_2 \equiv f''(u_h)$ in order to keep the expressions compact.
Fixing $\varrho=2/f_2$, we obtain the final metric:
\begin{equation}
ds^2\,=\,\frac{2}{u_h^2\,f_2}\,\frac{-dt^2\,+\,d \Upsilon^2}{\Upsilon^2}\,+\,\frac{1}{u_h^2}\,d \vec{x}^2\,=\,\frac{1}{L^2_2}\,\frac{-dt^2\,+\,d \Upsilon^2}{\Upsilon^2}\,+\,d \vec{\tilde{x}}^2\label{ads2}
\end{equation}
which is indeed the form of an $AdS_2 \times R^2$ spacetime with the AdS$_2$ radius given by $L_2^2=2/(u_h^2\,f_2)$.\\
To understand the behaviour of the green function at zero temperature, we have to study the solution of the equation \eqref{spin2} on top of the $AdS_2 \times R^2$ spacetime. Using this technique, we will be able to prove that at low temperature the viscosity ratio scales like
\begin{equation}
\frac{\eta}{s}\,\sim\,T^{2\,\alpha}\qquad\text{as}\qquad T\rightarrow 0
\end{equation}
where $\alpha$ is determined by the conformal dimension of the $T^{xy}$ operator $\Delta=1+\alpha$ in the extremal $AdS_2\,\times\,\mathbb{R}^2$ geometry.\\
We are going now to sketch the proof; more details can be found in \cite{Ling:2016ien} and more generally in \cite{Faulkner:2009wj}. Assuming a power law ansatz for the bulk field $h$ of the type:
\begin{equation}
h(u)\,\sim\,\left(u\,-\,u_h\right)^\nu
\end{equation}
the equation for the shear perturbation around the near-horizon geometry simplifies into
\begin{equation}
f''(u_h)\,\nu\,\left(1\,+\,\nu\right)\,-\,2\,m^2(u_h)\,=\,0\,\label{ss2}
\end{equation}
where we defined the generic graviton mass:
\begin{equation}
    m^2(u)\,=\,g^{xx}\,T_{xx}\,-\,\frac{\delta\,T_{xy}}{\delta g_{xy}}
\end{equation}
Inverting the previous expression we find
\begin{equation}
\nu\,=\,-\,\frac{1}{2}\,\pm\,\frac{1}{2}\sqrt{\frac{8 \,m^2(u_h)}{f''(u_h)}+1}\,\equiv -\,\frac{1}{2}\,\pm\,\xi
\label{nuformula}
\end{equation}
where $\xi>0$. Following with the computations, we can write the generic solution as
\begin{equation}
    h(u)\,=\,\#\,\left(u^{-1/2-\xi}\,+\,\mathcal{G}_{xyxy}(\omega)\,u^{-1/2+\xi}\right)
\end{equation}
Now, we know that the frequency $\omega$, simply because dimensional analysis (see \cite{Faulkner:2009wj} for a much more rigorous derivation), scales like the inverse of the radial coordinate $\omega \sim u^{-1}$. This implies that:
\begin{equation}
    \omega^{1/2+\xi}\,\sim\,\mathcal{G}_{xyxy}(\omega)\,\omega^{1/2-\xi}\,\quad \longrightarrow \quad \mathcal{G}_{xyxy}(\omega)\,\sim\,\omega^{2\,\xi}
\end{equation}
and therefore:
\begin{equation}
    \eta(T)\,\sim\,T^{2\,\xi\,-\,1}
\end{equation}
The last step consists in realizing that the entropy corresponding to an AdS$_2$ spacetime is constant at zero temperature\footnote{ There are various ways to obtain this result. One possibility is to realize that AdS$_2$ is hiddenly a Lifzhitz spacetime with infinite dynamical exponent $z=\infty$ and that the entropy scales like $s\sim\,T^{d/z}\sim cost.$! This property is recently very discussed \cite{Sachdev:2019bjn}, because of its relation with the SYK model \cite{Sarosi:2017ykf}, the strange metals phenomenology and the physics of glasses \cite{Facoetti:2019rab}. Have a look!}, such that we finally obtain:
\begin{equation}
    \frac{\eta}{s}\,\sim\,T^{2\,\xi\,-\,1}\,\quad \,\text{in the limit of}\quad T \,\rightarrow\,0
\end{equation}
Notice that in the limit of zero graviton mass $\xi=1/2$, and therefore the ratio is constant, as we know. Notice also that the important point is the value of the graviton mass at the horizon! One could build models (and they exist \cite{Donos:2013eha}) in which there is a finite graviton mass, but it becomes zero at the horizon. In those models. the value of $\eta/s$ interpolates between $1/4\pi$ at high temperature to another, and smaller, constant value at zero temperature.\\
In order to understand the $\eta/s \sim T^2$ scaling, which we observe in the numerics, we have to consider an additional ingredient. Concretely, the Einstein's equations projected on the extremal horizon give the non-trivial relation:
\begin{equation}
    f''(u_0)\,=\,m^2(u_0)
\end{equation}
where $u_0\equiv u_h(T=0)$ is the position of the extremal horizon. Using the above relation into eq.\eqref{nuformula}, we obtain that $\xi=3/2$ and therefore:
\begin{equation}
    \frac{\eta}{s}\,\sim\,T^2\,\quad \,\text{in the limit of}\quad T \,\rightarrow\,0
\end{equation}
\begin{mdframed}[style=MyFrame3]
\begin{center}
    \textbf{Exercise \#9 : The effects of charge on $\eta/s$.}
    \end{center}\vspace{0.15cm}
In translational invariant systems the ratio $\eta/s$ does not depend on the charge density of the system. You can convince yourself by using the membrane paradigm on the Reissner Nordstrom solution.\\
Once translations are broken, the charge density might have an effect, and in fact it does. Try to compute the $\eta/s$ ratio numerically in the same model considered in this section but at finite charge density.\\
What you will discover is that the presence of a finite charge will slow down the power law decay at $T \rightarrow 0$. You can also derive this result using the same analytic argument of this section. It boils down to the fact that the Einstein's equations at the extremal horizon now implies a relation of the type:
\begin{equation}
     f''(u_0)\,=\,m^2(u_0)\,+\,\#\,\rho^2\,,\quad \quad \#>0
\end{equation}
This was the reason that led the authors of \cite{Hartnoll:2016tri} to claim that $T^2$ was the fastest decay possible. It turns out that it is not true anymore if one considers more complicated IR fixed points \cite{Ling:2016ien}.
\end{mdframed}

%
%
%
\chapter{Holographic Transport via analytic and numerical techniques}
\label{intro4} 
\hspace{0.2cm} \includegraphics[width=0.5\textwidth]{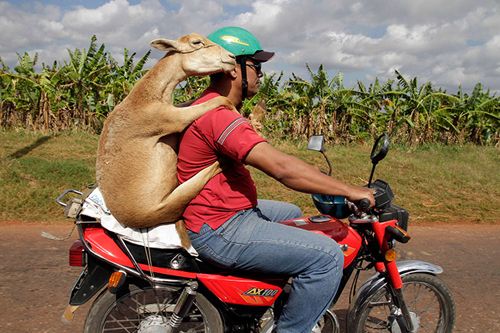}\\
\epigraph{A wise man can learn more from a foolish question than a fool can learn from a wise answer.}{\textit{Bruce Lee}}
\subsection{Linear response, another example}
Let us insist on linear response theory and the definition of the Green function. Take a system and apply an external source to it, \textit{i.e}  $\text{source}(\omega,k)$, which is frequency and momentum dependent. Read the response of the system: $\text{response}(\omega,k)$. At leading order, the two are related by a linear map
\begin{equation}
   \text{response}(\omega,k)\,=\,\text{Green Function}(\omega,k)\,\,\text{source}(\omega,k)\,+\,\mathcal{O}(\text{source}^2)
\end{equation}
whose frequency and momentum dependent ''coefficient'' is the Green function. This is an entire industry, whose foundations can be found in \cite{1987imsm.book.....C,PhysRev.86.702,doi:10.1002/9783527638154.ch2,doi:10.1143/JPSJ.12.570}.\\
Here, we introduce another very simple example: the electric conductivity. We apply an external electric field $E$, which will be our source, and we measure the produced electric current $J$, which is going to be our response. We do not have to explain that the conductivity is defined as:
\begin{equation}
    J\,=\,\sigma\,E
\end{equation}
which is simply Ohm's law. The current $J^\mu$ is our operator and it is sourced by a vector field $A^\mu$. The electric field relates to the external vector field via:
\begin{equation}
    E_i\,=\,-\,i\,\omega\,A_i
\end{equation}
where, to be general, we indicate with $i$ the direction of the electric field. The previous relation comes directly from the definition of electric field $E_i \equiv F_{ti}$, and the observation that $\partial_t \sim\,i \omega$. Therefore, we can immediately write down:
\begin{equation}
    \underbrace{J_i(\omega)}_{\text{response}}\,=\,\underbrace{-\,\sigma(\omega)\,i\,\omega}_{\text{Green Function}}\,\underbrace{A_i(\omega)}_{\text{source}}
\end{equation}
which implies that:
\begin{equation}
    \sigma(\omega)\,=\,\frac{i}{\omega}\,\mathcal{G}_{JJ}(\omega)
\end{equation}
In other words, the electric conductivity relates directly to the Green Function of the $U(1)$ conserved current $J$. In this section, we will play with various methods and we will discuss how to extract the electric conductivity from a charge black hole using the holographic dictionary.
\subsection{Charged Black holes and a Superconducting instability: an excuse to play with near-horizon geometries}
\begin{figure}[h!]
    \centering
    \includegraphics[width=0.7 \linewidth]{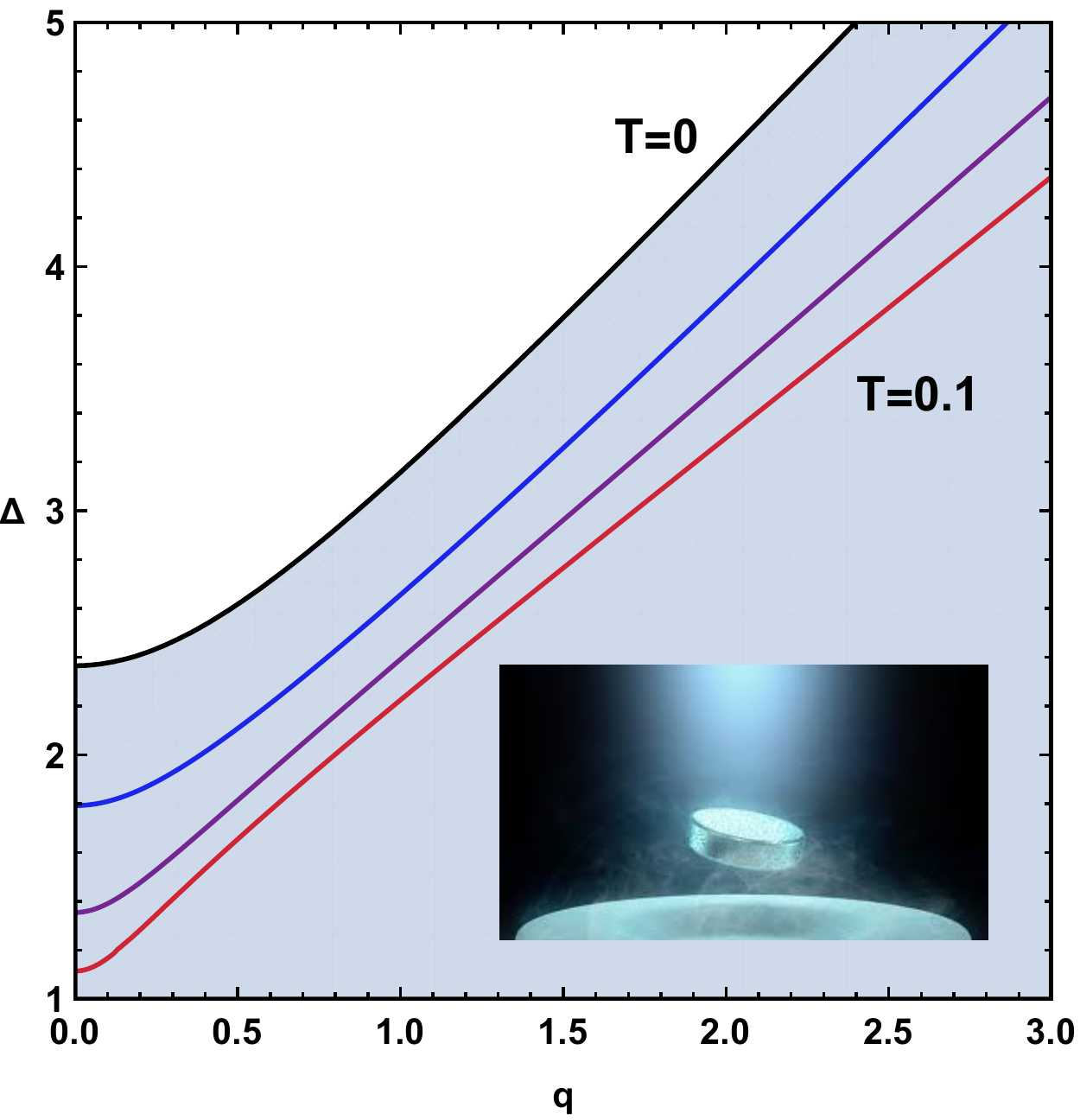}
    \caption{The onset of instability for the holographic superconductor model. The black line is the zero temperature result. The other lines are obtained numerically as described in the main text. Similar results can be found for example in \cite{Denef:2009tp}.}
    \label{fig:instability}
\end{figure}

\begin{mdframed}[style=MyFrame3]
\begin{center}
    \textbf{Exercise \#10 : The RN Black Hole, the soul of AdS-CMT.}
\end{center}\vspace{0.15cm}
Derive the Reissner-Nordstrom solution shown in this section.\\
You can play and deform the theory as:
\begin{equation}
     S\,=\,\int d^4x \left[\,R\,-\,2\,\Lambda\,-\,\mathcal{K}\left(\frac{1}{4}\,F^2\right)\right]
\end{equation}
where $\mathcal{K}$ is an arbitrary function.
How is the background solution modified? Have a look at \cite{Baggioli:2016oju}.
\end{mdframed}
In this section, we consider Einstein-Maxwell theory defined via the action
\begin{equation}\label{EMax}
    S\,=\,\int d^4x \left[\,R\,-\,2\,\Lambda\,-\,\frac{1}{4}\,F^2\right]
\end{equation}
Assuming the following geometry for the background
\begin{equation}
    ds^2\,=\,\frac{1}{u^2}\,\left[-f(u)\,dt^2\,+\,\frac{du^2}{f(u)}\,+dx^2\,+\,dy^2\,\right] \label{mm}
\end{equation}
the most general solution is given by
\begin{equation}
    A_t\,=\,\mu\,-\,\rho\,u\,,\quad f(u)\,=\,u^3\,\int_{u_h}^u\,\left(-\frac{3}{y^4}\,+\,\frac{\rho^2}{4}\right)\,dy
\end{equation}
where $\mu$ is the chemical potential and $\rho$ is the charge density of the dual field theory. \begin{mdframed}[style=MyFrame4]
\begin{center}
    \textbf{The chemical potential and the charge density.}
\end{center}\vspace{0.15cm}
We have just stated that the background solution for the gauge field is:
\begin{equation}
    A_t\,=\,\mu\,-\,\rho\,u\,,
\end{equation}
where $\mu$ is the chemical potential and $\rho$ is the charge density of the dual field theory. Why is that?\\
It is simply the application of the dictionary. The subleading term for the above solution is the VEV of the dual operator $J_t$ an it corresponds indeed to a finite charge density. The first term is the associated source, which is this time the chemical potential. It is straightforward to test that such solution implies a boundary deformation of the type:
\begin{equation}
    \delta \mathcal{S}\,=\,\int\,d^dx\,\mu\,\rho
\end{equation}
A more sophisticated analysis shows that also all the known thermodynamic properties hold. For example, the computation on the on-shell action gives the known thermodynamic relation:
\begin{equation}
    d\,\epsilon\,=\,T\,ds\,+\,\mu\,d \rho
\end{equation}
See for example the appendix in \cite{Baggioli:2015zoa} for a proper derivation.
\end{mdframed}
\begin{mdframed}[style=MyFrame3]
\begin{center}
    \textbf{Exercise \#11: Why we are obsessed with dimensionless quantities?}
\end{center}\vspace{0.15cm}
Most of the systems we consider in holography are scale invariant. The simplest example is pure Einstein-Hilbert gravity on AdS spacetime, which is dual to a conformal field theory. Why scale invariance can be tricky and what is it? Let us provide two simple examples.\\[0.15cm]
\textbf{The temperature of AdS.}\\
Consider the simple Schwarzschild solution in $d=4$ coming from:
\begin{equation}
    S\,=\,\int d^4x\,\sqrt{-g}\,\left[R\,+\frac{6}{L^2}\right]
\end{equation}
and given by:
\begin{equation}
    ds^2\,=\,\left[-f(r) \,dt^2\,+\,\frac{dr^2}{f(r)}\,+\,r^2\,d\vec{x}^2\right]\,,\quad f(r)\,=\,r^2\,\left[1-\,\left(\frac{r_h}{r}\right)^3\right]
\end{equation}
The temperature of the BH solution and of the dual field theory is just:
\begin{equation}
    T\,=\,\frac{3\,r_h}{4\,\pi\,L^2}
\end{equation}
Is this meaningful? Is $T=100$ warmer than $T=10$? No! The value of the temperature is completely meaningless and the only thing that matters is if it is zero or finite. Why?\\
First, let us write down the Einstein's equations:
\begin{equation}
    R_{\mu\nu}\,-\,\frac{1}{2}\,g_{\mu\nu}\,R\,-\,\frac{3}{L^2}\,g_{\mu\nu}\,=\,0
\end{equation}
The system possesses various scaling symmetries, which leave the system invariant. One of them (find the others!) is the scaling transformation:
\begin{equation}
    r\,\rightarrow \lambda\,r\,,\quad t\,\rightarrow \lambda\,t\,,\quad L\,\rightarrow \lambda\,L \label{tt}
\end{equation}
under which the metric and the geometrical objects transform as:
\begin{equation}
    g_{\mu\nu}\,\rightarrow \lambda^2\,g_{\mu\nu}\,,\quad  R_{\mu\nu}\,\rightarrow \,R_{\mu\nu}\,,\quad  R\,\rightarrow \lambda^{-2}\,R
\end{equation}
such that the Einstein's equation is left invariant.\\
The transformation in eq.\eqref{tt} connects equivalent solutions, but it shifts the temperature as:
\begin{equation}
    T\,\rightarrow T/\lambda
\end{equation}
This means that all the finite temperatures are equivalent, because they can be connected to each other by the above scaling transformation. Physically, this means that, in a scale invariant system, no dimensionful scale can appear; therefore, in presence of only one dimensionful quantity, the temperature $T$, its value is not physical.\\[0.15cm]

\textbf{Reissner-Nordstrom solution}\\
Now, let us make a step forward and consider the following system:
\begin{equation}
    S\,=\,\int d^4x\,\sqrt{-g}\,\left[R\,+\frac{6}{L^2}\,\,-\,\frac{1}{4\,g^2}\,F^2\right]
\end{equation}
The additional profile for the gauge field is given by:
\begin{align}
    A\,=\,A_t\,dt\,,\quad A_t\,=\,\mu\,\left(1\,-\,\frac{r_h}{r}\right)
\end{align}
The scaling transformation of eq.\eqref{tt} is still a symmetry of the system, but only if we properly rescale the chemical potential as follows:
\begin{equation}
    \mu\,\rightarrow \mu/\lambda
\end{equation}
This time, we can construct a dimensionless ratio $T/\mu$, which is left invariant under the above scaling transformation. This means that solutions with different $T/\mu$ are now different solutions and they cannot be connected with any scaling transformation. The value of $T/\mu$ is physical! This is the reason why in scale invariant system we always considered dimensionless quantities, which do not scale under scaling transformations. Another typical example is to consider $\omega/T$ or $k/T$.\\

\textbf{Exercise.} Scaling symmetries can be also always used to set the value of the horizon radius to $1$. This simplifies lot of the computations and it is perfectly legitimate as far as you consider only ''meaningful'' dimensionless quantities. Find out which are the other scaling symmetries. If you get stuck have a look at \cite{Hartnoll:2008kx} for help.
\end{mdframed}
\vspace{0.2cm}
As explained before, the RN black hole, at $T=0$, is a geometry interpolating between an AdS spacetime in the UV to an AdS$_2$ in the deep IR. This last manifold is very often encountered in applied holography, since it represents the near horizon geometry of every charged black hole. It has very interesting properties and it is very delicate, in the sense that is higly unstable towards all deformations. As a pratical example, let us imagine to add to the Einstein-Maxwell action in \eqref{EMax} a massive complex scalar charged under the bulk $U(1)$ symmetry:
\begin{equation}
    \mathcal{L}_{\psi}\,=\,|D \psi|^2\,+\,M^2\,\psi^2
\end{equation}
where $D_\mu \equiv \partial_\mu-i q A_\mu$ is the typical covariant derivative.\\
The equation for the scalar on the background is given by:
\begin{equation}
    \psi''+\left(-\frac{2}{u}+\frac{f'}{f}\right)\psi'
+\left(\frac{q^2 A_t^2}{f^2}-\frac{M^2 L^2}{u^2 f}\right)\psi\,=0
\end{equation}
and it can be re-written in the following more illustrative form
\begin{equation}
    \Box \,\psi\,=\,M^2_{eff}(u) \,\psi
\end{equation}
whith the effective mass given by
\begin{equation}
    M_{eff}^2\,L^2 =\,M^2\,L^2+q^2\,A_t^2\, g^{tt}\,L^2 \label{massmass}
\end{equation}
and $\Box$ defined as the Laplacian operator \footnote{Its definition is given by:
\begin{equation}
    \Box\,\mathrm{F}\,\equiv\,\frac{1}{\sqrt{-g}}\,\partial_\sigma \sqrt{-g}\,\partial^\sigma\,\mathrm{F}
    \end{equation}} on the curved metric in \eqref{mm}.\\
    What is the interesting thing here? The second term in the effective mass \eqref{massmass} is negative and it grows towards the horizon! Physically, this means that going towards low energy, or equivalently to low temperature, the mass of the scalar becomes smaller and eventually negative!
\begin{mdframed}[style=MyFrame4]
\begin{center}
    \textbf{Breitenlohner Freedman stability condition.}
\end{center}\vspace{0.1cm}
We give a sketchy derivation of the BF bound, which ensures the stability of a curved spacetime under a specific deformation driven by a massive bulk field. More specifically, the Breitenlohner Freedman stability condition requires that the mass of a free scalar field on a curved spacetime has to satisfy a lower bound:
\begin{equation}
    m^2\,>\,m_{BF}^2
\end{equation}
which depends crucially on the curvature of the geometric background. For all the details we refer to the original works \cite{1982AnPhy.144..249B,BREITENLOHNER1982197} or the more modern discussion of \cite{Moroz:2009kv}.\\
Let us consider an AdS$_{d+1}$ spacetime defined by the metric:
\begin{equation}
        ds^2\,=\,\frac{dz^2\,+\,\eta_{\mu\nu}\,dx^\mu dx^\nu}{z^2}
\end{equation}
where the $z$ coordinate span from $z=0$ to $z=\infty$. The field equation for a scalar $\phi$ with effective mass $m_{eff}^2$ on this background can be written as:
\begin{equation}
    \partial_z^2\,\phi\,-\,\frac{d-1}{z}\,\partial_z\,\phi\,-\,\frac{m_{eff}^2}{z^2}\,\phi\,+\,\left(\omega^2\,-\,k^2\right)\,\phi\,=\,0
\end{equation}
Using the field redefinition $\phi=z^{(d-1)/2}\,\psi$ we can write the equation above in the suggesting form:
\begin{equation}
    -\,\partial_z^2\,+\,\frac{m_{eff}^2\,+\,\frac{d^2-1}{4}}{z^2}\,\psi\,=\,(\omega^2\,-\,k^2)\,\psi
\end{equation}
which corresponds to a Schrodinger equation with potential:
\begin{equation}
    V(z)\,=\,\frac{m_{eff}^2\,+\,\frac{d^2-1}{4}}{z^2}
\end{equation}
This type of problem has been studied extensively and it is known that it admits a stable solution only if:
\begin{equation}
    V\,>\,-\frac{1}{4}
\end{equation}
This converts the stability condition into a bound for the mass which reads:
\begin{equation}
    m_{eff}^2\,>\,-\,\frac{d^2}{4}
\end{equation}
which is the already mentioned BF bound. Interestingly enough, the mass squared of the bulk field can be negative without spoiling the stability of the system. This is a result of the curvature of the spacetime, which accounts for an additional positive contribution to the eigenvalues energies.\\
From the holographic point of view, the BF bound comes directly from solving the indicial equation for a power law ansatz. Requiring the power to be real gives immediately the BF condition (check it).
\end{mdframed}
\vspace{0.3cm}
The stability requirement around the AdS$_2$ near-horizon geometry is given by
\begin{equation}
    M_{eff}^2\,L^2_{2}< -\frac{1}{4}
\end{equation}
It is simple to determine analytically this condition in terms of the charge $q$ and the conformal dimension $\Delta$ of the charged scalar, see fig.\ref{fig:instability}. The BF condition only tells you if the background at zero temperature is stable or not, but it does not tell you any information about the temperature at which the instability appears. To find that, a bit more of work is needed. First, let us connect the bare mass of the complex scalar with its conformal dimension and with its asymptotic behaviour at the boundary:
\begin{equation}
    M=\frac{1}{L}\sqrt{\Delta(\Delta-3)}\,,\quad \psi(u)=\frac{\psi_1}{L^{3-\Delta}}u^{3-\Delta}+\frac{\psi_2}{L^\Delta}u^\Delta\,
\end{equation}
Using the holographic dictionary we identify the coefficient $\psi_1$ as the source for the dual operator. We expect that, decreasing the temperature from $T=\infty$, an instability might occur at a certain critical temperature $T=T_c$. Close to the instability, the value of $\psi$ is small, and therefore one can safely neglect its backreaction on the geometry. This means that the only equation we have to consider is:
\begin{equation}
    \psi''+\left(-\frac{2}{u}+\frac{f'}{f}\right)\psi'+\left(\frac{q^2\rho^2}{f^2}-\frac{M^2 L^2}{u^2 f}\right)\psi\,=0\, \label{tosol}
\end{equation}
How are we going to find the onset of instability?
Imagine to collect the excitations of the system at zero momentum using their Fourier decomposition:
\begin{equation}
    e^{-i\,\omega_1\,t}\,,e^{-i\,\omega_2\,t},\,\dots\,,e^{-i\,\omega_n\,t}
\end{equation}
where $\omega_n$ are a priori complex frequencies. These excitations are called quasi-normal modes\footnote{In contrast with the normal frequencies with $\omega_n$ real.} and they decay in time within a timescale determined by their imaginary parts. Now, let us imagine that one of those modes has a eigenfrequency with a positive imaginary part; this corresponds to an excitation:
\begin{equation}
    e^{-i\,\omega^*\,t}\,=\,e^{-i\,Re[\omega^*]\,t}\,e^{Im[\omega^*]\,t}
\end{equation}
This mode is growing exponentially in time! It is a bomb and it is going clearly to destabilize the system. This is exactly the reason why the requirement of stability of a system can be translated in the absence of quasinormal modes with positive imaginary part.\\
Now two comments are in order:
\begin{itemize}
    \item The quasinormal mode are eigen-frequencies of the system. They correspond to bulk solution with zero source.
    \item The onset of instability, which follows from the previous picture, can be determined directly by looking when a specific mode with positive imaginary part appears. This means that, exactly at $T_c$, there will be a quasinormal mode crossing the real axis and lying exactly at the origin of the complex plane\footnote{This has been verified by explicit computations in several examples. The most famous one is that of holographic superconductors studied in \cite{Amado:2009ts}.}.
\end{itemize}
Collecting all this new information, our task is clear. We have to solve equation \eqref{tosol} numerically and search for solution with zero source, $\psi_1=0$, and zero frequency. This can be done easily with a \textit{FindRoot} routine (\url{https://reference.wolfram.com/language/ref/FindRoot.html}) and it is the topic of one of the notebooks available with our lectures. The numerical results are shown in fig.\ref{fig:instability}.
\subsection{Holographic Conductivity: an excuse to learn some numerics (and some physics)}
Now, let us forget about the superconducting instability and let us consider the system in the normal phase, where the complex scalar is trivially zero. What we are interested in is to compute the electric conductivity of the normal phase using the holographic methods. In order to make the computations slightly more interesting we consider  a generic massive gravity model, which was considered first in \cite{Baggioli:2014roa} and whose action reads:
\begin{equation}\label{action}
S\,=\, M_P^2\int d^4x \sqrt{-g}
\left[\frac{R}2+\frac{3}{\ell^2}- \, m^2 V(X)\,-\,\frac{1}{4}\,F^2\right]\, ,
\end{equation}
with $X \equiv \frac12 \, g^{\mu\nu} \,\partial_\mu \phi^I \partial_\nu \phi^I$ and $F^2=F_{\mu\nu}F^{\mu\nu}$, $F=dA$.\\
The background solution is the usual black hole geometry:
\begin{equation}
    ds^2\,=\,\frac{1}{u^2}\,\left[-f(u)dt^2\,+\,\frac{du^2}{f(u)}\,+\,dx^2\,+\,dy^2\right]
\end{equation}
where the blackening factor $f(u)$ is given by
\begin{equation}
f(u)= u^3 \int_u^{u_h} dv\;\left[ \frac{3}{v^4} -\frac{\rho^2}{2}-\frac{m^2}{v^4}\, 
V\left({v^2}\right) \right]
\end{equation}
and $u_h$ stands for the location of the event horizon. Additionally, the background profiles for the gauge and scalar bulk fields are given by:
\begin{equation}
    A_t\,=\,\mu\,-\,\rho\,u\,,\quad \phi^I\,=\,\alpha\,x^I
\end{equation}
Regularity of the gauge field on the horizon requires  $\rho = {\mu \over u_h}$, and it implies that the temperature is given by
\begin{equation}
T=-\frac{f'(u_h)}{4\pi}=\frac{6 - {\mu^2 u_h^2} -  2m^2 V\left(\alpha^2 u_h^2 \right) }{8 \pi u_h}~.
\end{equation}
For all the physical motivations and the technical details we refer to the original paper. Here, we only repeat the ingredients which are necessary for our computation.\\
In order to compute the conductivity, we have to discuss the behaviour of small excitations around the background. We perturb the previous solution by setting:
\begin{equation}
    A_\mu=\bar A_\mu+a_\mu,\quad g_{\mu\nu}=\bar g_{\mu\nu}+h_{\mu\nu},\quad \Phi^I=\bar \Phi^I+\phi^I
\end{equation}
where we indicate with bars the background profiles. We concentrate on the transverse sector of the fluctuations , which contains the following perturbations
$ a_i, \, \phi_i, \, h_{ti}, \, h_{ui} \; {\rm and} \;  h_{ij}\equiv \frac{1}{u^2}\partial_{(i} b_{j)}$. 
Aside from $a_i$, we use the gauge-invariant variables 
\begin{equation}
T_i \equiv u^2 h_{ti} - \partial_t \phi_{i}  \; , \;
U_{i}  \equiv  f(u)\big[h_{ui} -  \frac{\partial_u \phi_i} {u^2}\big] \; , \;
B_i \equiv   b_i -{\phi_i}\,\,
\end{equation}
In terms of these variables, and by restricting ourselves to the homogeneous perturbations (by fixing the momentum $k=0$), we are left with a system of two coupled differential equations
\begin{align}
    &{\partial_u}\left( f \, \partial_u \;a_i \right)+\Big[\frac{\omega^2}{f}-2 \,u^2\, \rho^2\Big] a_i =
 \frac{i \,\rho\, \,u^2\, 2\,\bar{m}^2  }{\omega} U_i \, \nonumber\\
    &\frac{1}{u^{2}}\partial_u\left[ \frac{f u^2}{\bar{m}^2}  \partial_u \left(\bar{m}^2 U_i\right)\right]
+\left[\frac{\omega^2}{f} -2 \bar {m}^2\right] U_i  = 
 - 2 \,i\, \,\rho \,\omega \,a_i \label{condeq}
\end{align}
which has to be solved numerically. For simplicity, we introduced the quantity $\bar{m}^2(u) = m^2 V'( u^2)$, which relates directly to the graviton mass.
\begin{mdframed}[style=MyFrame]
\begin{center}
    \textbf{Trick \#2 + Exercise : Gauge-Invariant variables}
\end{center}
\vspace{0.15cm}
What are they? How can we obtain and use them?  Let us learn it by finding a typo in the Arxiv version of my paper \cite{Baggioli:2014roa}. \\
First, let us define the GI variables.
Gauge invariance is simply a redundancy in the mathematical description of a physical theory which has nevertheless strong implications on the number of physical degrees of freedom. In presence of a gauge symmetry, in our case diffeomorphism invariance, a very common practice is to fix (partially) the gauge by using the so-called radial gauge and kill all the perturbations with have at least one index in the holographic radial direction:
\begin{equation}
    h_{ui}\,=\,0\,\quad A_u\,=\,0\,,\quad \dots
\end{equation}The idea is that those perturbations do not have a clear interpretation from the boundary field theory and therefore it might be a good practice to use the gauge freedom to eliminate them from the game. A more precise way to proceed is to avoid gauge fixing and work in variables which are invariant under the gauge transformations. This is what we are going to do.\\[0.1cm]
\textbf{Exercise.}\\
Consider the action in \cite{Baggioli:2014roa}
\begin{equation}
    S\,=\int d^4x\,\sqrt{-g}\,\left[R\,-\,2\,\Lambda\,-\,m^2\,V(X)\,-\,\frac{1}{4}\,F^2\,\right]
\end{equation}
together with the metric ansatz
\begin{equation}
    ds^2\,=\,\frac{1}{u^2}\,\left(-f(u)\,dt^2\,+\,\frac{du^2}{f(u)}\,+\,dx^2\,+\,dy^2\,\right)
\end{equation}
and the background profile for the scalar fields:
\begin{equation}
    \phi^I\,=\,\alpha\,x^I
\end{equation}
Perturb the background solution with:
\begin{equation}
    a_i, \,\,\, \phi_i, \,\,\, h_{ti}, \,\,\, h_{ui} ,\,\,\,  h_{ij}\equiv \frac{1}{u^2}\partial_{(i} b_{j)}
\end{equation}
and use the gauge invariant variables:
\begin{equation}
    T_i \equiv u^2 h_{ti} - {\partial_t \phi_{i} \over \alpha} \; , \;
U_{i}  \equiv  f(u)\big[h_{ui} -  {\partial_u \phi_{i}\over \alpha u^2}\big] \; , \;
B_i \equiv   b_i -{\phi_i\over\alpha}
\end{equation}
1) Check explicitly that these variables are gauge invariant. In particular, they must be invariant under a diffeomorphism transformation:
\begin{equation}
    \tilde{h}_{\mu\nu}\,=\,h_{\mu\nu}\,-\,\nabla_{(\mu}\,\xi_{\nu)}\,,\quad \tilde{a}_\mu\,=\,a_\mu\,-\,\left(\xi^{\nu}\partial_\nu A_\mu+A_\nu \partial_\mu \xi^\nu \right)\,,\quad \tilde{\phi}^I\,=\,\phi^I\,-\,\xi^I
\end{equation}
where $\xi^\mu$ is the parameter of the infinitesimal transformation.\\
2) Obtain the equations for the pertubations at finite frequency $\omega$ and finite momentum $k$. Did you find the typo in \cite{Baggioli:2014roa}?\\
If not, check the corresponding notebook available with these lectures.
\end{mdframed}
\vspace{0.3cm}
Now, given the equations in \eqref{condeq}, there is a brute force and simple way to obtain the conductivity, which works as follows:
\begin{enumerate}
    \item Analyze the expansion of the two gauge invariant fields close to the boundary and obtain:
    \begin{align}
        & a_i\,=\,a_i^{(0)}\,+\,a_i^{(1)}\,u\,+\,\dots\\
        & U_i\,=\,\frac{1}{u}\,U_i^{(0)}\,+\,U_i^{(1)}\,+\,\dots
    \end{align}
    and identify the sources for the dual operators with the coefficients $a_i^{(0)},U_i^{(0)}$. (Trick: it is convenient to redefine all the fields such that the boundary expansion starts always with a constant term);
    \item Solve the equations perturbatively close to the horizon by imposing ingoing boundary conditions (or equivalently by imposing regularity after subtracting the ingoing part of each field). You will obtain an expansion in terms of only one undetermined coefficient, $c_0$.
    \item Use a \textit{FindRoot} routine to impose that the source of the $U_i$ field vanishes, $U_i^{(0)}=0$. This will fix the parameter $c_0$ uniquely.
    \item Use the solution with zero $U$ source and read:
    \begin{equation}
        \mathcal{G}_{\mathcal{J}\mathcal{J}}\,\sim\,\frac{a_i^{(1)}}{a_i^{(0)}}
    \end{equation}
    This will be your Green Function for the $U(1)$ current, from which you can obtain the conductivity.
\end{enumerate}
Try it and compare with the results obtained in the notebook available with the lectures.\\
This last method can work for two bulk fields, but it becomes immediately hopeless in presence of more fields. Let us introduce a more powerful method, which is widely used to read correlators and to identify quasinormal modes. The main ideas appeared in \cite{Kaminski:2009dh} and \cite{Amado:2009ts}. Here, we explain it using the simple problem of only two mixed bulk fields, as that of equations \eqref{condeq}.\\
Consider two bulk fields $\Psi_a,\Psi_b$  which obey a system of coupled differential equations and which are gauge invariant\footnote{If you do not use gauge invariant variables there are additional complications in the procedure. See \cite{Kaminski:2009dh,Amado:2009ts} for details. My advise is: use GI variables!}.
At the horizon, we impose infalling boundary conditions for both the fields:
\begin{equation}
    \Psi_{a,b}(u)\,=\,(1-u)^{-\,i\,\omega/4\pi\,T}\,\left(\Psi_{a,b}^0\,+\,\Psi_{a,b}^1\,(u-1)\,+\,\dots\right)
\end{equation}
where $\Psi_{a,b}^0$ are undetermined coefficients.\\
Now, we can choose a basis of independent solutions. In our case, they correspond to the following independent horizon shooting conditions:
\begin{equation}
    \left(\Psi_a^0,\Psi_b^0\right)\,=\, \left(0,1\right)\,,\left(1,0\right)
\end{equation}
Suggestion: it turns out\footnote{We thank Amadeo Jimenez for pointing this out.} it is much better to choose a basis without zeros, such as:
\begin{equation}
    \left(1,1\right)\,,\left(1,-1\right)
\end{equation}
We will call these two independent solutions $\Psi^I,\Psi^{II}$. At the boundary, these solutions will have the usual asymptotic form:
\begin{align}
   & \Psi^I_{a}\,=\,\Psi^I_{a,(L)}\,u^{\Delta_L}\,+\,\Psi^I_{a,(S)}\,u^{\Delta_S}
\end{align}
where $a$ label the two fields, and $I$ the two solutions with independent horizon b.c.s. as indicated above. We are ready to define the following matrices:
\begin{equation}
    L\,=\,\begin{pmatrix} 
\Psi^I_{a,(L)} & \Psi^{II}_{a,(L)} \\
\Psi^I_{b,(L)} & \Psi^{II}_{b,(L)}
\end{pmatrix} \quad 
S\,=\,\begin{pmatrix} 
\Psi^I_{a,(S)} & \Psi^{II}_{a,(S)} \\
\Psi^I_{b,(S)} & \Psi^{II}_{b,(S)}
\end{pmatrix}
\end{equation}
which determine the subleading and leading coefficients of the various fields on the (two in this case) independent solutions.\\
Before continuing, we have to be careful about one detail. We always claim that the Green function of an operator $\mathcal{O}$ is given by :
\begin{equation}
    G_{\mathcal{O}\mathcal{O}}\,=\,\frac{\Psi^\mathcal{O}_{(S)}}{\Psi^\mathcal{O}_{(L)}}
\end{equation}
where the two terms are just the leading and subleading coefficients in the expansion of the dual bulk field close to the UV boundary. Strictly speaking, this statement holds true as far as the renormalized boundary action takes the form:
\begin{equation}
    S_{\text{boundary}}\,=\,\int d^dx \sqrt{-\gamma}\,\Psi^\mathcal{O}_{(L)}\,\Psi^\mathcal{O}_{(S)}
\end{equation}
which might not necessarily be the case, especially with multiple fields. Generically, we have a more complicated expression of the type:
\begin{equation}
      S_{\text{boundary}}\,=\,\int d^dx \sqrt{-\gamma}\,\Psi^{\mathcal{O}_i}_{(L)}\,B_{ij}\,\Psi^{\mathcal{O}_j}_{(S)}
\end{equation}
with a generic matrix $B_{ij}$.
In the simple case the matrix $B$ is diagonal, the problem is trivial. Nevertheless, here it is an explicit counterexample:
\begin{equation}
    S_{\text{boundary}}\,=\,\int d^dx \sqrt{-\gamma}\,\left[\,\Psi^1_{(L)}\,\Psi^1_{(S)}\,+\,\Psi^2_{(L)}\,\Psi^2_{(S)}\,+\,\Psi^1_{(L)}\,\Psi^2_{(S)}\right]
\end{equation}
Now if we use the correct variational definition for the Green function:
\begin{equation}
   G_{\Psi^1\Psi^1}\,\sim\, \frac{\delta S_{\text{boundary}}}{\delta \Psi^1_{(L)}}\,=\,\Psi^1_{(S)}\,+\,\Psi^2_{(S)}\,\neq\,\Psi^1_{(S)}
\end{equation}
we obtain that:
\begin{equation}
    G_{\Psi^1\Psi^1}\,\neq\,\frac{\Psi^1_{(S)}}{\Psi^1_{(L)}}
\end{equation}
The main message is: be careful!\\
Taking into account this possible caveat, we finally get to the conclusive answer for our correlators which is:
\begin{equation}
    G\,=\,B\,\cdot\,S\,\cdot L^{-1}
\end{equation}
where $G$ is intended now as the matrix of the Green functions. In the case of our two fields:
\begin{equation}
    G\,\equiv\,\begin{pmatrix} G_{aa} & G_{ab}\\ G_{ba}& G_{bb} \end{pmatrix}
\end{equation}
This method is exactly what we used for the conductivity in the notebook available with the lectures. You can convince yourself that, even for only two bulk fields, this procedure is much faster and efficient. Additionally, this method that we outlined is very powerful to compute the quasinormal modes \cite{Berti:2009kk}. In this language, they come directly from imposing:
\begin{equation}
    det(L)\,=\,0
\end{equation}
and solving it for $\omega$. This procedure takes the name of the \textit{determinant method}, see \cite{Amado:2009ts}.

\begin{mdframed}[style=MyFrame2]
\begin{center}
    \textbf{Warning \# 2: EF are nice but be careful}
\end{center}\vspace{0.1cm}
We have stated several times that it is very convenient to redefine the bulk fields by removing the ingoing part at the horizon or by using EF coordinates.\\
This is certainly true, but (once more) you have to proceed cautiously. For example, let us take our equations and redefine the fluctuation of the gauge field $a_i$ as:
\begin{equation}
    a_i(u)\,=\,e^{-\,i\,\omega\,\int \frac{1}{f(y)}\,dy}\,\tilde{a}_i \label{op}
\end{equation}
Now we could say the Green function, and therefore the conductivity, is simply given by:
\begin{equation}
    \sigma(\omega)\,=\,\frac{i}{\omega}\,G_{JJ}\,=\,\frac{i}{\omega}\,\frac{\tilde{a}_i^{(S)}}{\tilde{a}_i^{(L)}} \label{ww}
\end{equation}
If you follow this prescription, you would get the wrong result! You can try to compare the numerical results to the DC formula we will derive later. What you are missing?\\
The Green function of the current dual to the bulk gauge field $a_i$ is defined via the ratio between the leading and subleading coefficients of the $a_i$ bulk field; not of the transformed one $\tilde{a}_i$ obtained using \eqref{op}. More specifically, we can immediately recognize that:
\begin{equation}
    G_{JJ}\,=\,\frac{a_i^{(S)}}{a_i^{(L)}}\,=\,\frac{a'_i(u)}{a_i(u)}\Big|_{\text{boundary}}\,=\,-\,i\,\omega\,+\,\frac{\tilde{a}'_i(u)}{\tilde{a}_i(u)}\Big|_{\text{boundary}}\,=\,-\,i\,\omega\,+\,\frac{\tilde{a}_i^{(S)}}{\tilde{a}_i^{(L)}}
\end{equation}
As a consequence, if you use the naive formula \eqref{ww} in terms of the transformed bulk field $\tilde{a}_i$, you would miss completely the first term, which accounts for a shift in the conductivity $\sigma\rightarrow 1+\sigma$! Tricks are nice but have to be used carefully.
\end{mdframed}
\begin{figure}
    \centering
    \includegraphics[width=0.45\linewidth]{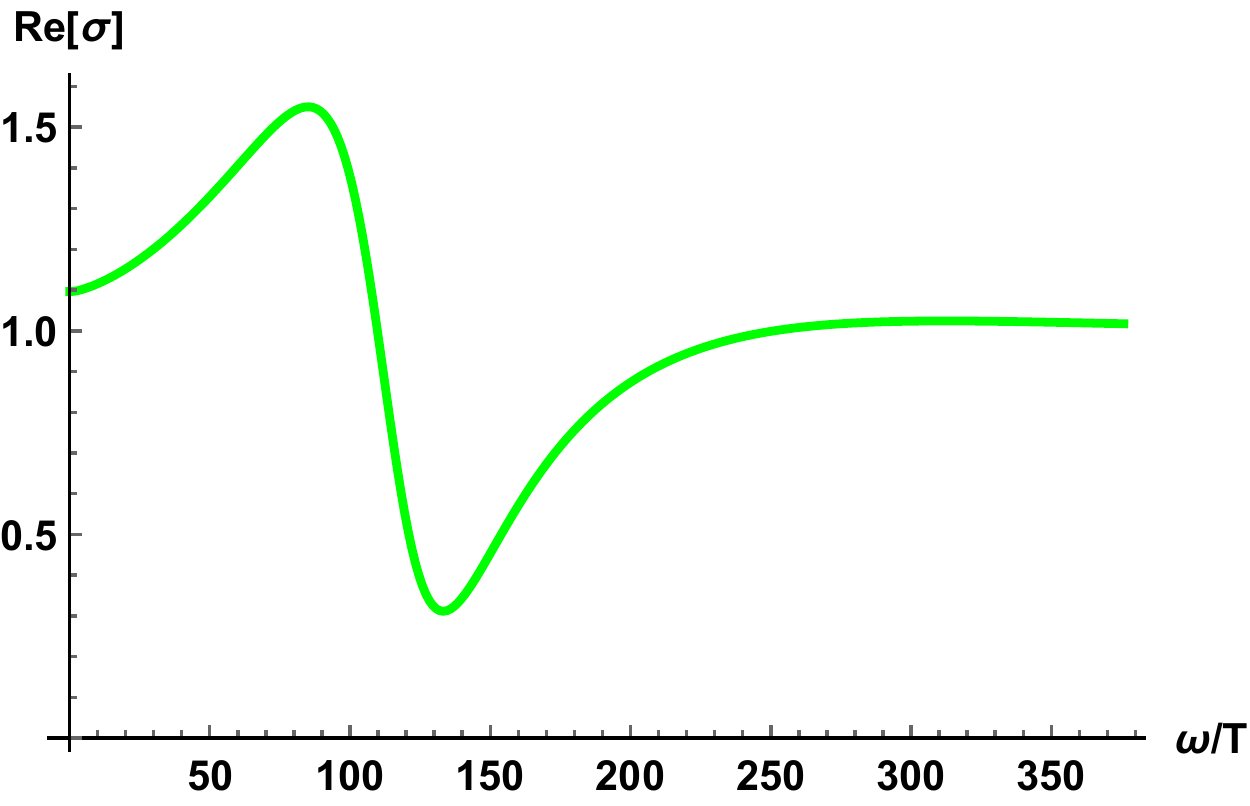}%
    \quad  \quad 
    \includegraphics[width=0.49\linewidth]{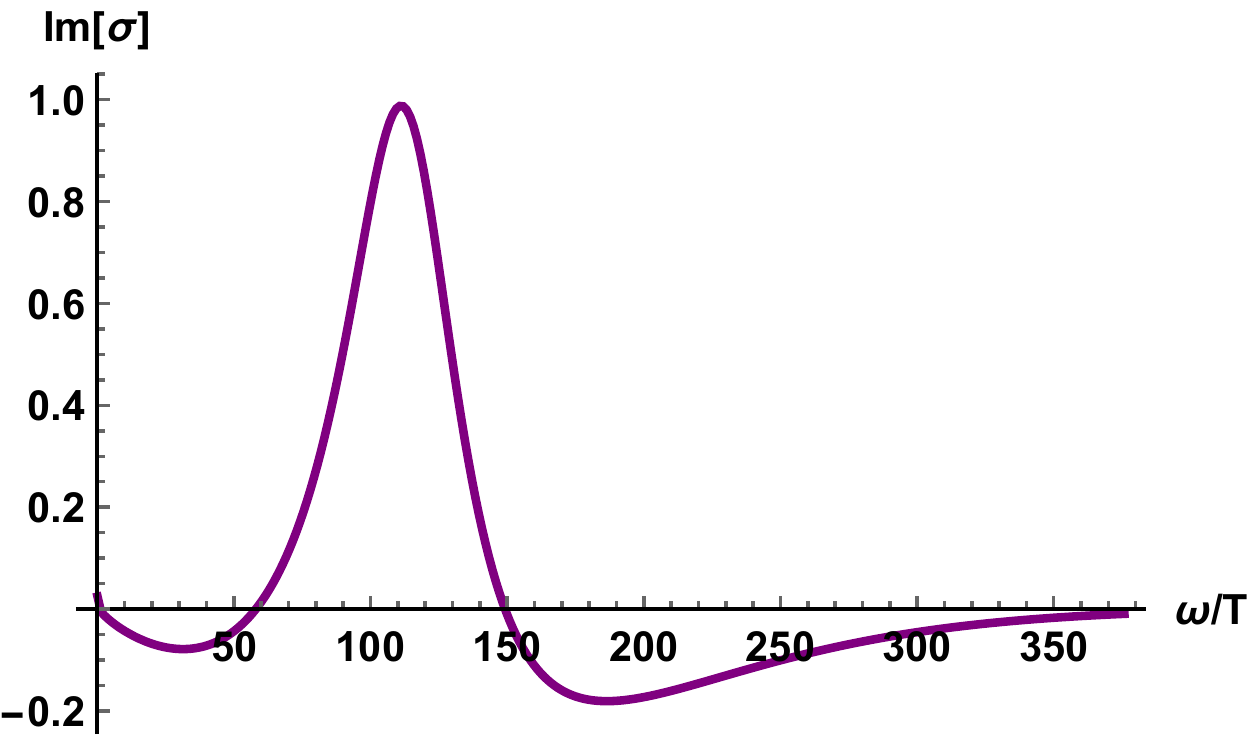}
    \caption{An example of the results for the optical conductivity obtained in the notebook available with the lectures. This plot is for the potential $V(X)=X+X^5$. You can try to reproduce the results in \cite{Baggioli:2014roa,Ammon:2019wci}.}
    \label{fig:AC}
\end{figure}
\subsection{DC conductivities: an example of analytic methods}
Numerical techniques are fine (and for somebody funny), but we can do more. More specifically, we are interested in what goes under the name of DC conductivity, which is simply the conductivity at zero frequency:
\begin{equation}
    \sigma_{DC}\,\equiv\,\sigma(\omega=0)
\end{equation}
This problem should sound familiar to you, after going through the computations of the KSS ratio in section \ref{secK}. Not surprisingly, the DC conductivity can be obtained using a generalization of the Membrane Paradigm technique. The original reference, which I invite you to check, is \cite{Donos:2014cya}.\\
Here, we focus on a simple model, which is a generalization of the  \textit{linear axion model} of \cite{Andrade:2013gsa}. It is a simple model which introduces momentum relaxation using a simple set of bulk scalar fields, and therefore produces finite values for the DC conductivities of the boundary theory\footnote{In absence of momentum relaxation the DC conductivity would be infinite. This can be understood by simply thinking at a collection of electrons accelerated by an electric field. If there is no way by which these electrons can lose their momenta, once accelerated, they will go forever and they will produce an infinite conductivity. Probably the best way to understand this is to look back at the famous Drude model \cite{kittel1976introduction}.}.\\
We are going to focus on the thermoelectric conductivities, which are defined by the following matrix of transport coefficients:
\begin{equation}
\left(
    \begin{array}{c}
      \mathcal{J} \\
      \mathcal{Q}
    \end{array}
  \right)\,=\,\left(
    \begin{array}{cc}
      \sigma & \alpha\,T\\
      \bar{\alpha}\,T & \bar{\kappa}\,T
    \end{array}
  \right)\left(
    \begin{array}{c}
      E \\
      -\frac{\nabla T}{T}
    \end{array}
  \right)
\end{equation}
where $\mathcal{J}^I$ is the electric current and $\mathcal{Q}^i= T^{ti}\,-\,\mu\,\mathcal{J}^i$ the thermal one. The various coefficients take the name of: electric conductivity ($\sigma$), thermal conductivity ($\bar{\kappa}$) and thermoelectric conductivities ($\alpha,\bar{\alpha}$). These four objects codify the response of the system under an external electric field $E$ and a thermal gradient $\nabla T/T$.\\
The theory which we consider is defined by the already mentioned action:
\begin{equation}
    S\,=\int d^4x\,\sqrt{-g}\,\left[R\,-\,2\,\Lambda\,-\,m^2\,V(X)\,-\,\frac{1}{4}\,F^2\,\right] \label{dudu}
\end{equation}
where the details have been already explained in the previous section and can be found in \cite{Baggioli:2014roa}. We consider a black hole solution in the following form
\begin{equation}
    ds^2\,=\,\frac{1}{u^2}\,\left(-f(u)\,dt^2\,+\,\frac{du^2}{f(u)}\,+\,dx^2\,+\,dy^2\,\right)
\end{equation}
and we assume for the scalars the solution:
\begin{equation}
    \phi^I\,=\,x^I
\end{equation}
which breaks the translational invariance of the boundary theory\footnote{The specific breaking pattern present several subtleties. For those of you interested, I suggest to read\cite{Alberte:2017cch,Alberte:2017oqx,Ammon:2019wci}.}.
\begin{mdframed}[style=MyFrame3]
\begin{center}
    \textbf{Exercise \#12 : Massive graviton}
\end{center}\vspace{0.15cm}
Consider the theory in \ref{dudu} and  show that the graviton is massive and what is the mass. If you want to play more with it, see \cite{Alberte:2015isw}. Notice that the presence of a graviton mass is not a mere technical detail, but it has fundamental phenomenological consequences \cite{Alberte:2016xja,Baggioli:2019abx}.
\end{mdframed}
Let us sketch now the derivation of the DC conductivities within this model. We start by applying an external electric field $E_x \equiv F_{xt}$ using the ansatz for the perturbations given by
\begin{align}
    &\delta g_{tx}(u,t,y)=h_{tx}(u)\,,\quad  \delta g_{tx}(u,t,y)=0\,,\quad   \delta g_{xu}(u,t,y)=h_{xu}(u)\,,\nonumber\\ &\delta \phi_x\,=\,\chi(u)\,,\quad \delta A_x\,=\,-\,E_x\,t\,+\,a_x(u)
\end{align}
which indeed account for a finite $F_{tx}\neq 0$.
The left relevant equations\footnote{You can prove by yourself that the others are redundant.} are:
\begin{align}
&E_x\, \mu\,  u^2+m^2 \,u_h \,f(u)\, V'(u^2)\, \left(\chi '(u)-u^2
  h_{xu}(u)\right) \label{touse}\\
    &u_h \left(f'(u) \,a_x'(u)+f(u)\, a_x''(u)\right)-\mu \, u
   \left(u \,h_{tx}'(u)+2\, h_{tx}(u)\,\right)\,=\,0
\end{align}
The fundamental point is to notice that the Maxwell equation can be written as:
\begin{equation}
    \partial_u\,\mathcal{J}(u)\,=\,0\,,\quad \quad \mathcal{J}(u)\,\equiv\,f(u)\,a_x'(u)\,-\,\frac{\mu}{u_h}\,u^2\,h_{tx}(u)
\end{equation}
and it corresponds to the conservation of a radially dependent quantity $\mathcal{J}(u)$. Luckily enough, at the boundary $u=0$, this is exactly the electric current of the dual CFT:
\begin{equation}
    \mathcal{J}\,\equiv\,\lim_{u\rightarrow 0}\,\sqrt{-g}\,F^{rx}\, =\,\lim_{u\rightarrow 0}\,\mathcal{J}(u)
\end{equation}
Here, we use the same trick of the membrane paradigm and since we have the freedom of computing this object everywhere we want in the bulk, we decide to compute it at the horizon $u=u_h$.\\
In order to do that, we need to know the behaviour of the various fields close to the horizon. The requirement of having the fields regular at the horizon implies the following relations
\begin{equation}
    a_x'(u)\,=\,\frac{E_x}{f(u)}\,,\quad h_{tx}(u)\,=\,f(u)\,h_{xu}(u)
\end{equation}
There are two ways of deriving them: (I) going to EF coordinates and impose regularity at the horizon, (II) compute the field strength square $F_{\mu\nu}F^{\mu\nu}$ at leading order in the perturbations and impose its regularity at the horizon $u=u_h$. You can convince yourself that you will obtain the same results.\\
Finally, using the constraint in eq.\eqref{dudu} evaluated at the horizon, together with the relations above, we can compute $\mathcal{J}(u)$ at the horizon :
\begin{equation}
   \mathcal{J}_e\,\equiv\, \mathcal{J}(u_h)\,=\,\left(1\,+\,\frac{\mu^2}{m^2\,V'\left(u_h^2\right)}\,\right)\,E_x
\end{equation}
From the latter, it is straightforward to obtain the electric DC conductivity of this model, which is given by
\begin{equation}
    \sigma_{DC}\,=\,1\,+\,\frac{\mu^2}{m^2\,V'(u_h^2)} \label{DCf}
\end{equation}
Let us make one step more and introduce a thermal gradient:
\begin{align}
    &\delta g_{tx}(u,t,y)=\,-\,t\,\zeta\,\frac{f(u)}{u^2}\,h_{tx}(u)\,,\quad  \delta g_{tx}(u,t,y)=0\,,\quad   \delta g_{xu}(u,t,y)=\,u^{-2}\,h_{xu}(u)\,,\nonumber\\ &\delta \phi_x\,=\,\chi(u)\,,\quad \delta A_x\,=\,t\,\left(-E_x\,+\,\zeta\,A_t(u)\right)\,+\,a_x(u)
\end{align}
encoded in the parameter $\zeta$.
\begin{mdframed}[style=MyFrame4]
\begin{center}
    \textbf{Thermal gradient}
\end{center}\vspace{0.15cm}
How to introduce an external thermal gradient?
We have used the following prescription:
\begin{equation}
    i\,\omega\,\delta g_{ti}\,=\,-\,\frac{\nabla_i\,T}{T}\,,\quad i\,\omega\,\delta A_j\,=\,\mu\,\frac{\nabla_i T}{T}\label{pre}
\end{equation}
Where does this rule come from? We follow the explanation of \cite{Hartnoll:2009sz}.\\
Consider a background Minkowski metric. The temperature is the inverse of the period of the euclidean time. Let us rescale the time coordinate to eliminate any temperature dependence in the period of the time circle:
\begin{equation}
    t\,\longrightarrow \bar{t}/T
\end{equation}
such that the time component of the metric becomes $g_{\bar{t}\bar{t}}=1/T^2$. Now let us perform a gradient in temperature:
\begin{equation}
    T\,\rightarrow\,T\,+\,x^i\,\frac{\nabla_i T}{T}\,\,\quad \Longrightarrow \quad \delta g_{\bar{t}\bar{t}}\,=\,-\,\frac{2\,x^i\,\nabla_i T}{T^3}
\end{equation}
The trick is to act with a gauge transformation which leaves the system invariant. In particular, we are using the infinitesimal parameters:
\begin{equation}
    \xi_{\bar{t}}\,=\,\frac{i\,x^I\,\nabla_i T}{T^3\,\bar{\omega}}\,,\quad \quad \xi_{i}\,=\,0
\end{equation}
together with the Fourier decomposition in terms of the frequency $\bar{\omega}$, associated to the time coordinate $\bar{t}$.\\
The action of this gauge transformation leads us to:
\begin{equation}
    \delta g_{\bar{t}\bar{t}}\,=\,0,\quad \delta g_{\bar{t}i}\,=\,\frac{i\,\nabla_i T}{\bar{\omega}\,T^3},\quad \delta A_i\,=\,-\,\frac{i\,\mu\,\nabla_i T}{\bar{\omega}\,T^3}
\end{equation}
which, after scaling back to the original time $t$, corresponds to the prescription given in eq.\eqref{pre}.
\end{mdframed}
\vspace{0.2cm}
With this new ansatz, the equations are slightly modified and they become:
\begin{equation}
    2 \,\mu\,  u^2 (E_x\, u_h\,+\,\zeta \, \mu \, (u-u_h)\,)\,-\,u_h^2\, \left(\zeta  f'(u)+2
   \,m^2\, f(u) \,V'\left(u^2\right) \left(h_{xu}(u)-\chi '(u)\right)\right)\,=\,0
\end{equation}
Following exactly the same procedure, we derive the value of the electric current at the horizon which reads
\begin{equation}
    \mathcal{J}_e\,\equiv\,\mathcal{J}(u_h)\,=\,\frac{\mu \, u^2\, \left(2 \,\mu  \,u^2\, (E_x\,u_h+\zeta  \,\mu  \,(u-u_h)\,)-\zeta 
  \, u_h^2 \,f'(u)\right)}{2 \,m^2\, u_h^5 V'\left(u^2\right)}+E_x
\end{equation}
Using linear response, we get the final result
\begin{equation}
    \alpha_{DC}\,=\,\bar{\alpha}_{DC}\,\equiv\,\frac{1}{T}\frac{d\mathcal{J}_e}{d \zeta}\,=\,\frac{2\,\pi\,\mu}{m^2\,V_X}\,\frac{1}{u_h}
\end{equation}
which is our thermoelectric DC conductivity.
\begin{figure}
    \centering
    \includegraphics[width=8.5cm]{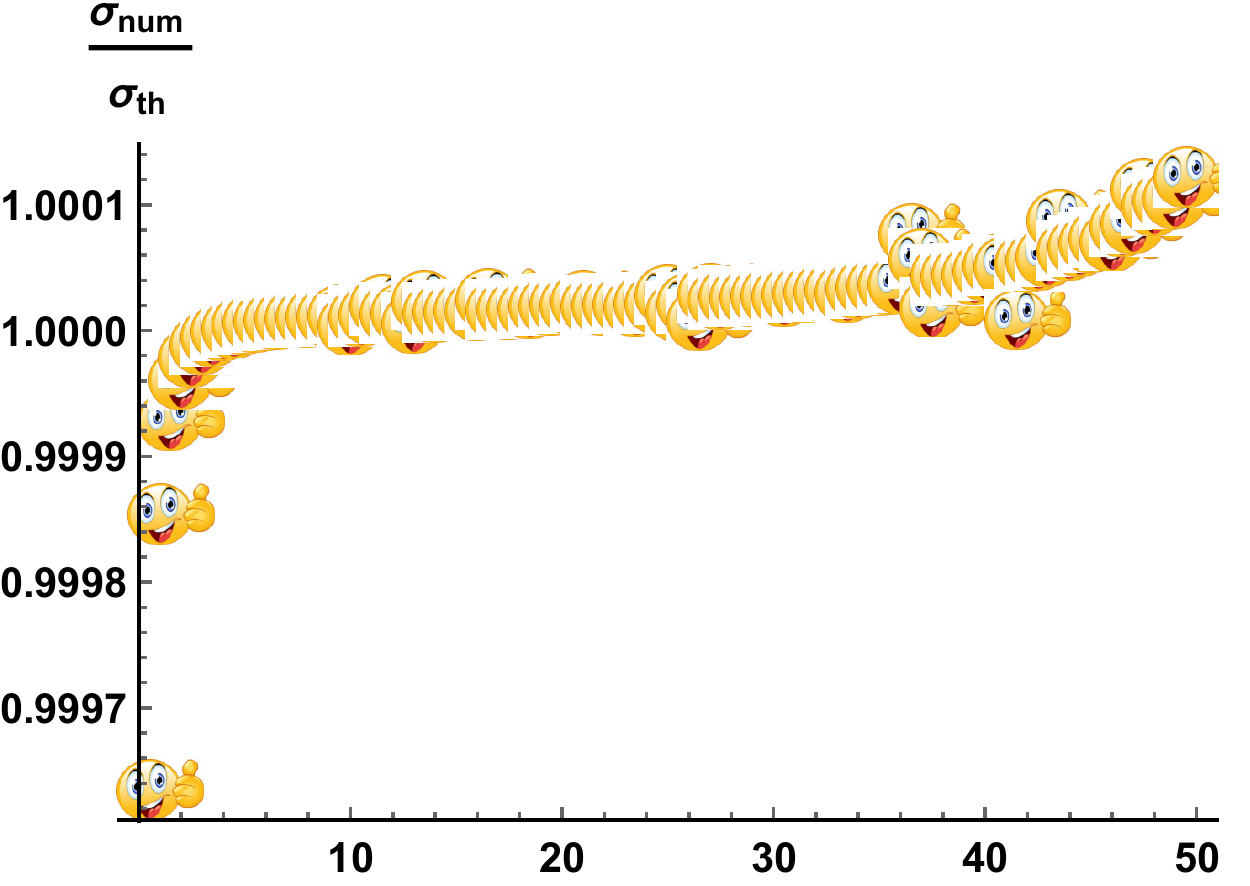}
    \caption{The comparison between the DC formula \eqref{DCf} and the numerical results. The plot is built with one of the notebooks available with the lectures and with a funny emoji found in the web.}
    \label{fig:DCnumeric}
\end{figure}
\begin{mdframed}[style=MyFrame3]
\begin{center}
    \textbf{Exercise \#13 : Your turn!}
\end{center}\vspace{0.15cm}
Try to obtain the DC conductivities for more complicated setups \cite{Baggioli:2016oqk,Amoretti:2016cad,Baggioli:2016oju,Baggioli:2016pia,Baggioli:2017ojd,Mokhtari:2017vyz}. Additionally, try to extract also the thermal conductivity. The fundamental point is to recognize that the equations of motion imply the conservation of an additional object $Q(u)$, which coincides at the boundary with the thermal current of the dual field theory. The definition of this object is not straightforward, see \cite{Donos:2014cya}. See the appendix in \cite{Baggioli:2017ojd} for an example explained step by step. Prove that, in absence of an external magnetic field, $\alpha=\bar{\alpha}$, and that this follows directly from the so-called Onsager relations.
\end{mdframed}
\begin{figure}
    \centering
    \includegraphics[width=0.48\linewidth]{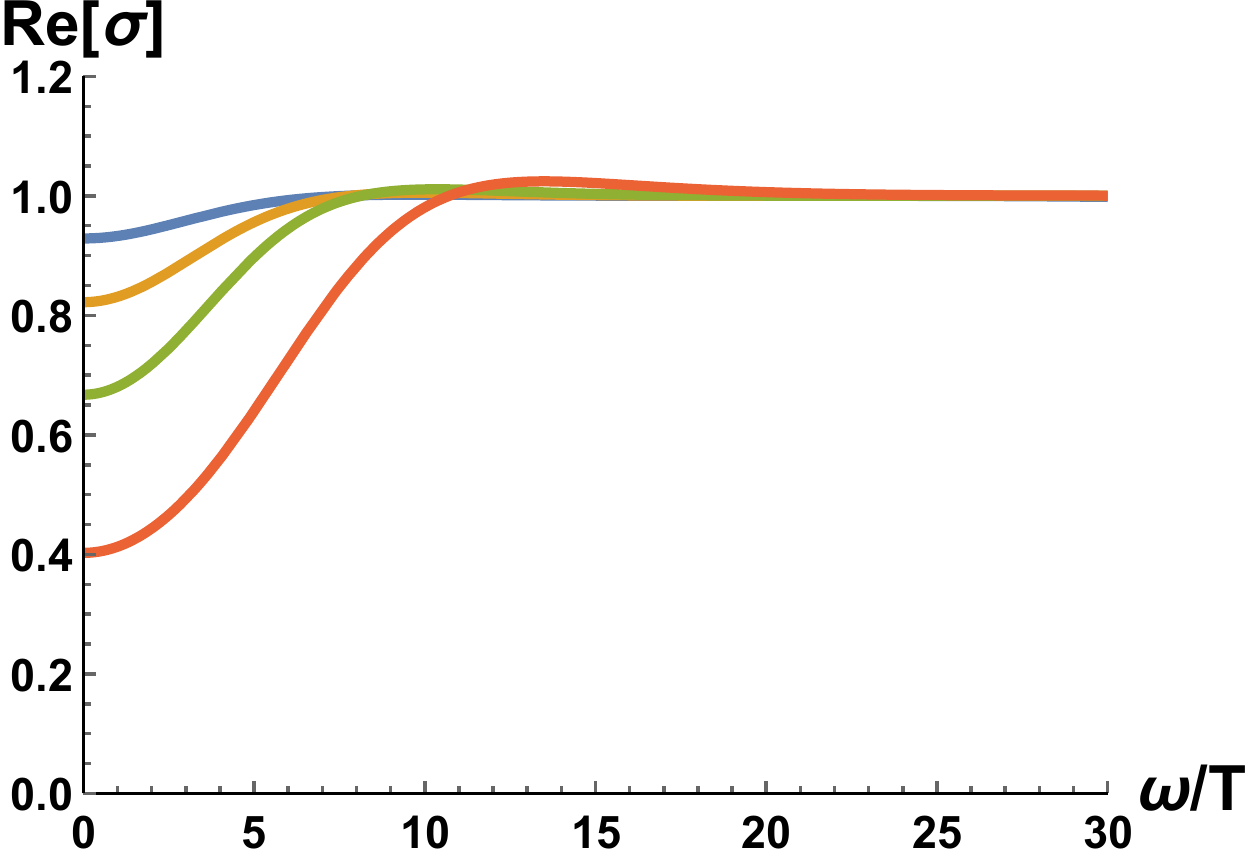}
    \quad
    \includegraphics[width=0.48\linewidth]{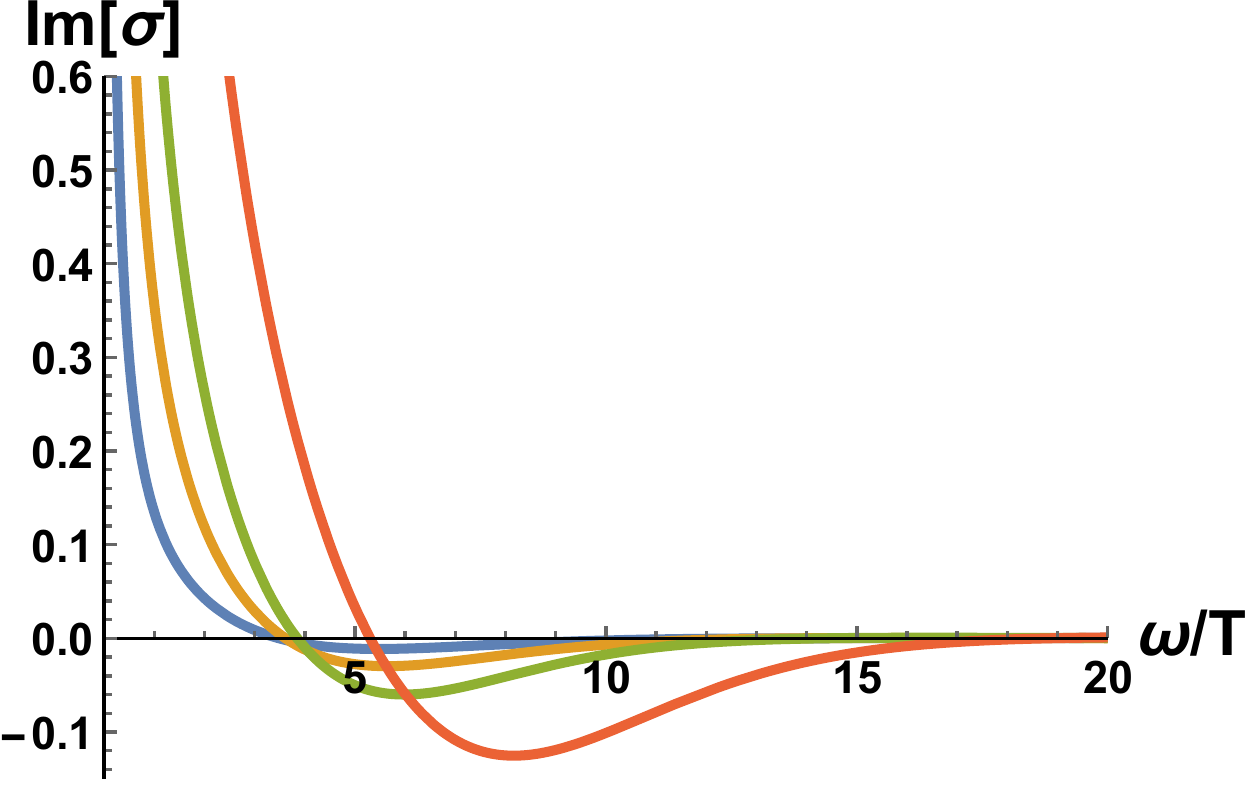}
    \caption{The optical electric conductivity $\sigma(\omega)$ for the model in eq.\eqref{sino}. The figures are taken from \cite{Alberte:2017oqx}.}
    \label{fig:ops}
\end{figure}
\begin{mdframed}[style=MyFrame2]
\begin{center}
    \textbf{Warning \#3 : Check the assumptions before using a tool and think.}
\end{center}\vspace{0.5cm}
Compute the optical conductivity, as explained in the main text, for the potential
\begin{equation}
    V(X)\,=\,X^3 \label{sino}
\end{equation}
You will obtain the results shown in fig.\ref{fig:ops}.
Why this is problematic/interesting?
First of all, if you apply naively the DC formula found in the main text you will realize that $\sigma_{DC}>1$ for every value of the parameter, but the numerics clearly show $Re[\sigma(0)]<1$, not in agreement with the theoretical expectations. The DC formula does not work! Moreover, the imaginary part of the conductivity shows a $1/\omega$ pole. What does that mean? In simple words \footnote{We thank Karl Landsteiner for suggesting this simple derivation.}, we have to remember that, when we consider the definition of the conductivity, what we have in mind is a \underline{retarded} correlator. This has to be thought always in terms of the $i\,\epsilon$ prescription, which means $i/\omega$ is actually
\begin{equation} \lim_{\epsilon\rightarrow 0}\,\frac{i}{\omega\,+\,i\,\epsilon}\,=\,\lim_{\epsilon\rightarrow 0}\,\left(\frac{\epsilon}{\omega^2\,+\,\epsilon^2}\,+\,i\,\frac{\omega}{\omega^2\,+\,\epsilon^2}\right)\,=\,\frac{i}{\omega}\,+\,\delta(\omega)
\end{equation}
In other words, every time we see a $1/\omega$ pole in the imaginary part of the conductivity, we should always think it as accompanied to a $\delta$ function in the real part. This is the signal that the DC conductivity in this case is infinite!\\
The physical reason behind is explained in \cite{Alberte:2017oqx} and in few words it is that the model in eq.\eqref{sino} does not break translational invariance explicitly but rather spontaneously. Momentum is a conserved quantity and therefore the electric conductivity is infinite. For some details about it see also \cite{Hartnoll:2012rj}.\\
The feature just discussed does not regard only the electric conductivity, but it is much more general. From a mathematical point of view, it follows from the  Kramers-Kronig theorem:
\begin{theorem}
Take a complex function analytic on the upper-half plane (and vanishes like $1/|\omega|$ or faster at infinity). We have:
\begin{equation}
    \mathcal{F}(\omega)\,=\,\mathcal{F}_1(\omega)\,+\,i\,\mathcal{F}_2(\omega)\,,\quad \mathcal{F}_2(\omega)\,=\,-\,\frac{1}{\pi}\,\mathcal{P}\,\int_{-\infty}^{+\infty}\,\frac{\mathcal{F}_1(\omega')}{\omega'\,-\,\omega}\,d\omega'
\end{equation}
\end{theorem}
You can prove that, using this theorem, you re-obtain what discussed before. More specifically, you can see that for this potential the electric conductivity takes the form:
\begin{equation}
    \sigma\,=\,\sigma_0\,+\,\left(\frac{i}{\omega}\,+\,\delta(\omega)\right)\,\frac{\rho^2}{\chi_{PP}} \label{ee}
\end{equation}
where $\sigma_0$ is sometimes referred to as the \textit{incoherent conductivity} (see \cite{Davison:2015taa}).\\[0.15cm]
\textbf{Exercise.} Compute numerically the conductivity for the model \eqref{sino} and check that the expression in \eqref{ee} is correct. You can also compute $\sigma_0$ analytically if you are brave enough and understand why the general formalism of \cite{Donos:2014cya} does not work in this case (tip: check well the boundary conditions for the various perturbations and the assumptions hidden behind the method). If you are interested in learning specific topic, I suggest to have a look at \cite{Amoretti:2017frz,Gouteraux:2018wfe,Donos:2018kkm}.
\end{mdframed}
\begin{figure}
    \centering
    \includegraphics[width=0.8\linewidth]{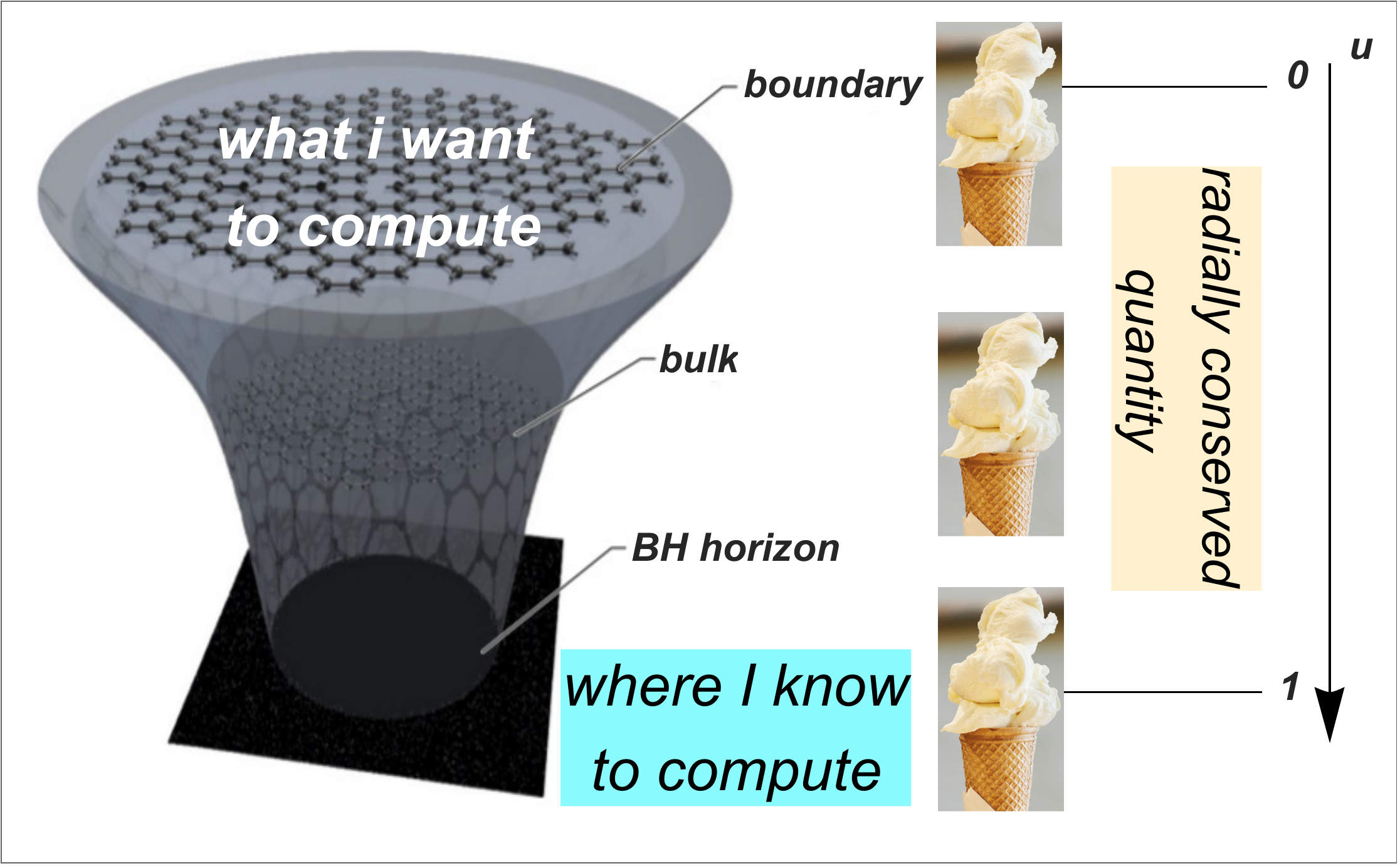}
    \caption{A cartoon of the idea behind the method to compute the DC transport coefficients introduced in \cite{Donos:2014cya}.}
    \label{fig:DC}
\end{figure}
\begin{figure}
    \centering
    \includegraphics[width=0.47\linewidth]{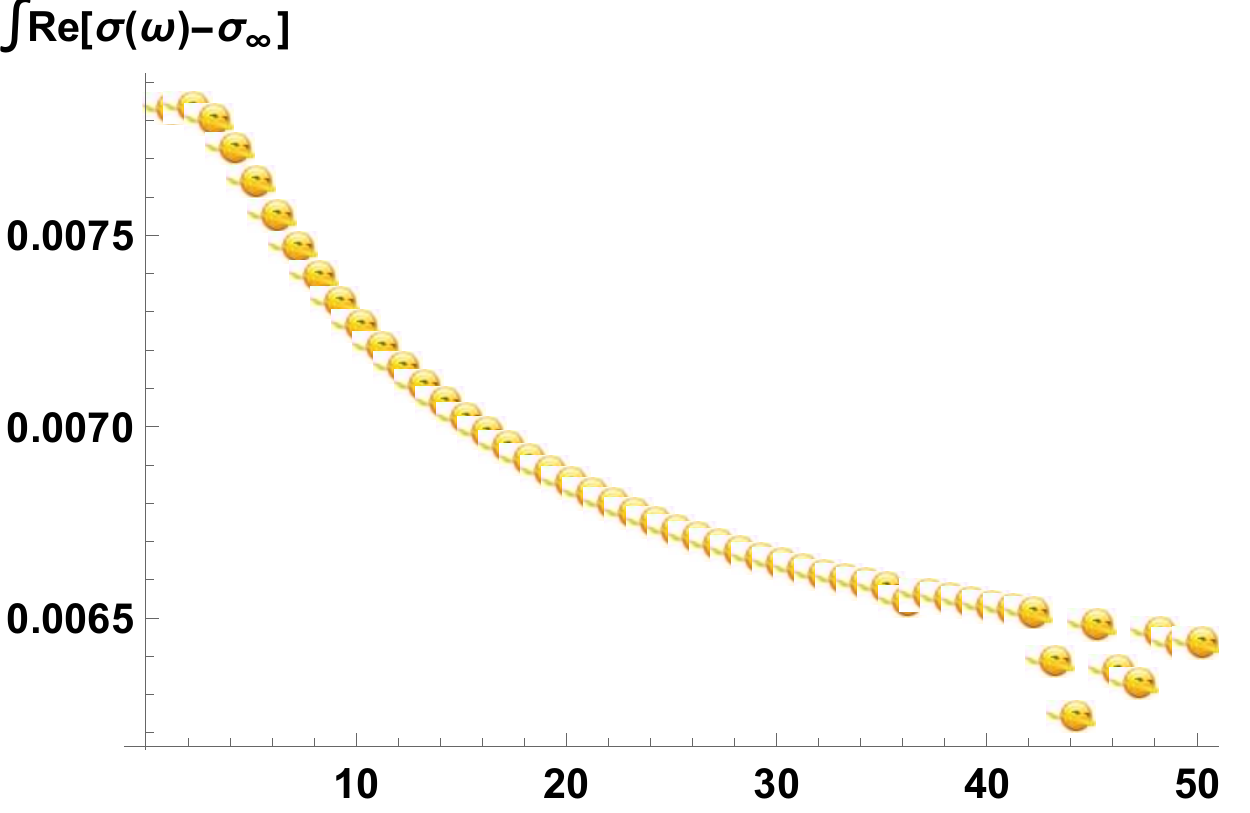}
    \quad 
     \includegraphics[width=0.47\linewidth]{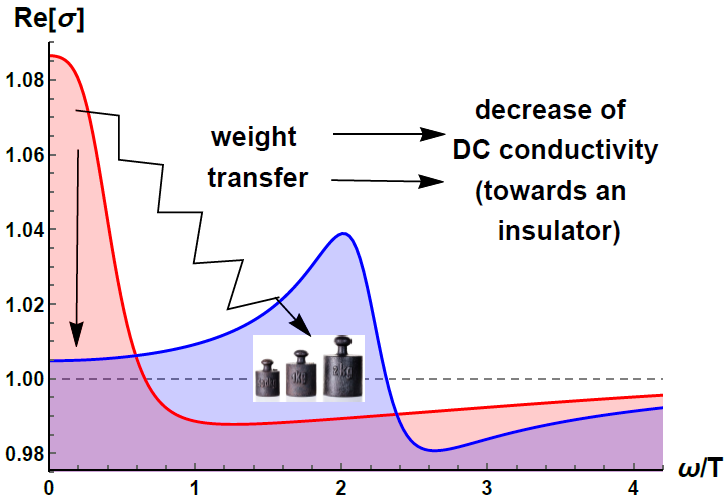}
    \caption{\textbf{Left: }The validity of the sum rule in eq.\eqref{asa} for the numerical computations of the electric conductivity. The plot is taken from one of the notebooks available with the lectures. \textbf{Right: }What the sum rule tells us in terms of physical mechanisms.}
    \label{fig:checksum}
\end{figure}
\begin{mdframed}[style=MyFrame]
\begin{center}
    \textbf{Trick \#4 : how to check your numerics?}
\end{center}\vspace{0.15cm}
Once you do some numerical calculations, especially if they are complicated, the first thing that you might want to do is to find a way to check they are right. In the context of numerical Green functions, a nice possibility is to use what is known as \textit{sum rules}. The sum rules come directly from the analytic properties of the Green functions. For some references see \cite{Kim:2015dna,Erdmenger:2015qqa,David:2011hy,David:2012cd}.\\
In the specific context of the electric conductivity, the useful sum rule is the following
\begin{equation}
    \int_0^{\infty}\,d\,\omega\,Re\left[\sigma(\omega)\,-\,\sigma_\infty\right]\,=\,0 \label{asa}
\end{equation}
which we tested as a check in fig.\ref{fig:checksum}. In the case of a superconducting state, there is an extension of the sum rule above known as Ferrell-Glover-Tinkham(FGT) sum rule:
\begin{equation}
    \int_0^\infty\,d\omega\,Re\left[\sigma_n(\omega)\,-\,\sigma_s(\omega)\right]\,-\,\frac{\pi}{2}\,K_s\,=\,0
\end{equation}
where $K_s$ is the Drude Weight for the superfluid part, namely the residue of the $1/\omega$ divergence in the imaginary part of the conductivity. See \cite{Kim:2015dna,Erdmenger:2015qqa} for an explicit use of the latter.\\
Notice that physically, the sum rule expresses the conservation of charged degrees of freedom, which are measured by the spectral weight, \textit{i.e.} the area under $Re[\sigma]$. This knowledge can be useful to predict several physical features.  An example is provided in fig.\ref{fig:checksum}. There the appearance of a growing peak in the mid-infrared already signals that the DC conductivity will decrease and the system will become an insulator.
\end{mdframed}
\vspace{0.3cm}
Before jumping to the next topic, let uso spend few words about the value of the conductivity at large frequency $\omega/T \gg 1$. That is the high energy limit of the Green function and it is totally governed by the UV part of the bulk geometry, which in our case is AdS. How can we understand the fact that:
\begin{equation}
    \sigma(\omega)\,=\,1\,,\quad \text{in the limit}\quad \omega/T\,\gg \,1
\end{equation}
using simple analytic arguments? We just need to consider the equation for the gauge field close to UV AdS geometry (which correspond to high energy physics and therefore the limit of high frequencies). The equation is:
\begin{equation}
    \left(f\,A_x'\right)'\,+\,\frac{\omega^2}{f}\,A_x\,=\,0
\end{equation}
whose solution, upon requiring infalling boundary conditions, reads:
\begin{equation}
    A_x(u)\,=\,e^{i\,\omega\,\int_1^u 1/f(q)\,dq}
\end{equation}
Given the bulk solution, and the fact that $f(0)=1$, we can read the conductivity as usual:
\begin{equation}
    \sigma(\omega)\,=\,\frac{G_{J_xJ_x}^R(\omega)}{i\,\omega}\,=\,\frac{A_x'(r)}{i\,\omega\,A_x(r)}\Big|_{u\rightarrow 0}\,=\,1
\end{equation}
which is indeed the value we observe in all our numerics. The fact that the conductivity is a constant at large frequency can be also understood by simple dimensional analysis in a CFT$_3$. You can try to compute the conductivity in a $5-$dimensional bulk spacetime which is asymptotically AdS$_5$. You will realize that in such case at large frequency:
\begin{equation}
    \sigma(\omega)\,=\,\omega\,,\quad \text{in the limit}\quad \omega/T\,\gg \,1
\end{equation}
This difference is due to the fact that the dimension of the conductivity in a CFT$_4$ is $[E]^1$. \subsection{Quasinormal modes: (modest) numerics and analytics}
In this last section, we consider the computation of the quasinormal mode frequencies for a single decoupled operator. The only necessary knowledge about the quasinormal modes required in this section is their definition as:
\begin{equation}
    \boxed{\text{QUASINORMAL MODE }}\quad  =  \quad \boxed{\text{POLE OF THE GREEN FUNCTION}}
\end{equation}
A quasinormal frequency is a complex frequency $\omega_{qnm}\,=\,\omega_R\,+\,i\,\omega_I$ for which the response displays a finite value even in absence of any source. In different words, it is an intrinsic excitation of the system, which can be simply found, using the holographic dictionary, by imposing:
\begin{equation}
    \text{leading}(\omega_{qnm})\,=\,0\,,\quad \text{subleading}(\omega_{qnm})\,\neq\,0
\end{equation}
The real part of the quasinormal mode is the ''normal'' part, which sets its frequency of propagation, while its imaginary part determines the lifetime of the excitation. More precisely, the relaxation time is given by:
\begin{equation}
    \tau_{rel}^{-1}\,\sim\,Im[\omega_{qmn}]
\end{equation}
This means that, at late time, the dynamics of the system is determined by the lowest quasinormal modes, those with lowest imaginary parts. In the situation, which is usually defined as \textit{coherent}, in which there is one (or few) pole whose imaginary part is much smaller than the rests,  one can do a consistent EFT-type truncation of the dynamics only in terms of that/those mode(s). Generally, this separation of scales is the consequence of a symmetry invariance or of a softly broken symmetry\footnote{See for example \cite{Grozdanov:2018fic}.}. The situation, in which there is no separation of scales, is more complicated; it is usually defined as \textit{incoherent}, and it necessitates to consider all the poles of the system. In other words, in this second case, an EFT description of the system is hard to obtain.\\
I invite you to read \cite{Jansen:2017oag,Denef:2009kn,Berti:2009kk,Kokkotas:1999bd,Birmingham:2001pj} if you want more details about quasinormal modes.\\
Here, we consider a simple example, which comes from \cite{Ammon:2019apj}. Take the equation for the scalar fluctuations
\begin{equation}
    0=\,\left(-k^2 \,N\, u+2 \,i \,N \omega -4 i\, \omega \right)\phi+ \left(u \,f'+2\, N\, f-4\, f+2 \,i \,u\,\omega \right)\phi'+u\, f \,\phi'' \label{eqQNM}
\end{equation}
defined on the Schwarzschild geometry:
\begin{equation}
    f(u)\,=\,1\,-\,u^3
\end{equation}
What we want to prove is that the lowest quasinormal mode for the scalar operator dual to this bulk field is a diffusive mode:
\begin{equation}
    \omega\,=\,-\,i\,D_\phi\,k^2\,+\,\dots\,,\quad D_{\phi}\,=\,\frac{N}{2\,N\,-\,3}\label{QNManal}
\end{equation}
where $N>3/2$ is an arbitrary parameter coming from the action.\\

\begin{figure}[H]
    \centering
    \includegraphics[width=0.42 \linewidth]{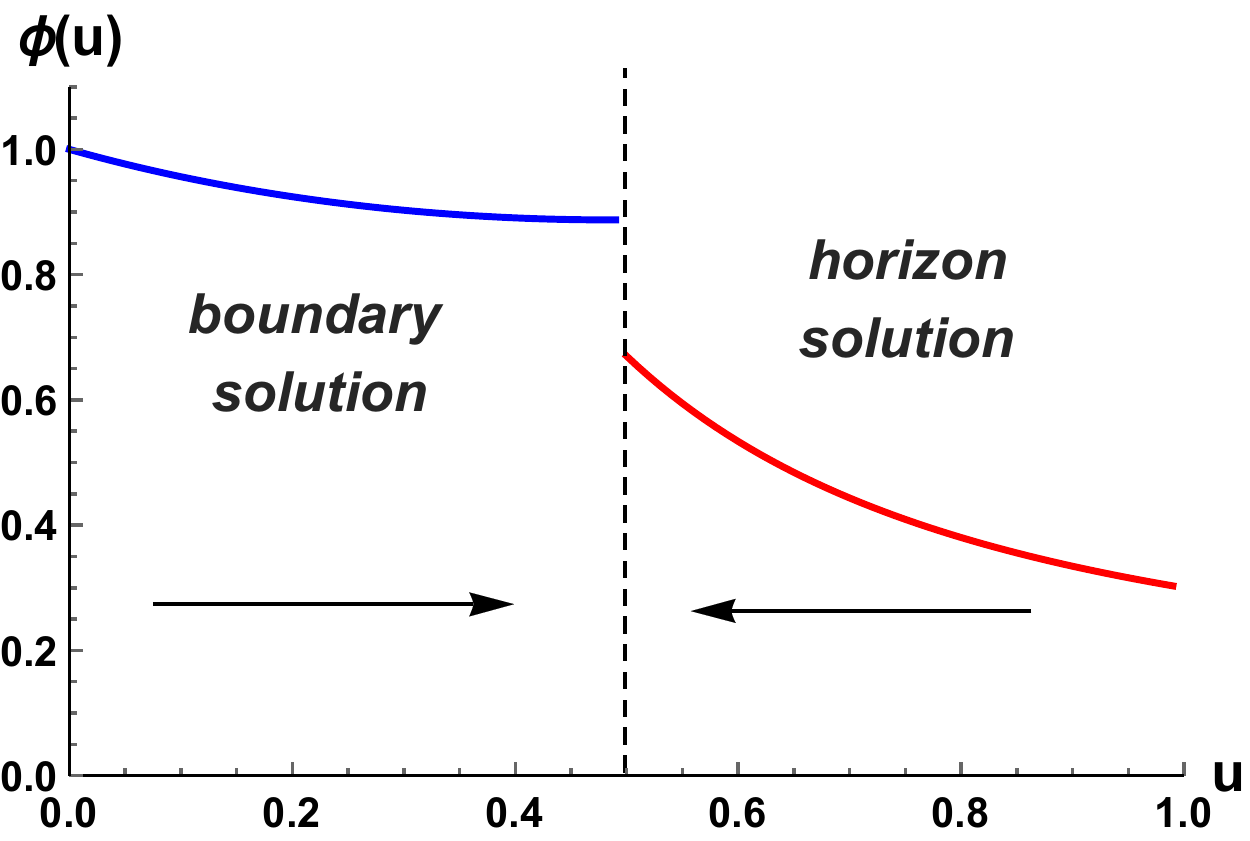}
    \quad 
    \includegraphics[width=0.5 \linewidth]{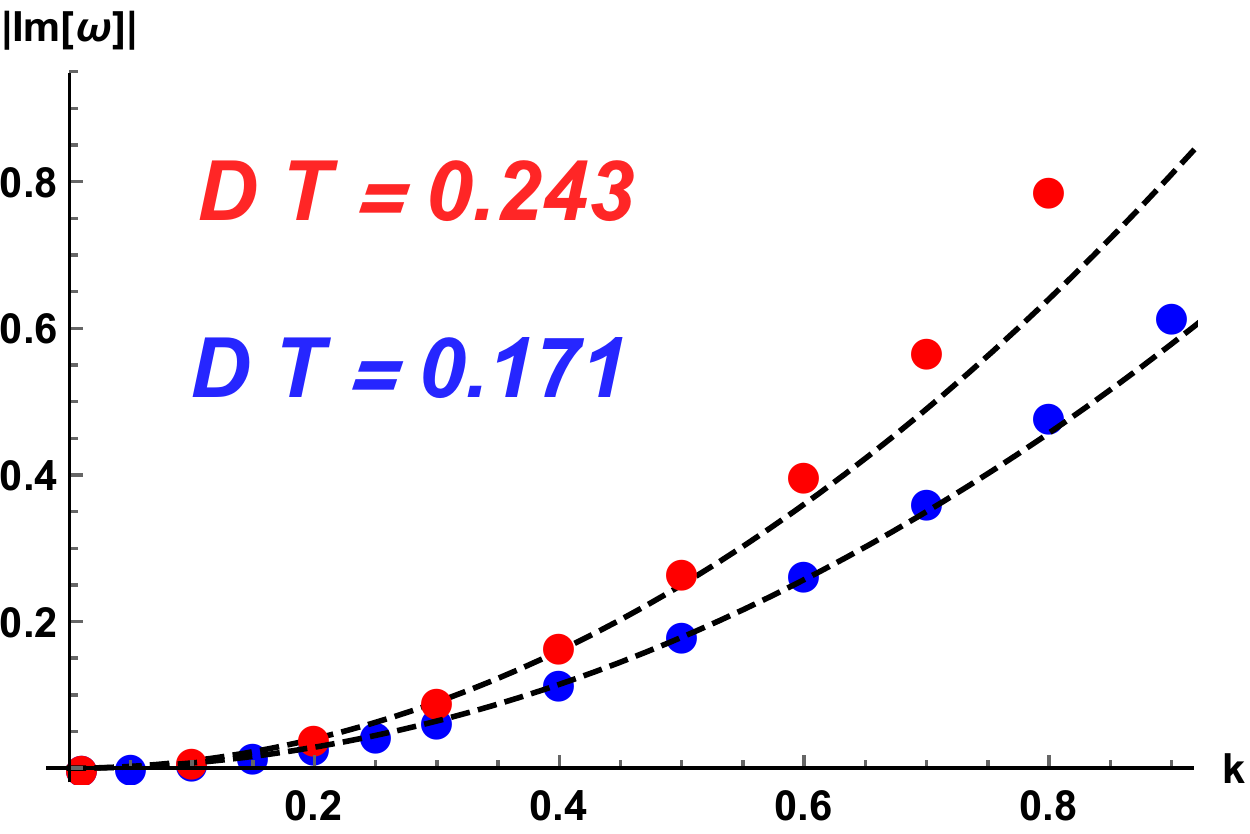}
    \caption{\textbf{Left: }Example of the two solutions. The full curve is clearly discontinuous at $u_m=1/2$. \textbf{Right: }The dispersion relation of the lowest diffusive QNM obtained from equation \eqref{eqQNM} for $N=3,5$ (red,blue).}
    \label{figex}
\end{figure}

\textbf{Numerics}\\[0.3cm]
Our first task is to obtain such quasinormal mode numerically, by using a simple procedure which involves a double shooting and a matching technique.
\begin{figure}
    \centering
    \includegraphics[width=0.85 \linewidth]{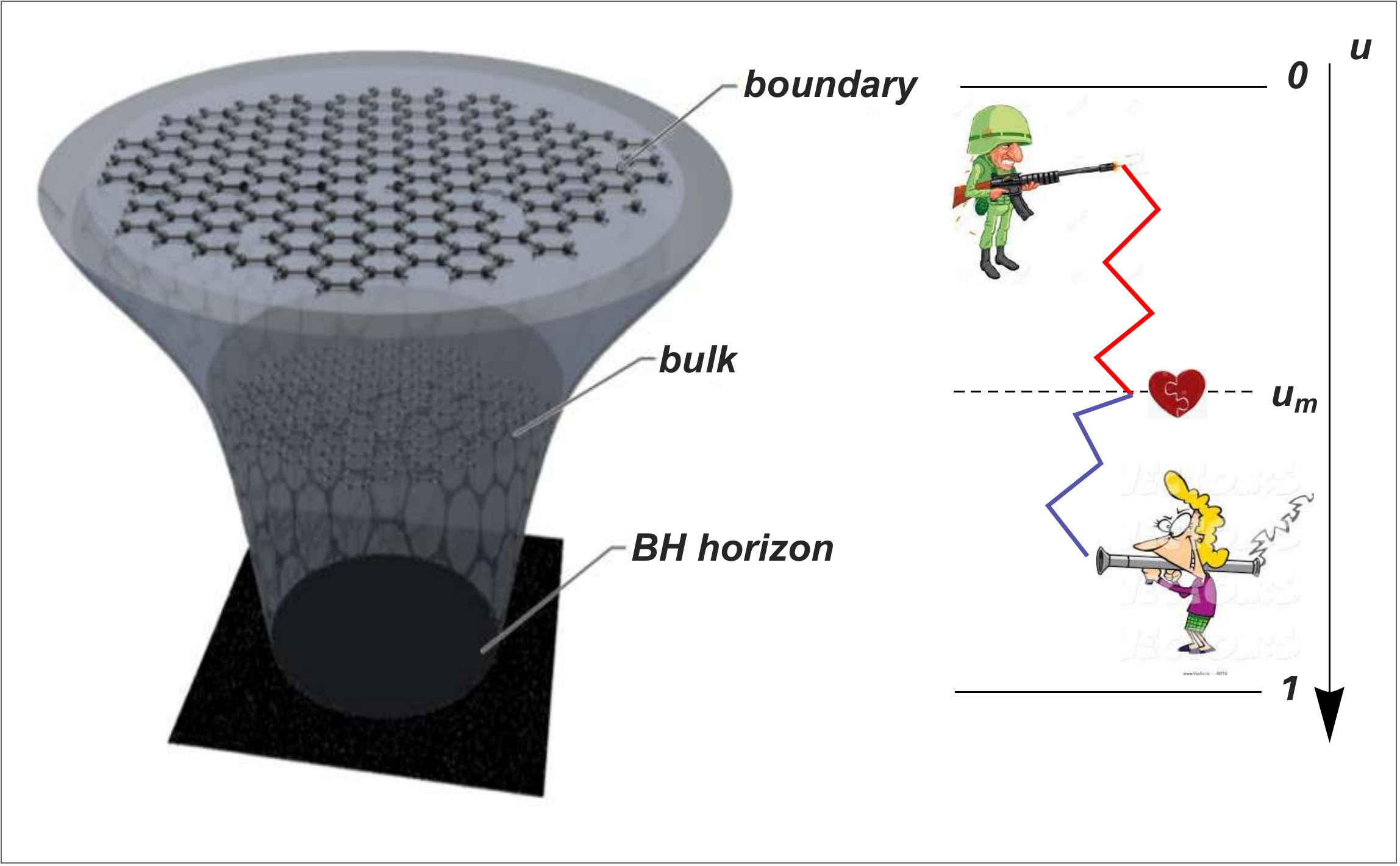}
    \caption{Sketch of the numerical procedure to obtain the quasinormal modes numerically from eq. \eqref{eqQNM}. The figure is adapted from \cite{Franz2018}.}
    \label{fig:cartoonshooting}
\end{figure}
The steps of the numerical procedure, which is sketched in fig.\ref{fig:cartoonshooting}, are as follows:
\begin{enumerate}
    \item Shoot from the boundary to an intermediate point, $\text{horizon}<u_m<\text{boundary}$, imposing as b.c.s. the vanishing of the source.
    \item Shoot from the horizon to an intermediate point, $\text{horizon}<u_m<\text{boundary}$, imposing regularity at the horizon (in EF coordinates).
    \item Generically, the two solutions will be discontinuous at the intermediate point $u_m$, as shown in fig.\ref{figex}.
    \item Match the two solutions (function value and first derivative value) at the intermediate point $u_m$ solving in terms of the free parameters and the (eventually complex) frequency $\omega$.
\end{enumerate}
The numerical results\footnote{For simplicity, we have taken a system with a purely imaginary quasinormal mode. The same method can be applied for a complex mode. In this case, every equation and every function have to be split in a real and an imaginary part. For example, in the case of a single field considered in the main text the matching conditions would be four now, instead of two.}, which can be found in one of the notebook available with the notes, are shown in fig.\ref{figex}.\\
\textbf{Analytics}\\[0.3cm]
Let us expand the solution for the field $\phi$ in powers of frequency and momentum as:
\begin{equation}
    \phi(u)\,=\,\phi_0(u)\,+\,\omega \,\phi_1(u)\,+\,k^2 \,\phi_2(u)\,+\,\dots
\end{equation}
and let us solve equation \eqref{eqQNM} order by order. For simplicity, we will do it only for $N=3$, the general result is left as an exercise.\\
For $N=3$, the equation at first order reads:
\begin{equation}
    \left(2-5 u^3\right) \phi_0'(u)-u \left(u^3-1\right) \phi_0''(u)\,=\,0
\end{equation}
The solution which is regular at the horizon is:
\begin{equation}
    \phi_0(u)\,=\,cost.
\end{equation}
(you can check that the other is logarithmically divergent) and we can fix the constant to $1$, thanks to the linearity of the system.\\ At second order, $\sim k^2$, we have:
\begin{equation}
    u \left(-\left(u^3-1\right) \phi_2''(u)-3\right)+\left(2-5 u^3\right) \phi_2'(u)\,=\,0
\end{equation}
whose regular solution is:
\begin{equation}
    \phi_2(u)\,=\,\frac{1}{u}\,+\,c_2
\end{equation}
Finally, at order $\sim \omega$, the equation reads:
\begin{equation}
    \left(u-u^4\right) \phi_1''(u)+\left(2-5 u^3\right) \phi_1'(u)+2 i\,=\,0
\end{equation}
and its regular solution is:
\begin{equation}
    \phi_1(u)\,=\,c_1-\frac{i}{u}-\frac{2\, i\, \tan ^{-1}\left(\frac{2 u+1}{\sqrt{3}}\right)}{\sqrt{3}}
\end{equation}
The final perturbative solution has the following form:
\begin{equation}
    \phi(u)\,=\,1+\omega \left(c_1-\frac{i}{u}-\frac{2 i \tan ^{-1}\left(\frac{2 u+1}{\sqrt{3}}\right)}{\sqrt{3}}\right)+k^2 \left(c_2+\frac{1}{u}\right)\,+\,\dots
\end{equation}
which is valid till this order in frequency and momentum. Close to the boundary, the solution does behave as:
\begin{equation}
    \phi(u)\,\sim\,\frac{k^2-i \omega}{u}+\left(c_2 k^2+c_1 \omega-\frac{i \pi  \omega}{3 \sqrt{3}}+1\right)+\mathcal{O}\left(u^1\right)
\end{equation}
Recognizing the leading term as the source for the dual operator $\mathcal{O}$, the quasinormal modes are given by its zeros. Therefore, we immediately get:
\begin{equation}
    \omega\,=\,-\,i\,k^2\,\quad \longrightarrow \quad D\,=\,1
\end{equation}
which is exactly the expected result for $N=3$. For your information, you just re-discover some of the results of \cite{Baggioli:2019aqf,Baggioli:2019abx,Ammon:2019apj}.
\begin{mdframed}[style=MyFrame3]
\begin{center}
    \textbf{Exercise \#13 : Your turn!}
\end{center}\vspace{0.15cm}
Try to obtain the quasinormal modes from eq. \eqref{eqQNM} for $N=7$ both numerically and analytically. Check the general result in \eqref{QNManal}.
\end{mdframed}
\begin{mdframed}[style=MyFrame]
\begin{center}
    \textbf{Trick \#5 : Master fields}
\end{center}
\vspace{0.15cm}
The trick of the master fields is a very efficient technique used to reduce the number of the equations to be solved. This can make your life simpler in several situations; let us see one together. This example is taken from \cite{Edalati:2010hk}; for the original discussion about the master fields see \cite{Kodama:2003kk}.\\
Consider the Einstein-Maxwell system defined by the action:
\begin{equation}
    S\,=\,\frac{1}{2\,\kappa_4^2}\int d^4x\,\sqrt{-g}\,\left[R\,-\,2\,\Lambda\,-\,\frac{1}{4}F^2\right]
\end{equation}
and assume the following ansatz for the background:
\begin{align}
   & ds^2\,=\,\frac{r^2}{L^2}\,\left(-f(r)dt^2\,+\,dx^2\,+\,dy^2\right)\,+\,\frac{L^2}{r^2\,f(r)}\,dr^2 \nonumber\\
   &A\,=\,A_t(r)\,dt
\end{align}
Let us consider the fluctuations:
\begin{equation}
    g_{\mu\nu}\,=\,\bar{g}_{\mu\nu}\,+\,h_{\mu\nu}\,,\quad A_\mu\,=\,\bar{A}_\mu\,+\,a_\mu
\end{equation}
and for simplicity let us take the radial gauge:
\begin{equation}
    a_r\,=\,0\,,\quad h_{r\nu}\,=\,0
\end{equation}
Although working in the radial gauge is convenient, it does not completely fix the gauge freedom. We still have the freedom determined by the diffeomorphism:
\begin{align}
   & \tilde{h}_{\mu\nu}\,=\,h_{\mu\nu}\,-\,\left(\nabla_\mu \xi_\nu\,+\,\nabla_\nu \xi_\mu\right)\\
   &\tilde{a}_\mu\,=\,a_\mu\,-\,\left(\xi^\nu \partial_\nu A_\mu\,+\,A_\nu\,\partial_\mu \xi^\nu\right)
\end{align}
After defining $u=r/r_h$ and $Q=k/\mu$ and $W=\omega/\mu$ and going to Fourier space, we define the gauge invariant modes
\begin{equation}
    X(u)\,=\,Q\,h^{y}_t(u)\,+\,W\,h^{x}_y(u)\,,\quad Y(u)\,=\,a_y(u)
\end{equation}
The statement is that using the master fields
\begin{equation}
  \small   \Phi_{\pm}\,=\,-\,\mu\,\frac{Q\,f(u)\,u^3}{W^2\,-\,f(u)\,Q^2}\,X'(u)\,-\,\frac{6}{u}\,\left[\frac{2\,f(u)\,Q^2}{W^2\,-\,f(u)\,Q^2}\,+\,u\,\left(1\,\pm\,\sqrt{1\,+\,Q^2}\right)\right]\,Y(u)
\end{equation}
the equations for the perturbations boil down to
\begin{equation}
    \left[u^2\,f\,\Phi'_\pm\right]'\,+\,\left[u\,f'\,+\,\frac{3}{u^2\,f}\,\left(W^2\,-\,f\,Q^2\right)\,-\,\frac{6}{u^3}\,\left(1\,\pm\,\sqrt{1\,+\,Q^2}\right)\right]\,\Phi_{\pm}\,=\,0
\end{equation}
which are two simple and (very importantly) decoupled equations.\\
Now, with these last decoupled equations you can directly applied the numerical methods you have learned and extract the QNMs and compare them with \cite{Edalati:2010hk}. Be careful, the relation between the physical green functions and the leading and subleading terms of the master fields is tricky. See for example \cite{Edalati:2010hk}. 
\end{mdframed}

\chapter{The geometrization process and holographic RG flows}
\label{intro5} 
\begin{figure}
    \centering
    \includegraphics[width=0.95\linewidth]{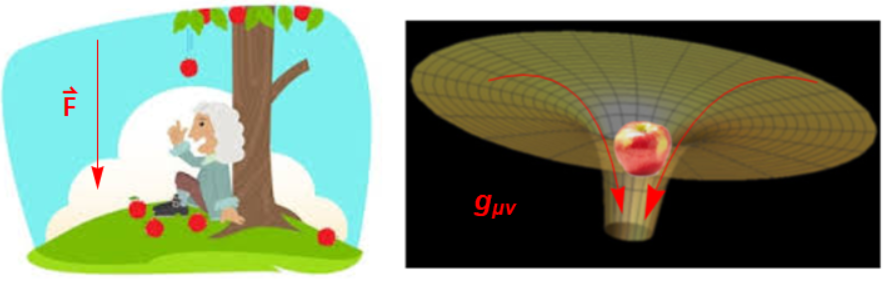}
    \caption{The old and new understanding of gravity. \textbf{Left:} The description of gravity as a force accelerating bodies. \textbf{Right:} The description of gravity in terms of curved geometry and bodies simply following geodesics on it.}
    \label{fig:geom}
\end{figure}

\begin{figure}
    \centering
    \includegraphics[width=0.6\linewidth]{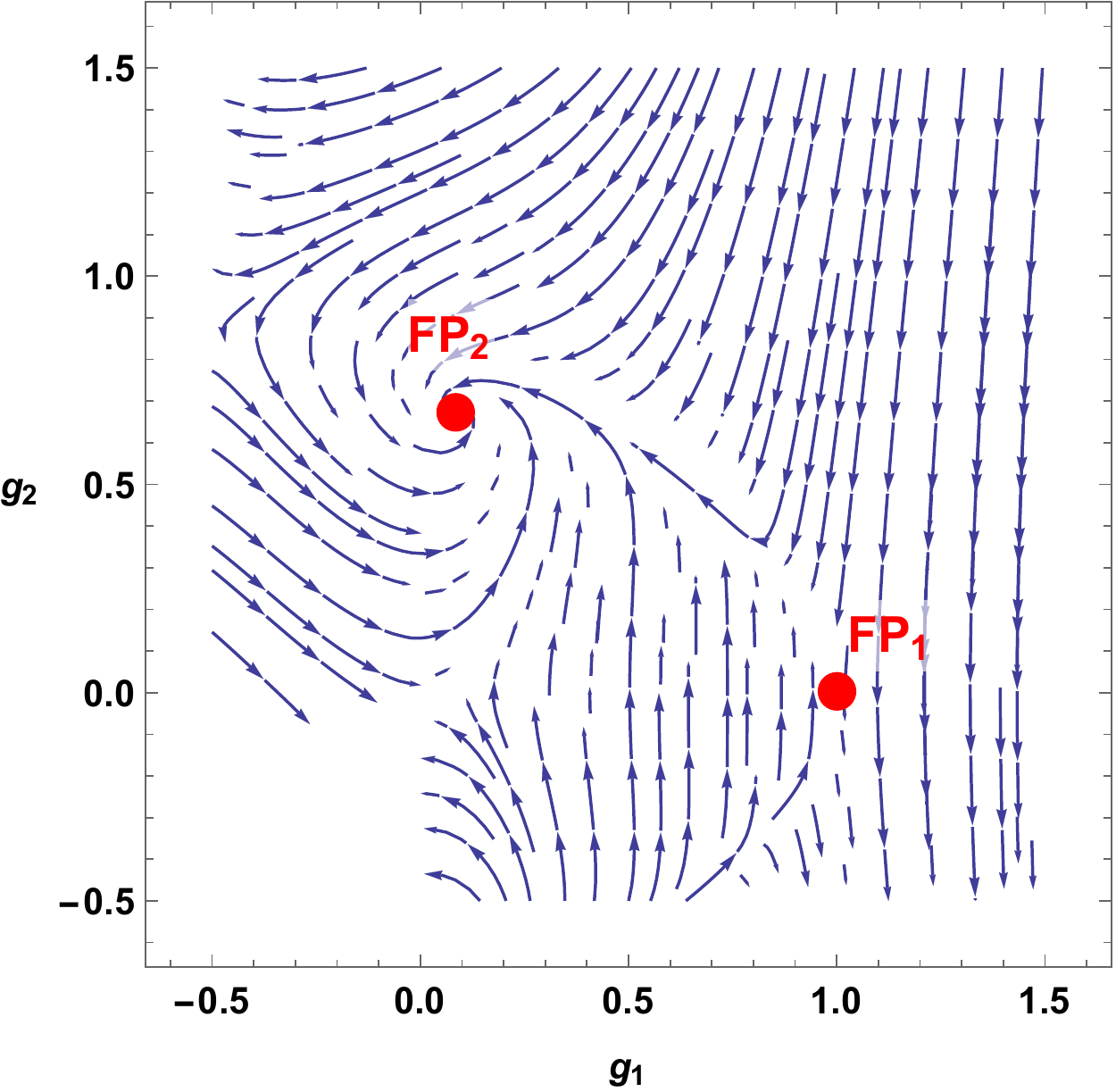}
    \caption{A graphic visualization of the RF flow trajectories between two fixed points FP$_{1,2}$ in function of two couplings of the theory $g_{1,2}$.}
    \label{figflow}
\end{figure}

\hspace{0.2cm} \includegraphics[width=0.3\textwidth]{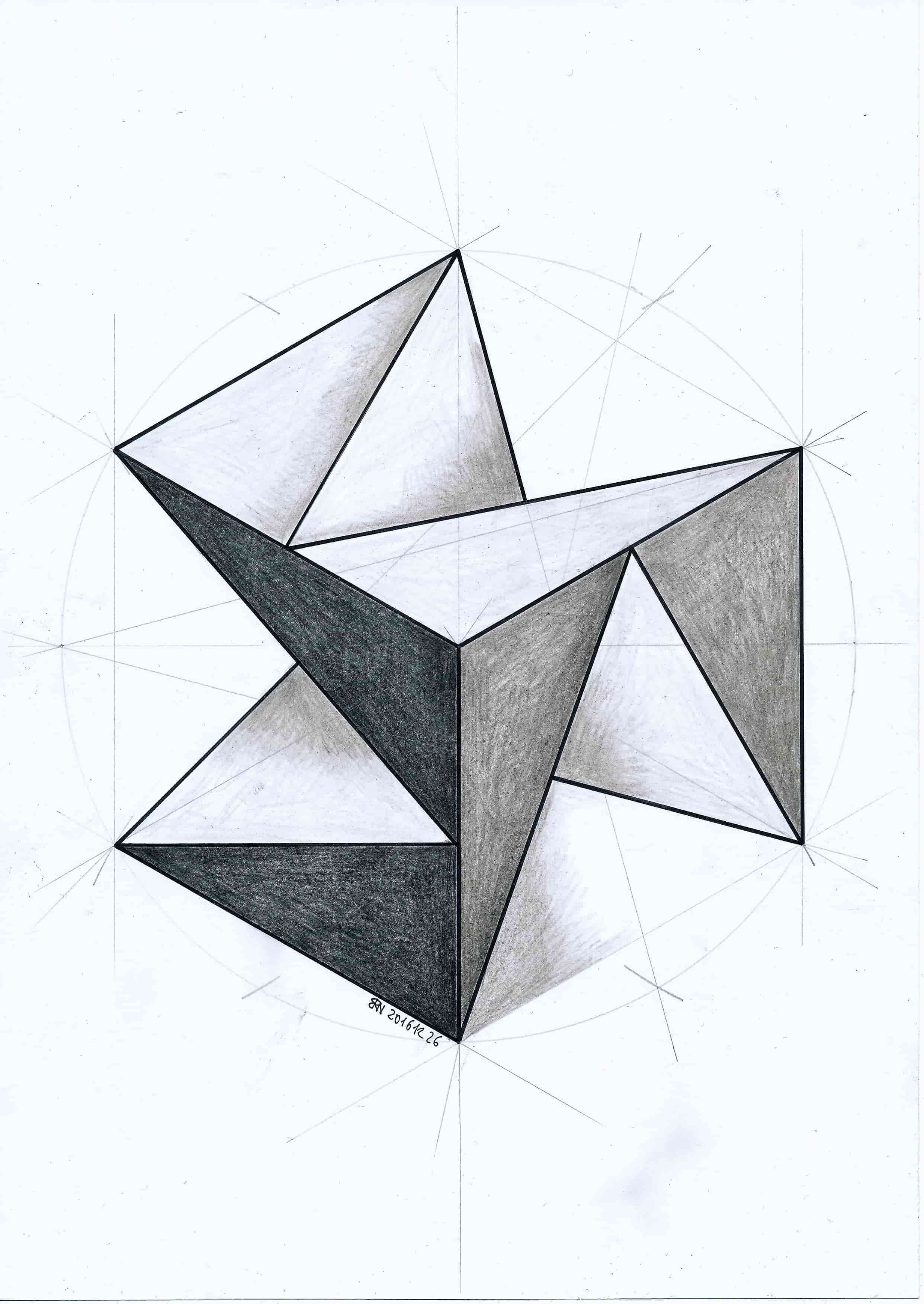}\\
\epigraph{Mighty is geometry; joined with art, resistless.}{\textit{Euripides}}
Since the early ages of Physics, Math has been identified and promoted to be the language of our Universe. Geometry, in particular, has been playing a fundamental role in the construction and understanding of modern theoretical physics \cite{morrison2018geometry}. Two beautiful examples are given by the Principle of least action and the theory of General Relativity (see fig.\ref{fig:geom}).
We can understand the refraction of light and the beauty of a rainbow by requiring that light rays follow the shortest possible geometrical path. We can describe the gravity that we experience every time an object falls on the ground by considering that we are leaving in a curved spacetime which is distorted by massive objects. Isn't that so elegant?\\
As we already mentioned in the previous section, Holography represents another wonderful example of \textit{geometrization}. The phase space of the dual field theory and its Renormalization Group (RG) Flow structure are beautifully encoded in the geometric properties of the dual gravitational spacetime.\\
The renormalization group is a key concept in quantum field theory and condensed matter and its foundations can be found in several textbooks. A brief but very nice set of lectures can be found in \cite{hollowood20096}. The main idea is that the physics of a specific system depends in a local way on the energy scale $\mathrm{E}$ through the so-called \textit{beta-function equation}. From a physical point of view, known as the \textit{Wilsonian approach}, the RG is a direct consequence of the coarse graining procedure, adopted when we look at our system at larger and larger length scales. Technically it arises from integrating out the heavy (or highly energetic) modes which are irrelevant at low energy. One of the main consequence of the RG idea is \textit{universality}, the fact that different microscopic UV theories can flow to the same IR fixed point, whose description necessitates only a small subset of the full spectrum of operators.\\
From a practical point of view, the definition of the RG flows comes from the beta-function equations, which constitute a dynamical system in terms of the coupling constant $g(\mathrm{E})$. In fig.\ref{figflow} we show an example of such flow between two different fixed points. The arrows indicate the direction of the flow and the stability of the fixed point, which are directly related to the relevance or irrelevance of the operators.
In this section, we analyze more in details this correspondence and we show some concrete and interesting examples of it.
\section{Why Anti-De Sitter and not something else?}
One concrete question related to this section (and which you should already have asked yourself) regards the privileged role that the Anti de Sitter spacetime plays in Holography. Why is that the case\footnote{This is even a more pressing question given the fact that our Universe is definitely not Anti de Sitter. All the cosmological observations indicate that our Universe is expanding; we live in a spacetime with small but positive cosmological constant (even if String Theory does not like it).}?\\
The symmetries of the dual field theory are visualized from the bulk point of view through the isometries of the spacetime geometry:
\begin{equation}
    \boxed{\text{symmetries of the field theory}}\,\,\longleftrightarrow  \boxed{\text{spacetime isometries}}
\end{equation}
Let us consider the simplest possible set of symmetries; a Lorentz invariant field theory with scaling symmetry. This means that our theory is invariant under the full Poincar\'e group, plus the scaling symmetry:
\begin{equation}
    x^\mu\,\quad \longrightarrow \quad \lambda \,x^\mu \label{ff}
\end{equation}
On the other side, the most generic Lorentz invariant $d+1$ dimensional geometry has the following form:
\begin{equation}
    ds^2_{(d+1)}\,=\,\Xi(u)\,\left(-dt^2\,+\,d\vec{x}^2\,+\,du^2\right)
\end{equation}
where at this point $\Xi(u)$ is an arbitrary function of the holographic extra-dimension. Notice how such function cannot depend on the boundary coordinates because of the requirement of invariance under time and space translations. Given the fact that the extra dimension, in our notations, has the dimension of the inverse of energy:
\begin{equation}
    u\,\,\sim\,\,\text{energy}^{-1}
\end{equation}
then the scaling symmetry in eq.\eqref{ff} has to be enhanced to a larger scaling transformation of the type:
\begin{equation}
    x^\mu\,\,\rightarrow \,\,\lambda\,x^\mu\,,\quad u\,\,\rightarrow \,\,\lambda\,u\ \label{trasf}
\end{equation}
Physically, this is simply the statement that a transformation with $\lambda>1$ is a coarse-graining operation, which corresponds to consider the physical system at lower energies. If we now require invariance under the scaling symmetry of eq.\eqref{trasf}, the function form of $\Xi(u)$ is unique and given by:
\begin{equation}
    \Xi(u)\,\,\rightarrow \,\,\lambda^{-1}\,\Xi(u)\,\,\Longrightarrow\,\,\Xi(u)\,=\,\frac{L}{u}
\end{equation}
As a result of this short argument, we are left with the following geometry
\begin{equation}
  \boxed{  ds^2\,=\,\frac{L^2}{u^2}\,\left(-dt^2\,+\,d\vec{x}^2\,+\,du^2\right)}\,,\quad u\,\in\,\left[0,\infty\right]
\end{equation}
which is indeed the AdS spacetime in Poincar\'e coordinates. The parameter $L$ is a dimensionful constant which determines the radius of the AdS spacetime and the corresponding negative constant curvature:
\begin{equation}
    R\,\sim\,\frac{1}{L^2}\,<\,0
\end{equation}
In these coordinates system, $u=0$ is the position of the conformal boundary, while $u=\infty$ is that of the ''horizon''\footnote{The horizon is defined as the hypersurface where the killing vector $\partial t$ has zero norm.}. For many more details about the AdS geometry and its properties have a look at \cite{Moschella2006}.\\
In summary, having our bulk geometry asymptotically AdS corresponds to have a dual field theory with a UV conformal fixed point. This is of course not always the case and there has been a lot of work about considering more generic situations. As an example, people considered field theories invariant under the more general scaling transformation
\begin{equation}
     \vec{x}\,\,\rightarrow \,\,\lambda\,\vec{x}\,,\, t\,\,\rightarrow \,\,\lambda^z\,t\,, \label{lifscal}
\end{equation}
which obviously breaks the Poincar\'e invariance of the system. Time and space are treated differently and the parameter $z$, encoding this difference, is usually called ''the dynamical exponent''. These field theories are defined as \textit{Lifshitz field theories} and they describe the quantum critical point of several condensed matter systems\footnote{Some concrete examples are: (I) $z=2$ -- onset of antiferromagnetism in
clean ‘itinerant’ fermion systems \cite{sachdevbook}; (II) $z=3$ --  onset of ferromagnetism in clean itinerant fermion systems \cite{Monthoux2007,PhysRevB.64.195109,PhysRevLett.88.217204}; $z=2.6$ --  CeCu$_{6-x}$Au$_x$ at critical doping \cite{Si2001}.}. Moreover, this type of scaling is very relevant for the Lorentz violating quantum gravity theory known as \textit{Horava-Lifshitz gravity} \cite{Horava:2009uw,Horava:2009if}.\\
The Lifshitz scaling in eq.\eqref{lifscal} defines a different class of fixed points. Those fixed points can be described using the following bulk geometries
\begin{equation}
    ds^2\,=\,\frac{L^2}{u^2}\,\left(-\,u^{2(1-z)}\,dt^2\,+\,d\vec{x}^2\,+\,du^2\right)
\end{equation}
which is indeed invariant under the scaling \eqref{lifscal}. You can convince yourself that these geometries are not solution of the Einstein-Hilbert action but they require some additional matter content. Naively, this matter content is the one which provides a fixed reference frame which breaks Lorentz invariance. We will construct these geometries in more details in the next sections.
\begin{mdframed}[style=MyFrame3]
\begin{center}
    \textbf{Exercise \#14 : AdS$_2$ as a Lifshitz geometry}
\end{center}
Prove that the AdS$_2 \times $R$^d$ geometry:
\begin{equation}
    ds^2\,=\,\frac{-dt^2+d\rho^2}{\rho^2}\,+\,d\vec{x}^2
\end{equation}
corresponds to a Lifshitz geometry:
\begin{equation}
    ds^2\,=\,-\,\frac{dt^2}{u^{2\,z}}\,+\,\frac{du^2+\,d\vec{x}^2}{u^2}
\end{equation}
with $z=\infty$.\\
\textbf{Tip}: use an appropriate change of coordinates such that the limit $z \rightarrow \infty$ is well defined.\\
\textbf{Curiosity}: The exercise of thinking about AdS$_2$ as a Lifshitz field theory with infinite dynamical scaling allows us to introduce the concept of \textit{semi-local  quantum criticality} \cite{Faulkner:2009wj,Iqbal:2011in}. The physical system separates in domains of size $\xi$. Inside those domains ($|x|<\xi$), because of this particular scaling symmetry, the physics is governed by a $0+1$ quantum mechanical model where the spatial coordinate plays no role and the Green function is simply $G(t) \sim t^{-2\delta}$. For more details see \cite{Iqbal:2011in}.

\end{mdframed}
\section{Walking around holographic fixed points}
The RG flows equations of a specific field theory can be written as a dynamical system:
\begin{equation}
    \mathrm{E}\,\frac{\partial g^I}{\partial \mathrm{E}}\,=\,\beta^I(g^J,\mathrm{E})
\end{equation}
where the unknowns are the couplings of the theory and the variable is the energy scale $\mathrm{E}$. From there, one can see that the flows are induced by a non trivial value for the beta functions $\beta^I$ and in particular the fixed points of such a dynamical systems are, by definition, those where the beta functions vanish:
\begin{equation}
    \beta(\mathrm{E})\,=\,0
\end{equation}
The vanishing of the beta function implies that the couplings of the theory do not depend on the energy scale $\mathrm{E}$ and therefore the theory is scale invariant. In other words, the fixed points of the RG flows coincide by definition with scale invariant theories (for more details see \cite{hollowood20096}).
\begin{figure}
    \centering
    \includegraphics[width=0.45 \linewidth]{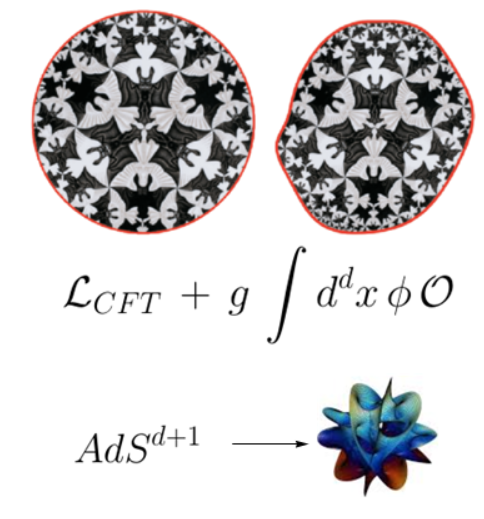}
    \quad \includegraphics[width=0.45 \linewidth]{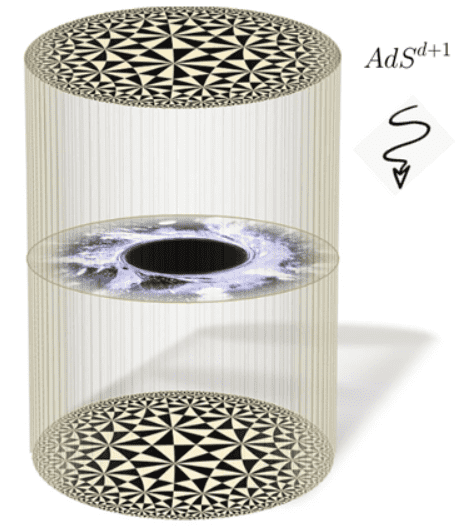}
    \caption{The geometrization of the RG flow. The deformation of a CFT by a relevant operator $\mathcal{O}$ is described by a geometric deformation of the UV AdS spacetime. The RG flow structure is encoded in the geometry of the bulk spacetime which now interpolates between a UV AdS geometry to something different in the IR. In the right panel we show the example of a temperature deformation, which corresponds to an asymptotically AdS black hole geometry. The right figure is adapted from \cite{Maldacena:2016upp}.}
    \label{fig:cftfig}
\end{figure}
How can we induce a non trivial flow? The idea is to deform a specific scale invariant\footnote{In the rest of this section we will use conformal invariant and scale invariant as a synonyms. Be aware that this is a highly non trivial statement which does not holds generically. For a review about this topic see \cite{Nakayama:2013is}.} fixed point with a relevant operator:
\begin{equation}
    \mathcal{L}_{CFT}\,+\,\int d^dx\,\phi_0\,\mathcal{O}
\end{equation}
where the conformal dimension of such operator is $\Delta_\mathcal{O}<d$. The relevant character of this operator implies that this deformation becomes more and more important decreasing the energy scale $\mathrm{E}$, and that the theory will run away from the original scale invariant (and from now UV) fixed point. In presence of another fixed point, where such deformation is irrelevant, the RG flow can end in this new point whose properties and symmetries might be very different from the original UV one. In particular:
\begin{itemize}
    \item[(I)] The properties of the IR fixed point where the RG flow ends at low energy could be completely independent of those of the UV fixed point. This leads to the concept of \textit{universality}, which is nothing else than the statement of insensitivity with respect to the UV (microscopic) details of the theory.
    \item[(II)] The symmetries of the IR fixed point are different from those of the UV one in general for two reasons. a) The deformation can break some of the original symmetry of the UV description. b) The IR fixed point can exhibit emergent symmetries which were not present at the UV level.
\end{itemize}
How is this situation realized from the gravitational bulk picture? The flow between two fixed points is beautifully geometrized as a complicated geometry which interpolates between two different asymptotic spacetime. As a concrete example, let us take in the UV a conformal field theory. Such field theory corresponds via the duality to an Anti de Sitter geometry. Once we deform the CFT with a relevant operator $\mathcal{O}$, the initial AdS geometry is deformed towards the interior of the bulk by the non-trivial profile of the bulk field dual to the $\mathcal{O}$ operator, $\phi$. Eventually, in the IR (typically the near-horizon region) the geometry acquires again a simple and scale invariant form (but not necessarily AdS), which signals the existence of an IR fixed point at which the RG flow ends. This geometrization process is summarized in fig.\ref{fig:cftfig}. Notice that the beta function which we discussed can be computed directly through holographic computations, see for example \cite{Megias:2014iwa}.
\section{An instructive example of holographic flows}
\begin{figure}[h]
    \centering
    \includegraphics[width=0.45\linewidth]{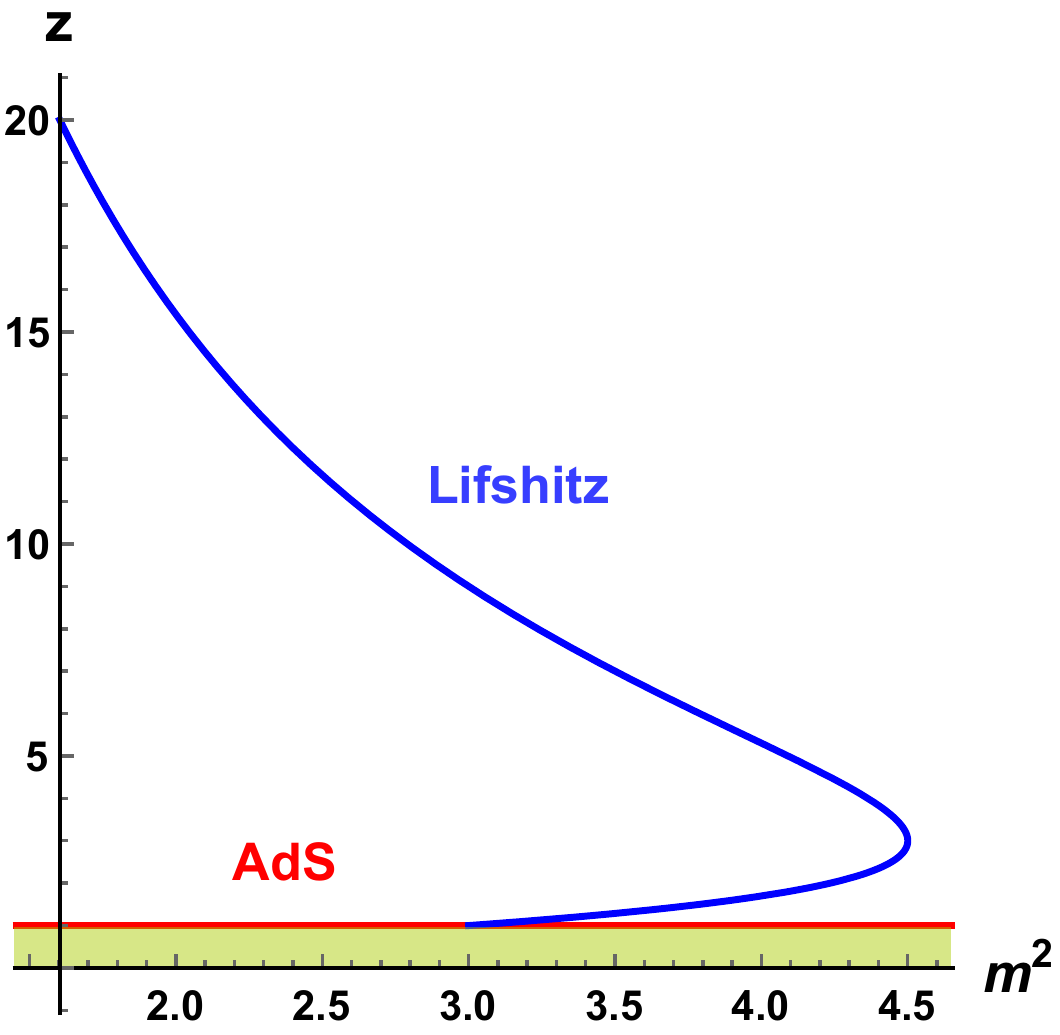}
    \quad
    \includegraphics[width=0.45\linewidth]{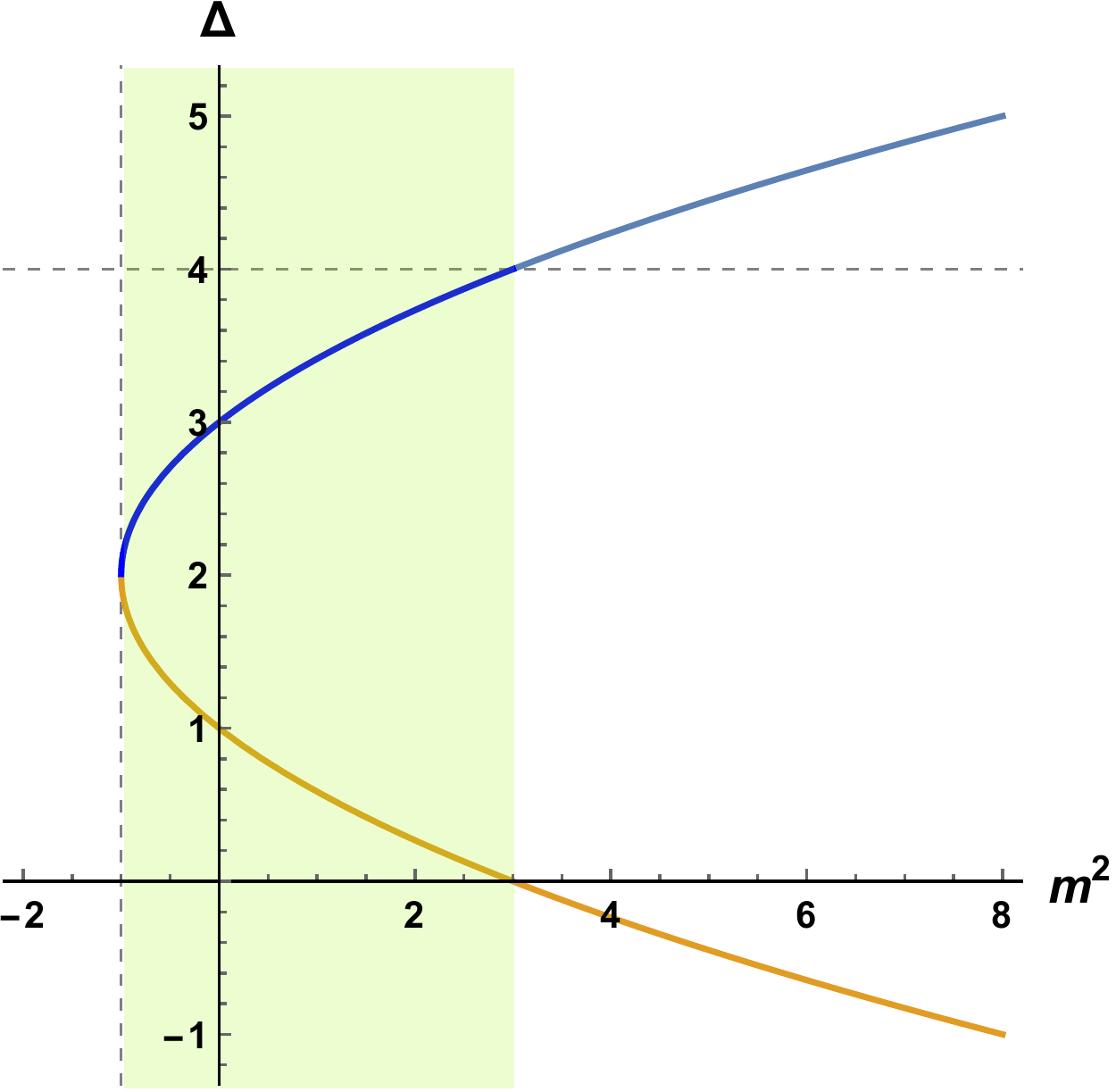}
    \caption{\textbf{Left: }The fixed points of the system in function of the Proca vector mass $m^2$. In red the AdS solution $z=1$; in blue the Lifshitz solution $z\neq 1$. The shaded region indicates the violation of the NEC. \textbf{Right: } The conformal dimension of the vector operator dual to the bulk field $B^\mu$ around the Lorentz invariant fixed points corresponding to the AdS geometry. The dark blue indicates the branch where the operator is relevant, $\Delta_\beta<d=4$.}
    \label{figdim}
\end{figure}

\begin{figure}
    \centering
    \includegraphics[width=0.3\linewidth]{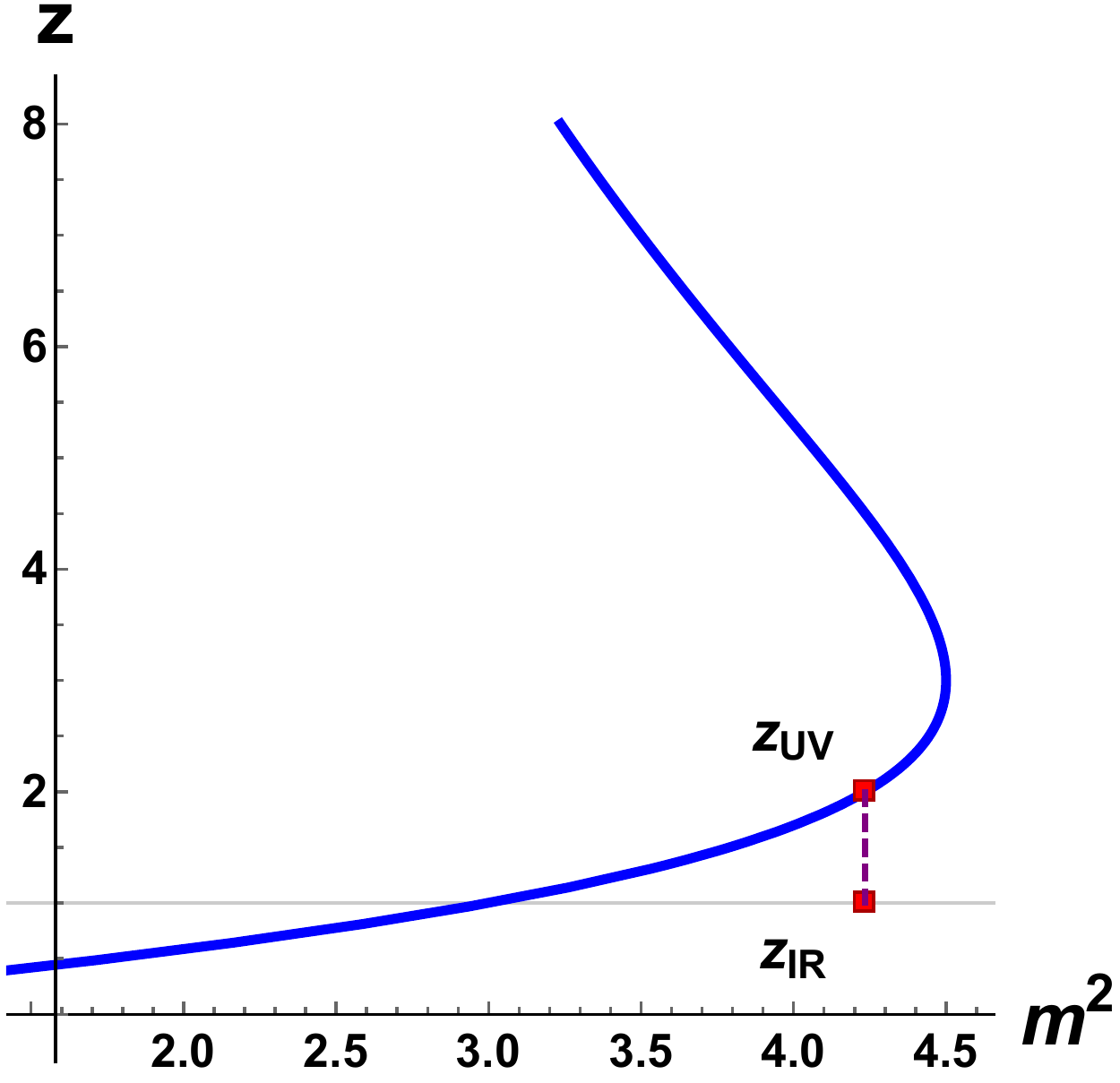}
    \quad
    \includegraphics[width=0.6\linewidth]{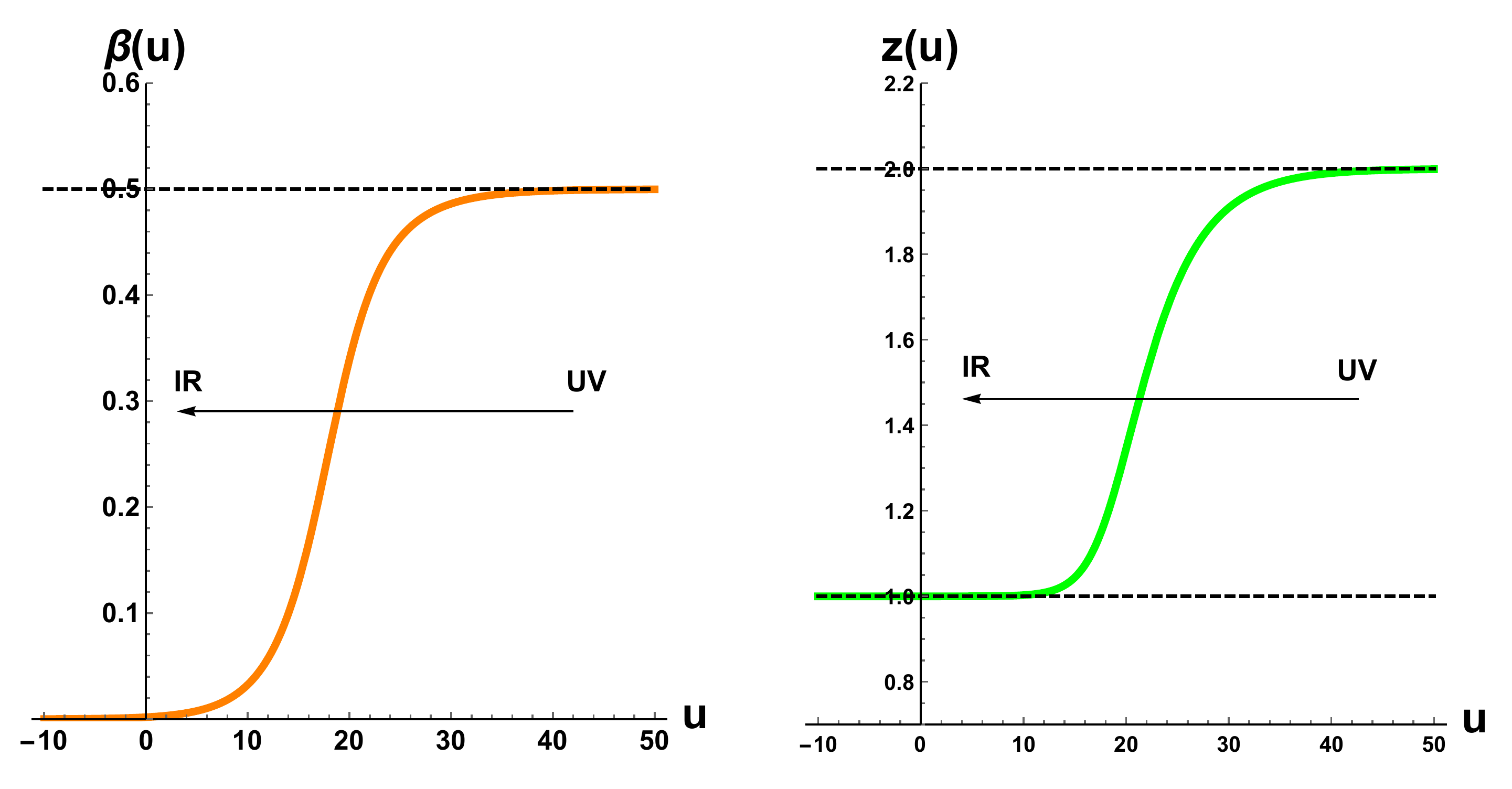}
    \caption{\textbf{Left: } A flow between a UV Lifshitz fixed point with $z=2$ and an IR AdS fixed point. \textbf{Right: }The profile of the metric and the massive vector in the bulk. The dashed lines indicate the fixed point values for $\beta_0$ and $w_0\equiv 1/z$.}
    \label{flow1fig}
\end{figure}

\begin{figure}
    \centering
    \includegraphics[width=0.95\linewidth]{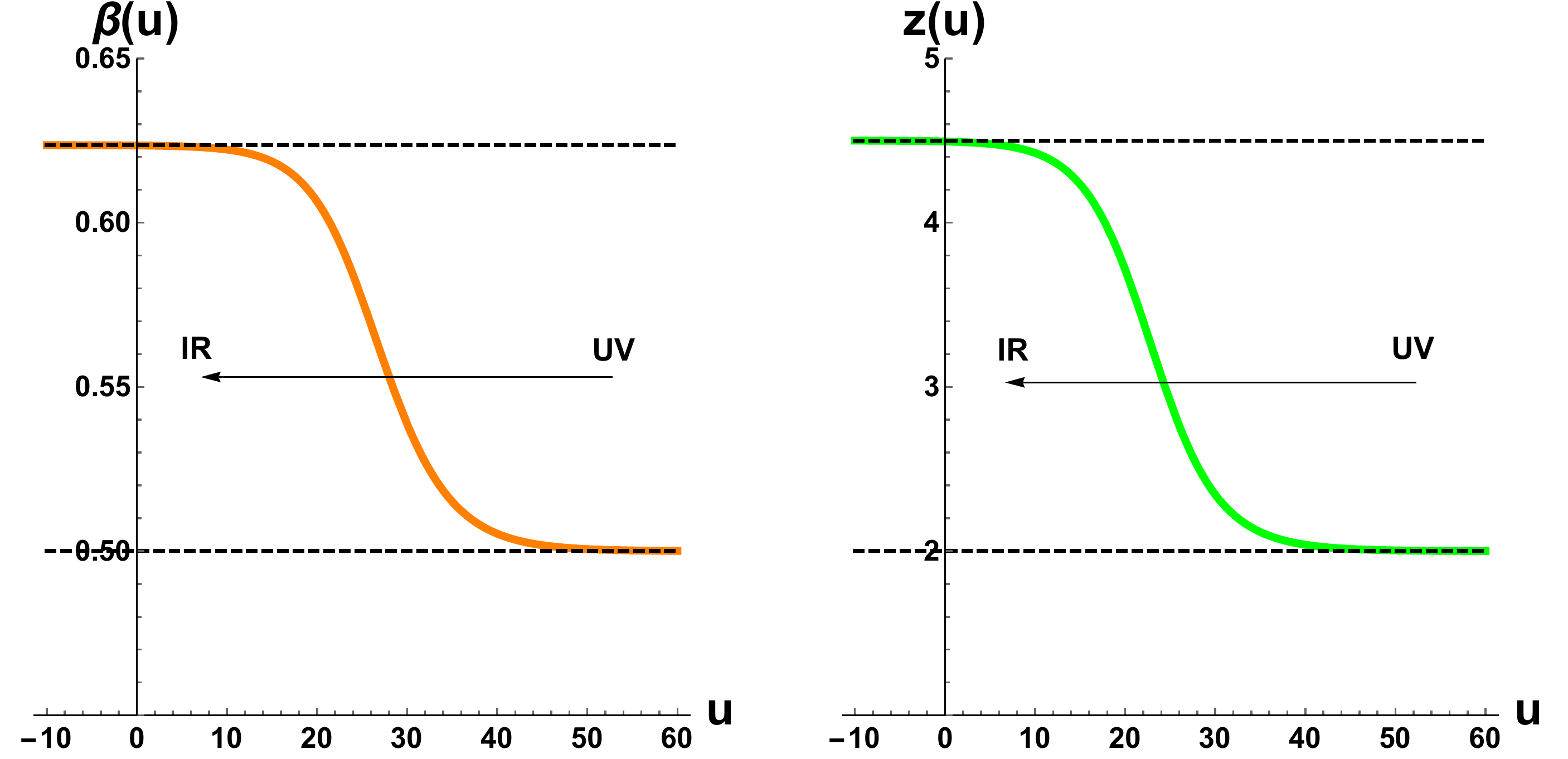}
    
    \vspace{0.5cm}
    
    \includegraphics[width=0.95\linewidth]{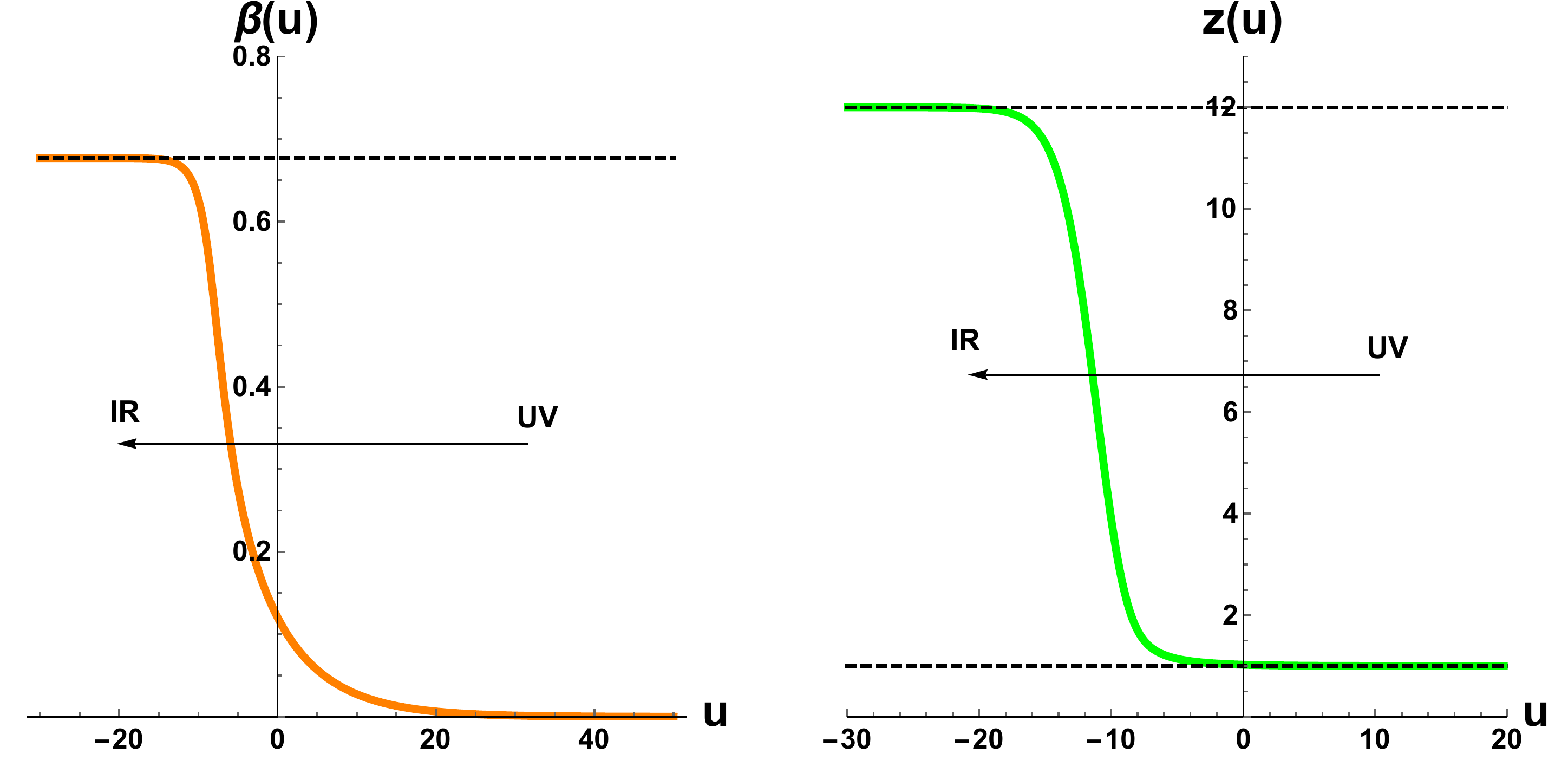}
    \caption{\textbf{Left: } A flow between a UV Lifshitz fixed point with $z=2$ and an IR Lifshitz fixed point with $z=4.5$. The dashed lines indicate the fixed point values for $\beta_0$ and $w_0\equiv 1/z$. \textbf{Right: } A flow between a UV AdS fixed point  and an IR Lifshitz fixed point with $z=12$. The dashed lines indicate the fixed point values for $\beta_0$ and $w_0\equiv 1/z$.}
    \label{flow2fig}
\end{figure}

Let us be less philosophical and more practical!\\
In this last section, we will construct holographic RG flows which interpolates between AdS fixed points and Lifshitz fixed points. From the dual field theory point of view, these geometries correspond to the RG flows between relativistic fixed point invariant under the scaling in eq.\eqref{ff} and non-relativistic fixed points obeying the Lifshitz scaling in eq.\eqref{lifscal}. The original paper presenting these solutions can be found in \cite{Braviner:2011kz}. See also \cite{Bednik:2013nxa,Kharuk:2015wga} for following discussions.\\
Since our flows require non-relativistic fixed points, we need to deform our Lorentz-invariant theory at least with a vector operator\footnote{A scalar clearly would not break Lorentz invariance.}. In order to do that, we consider the simple action
\begin{equation}
    \mathcal{S}\,=\,\int d^5x\,\sqrt{-g}\,\left[R\,-\,F^2\,-\,V(B^2)\right]\label{pp1}
\end{equation}
where $B^2 \equiv B_\mu B^\mu$ and $F\equiv dB$. For simplicity, we assume the following potential
\begin{equation}
    V(z)\,=\,-\,12\,+\,2\,m^2\,z \label{pp2}
\end{equation}
which fixes our cosmological constant to be $\Lambda=-6$. Within this choice, our system is known as the massive-Proca theory and it is the pillar of non-relativistic holography \cite{Taylor:2008tg,Taylor:2015glc,Korovin:2013bua}.\\
Following \cite{Megias:2014iwa}, we write down the metric in the so called
''RG-Gauge'':
\begin{align}\label{Ansatzg}
& ds^2\,=\,-e^{2\,u}\,dt^2+ e^{2\int w(\xi)\,d\xi} \,d\vec{x}^2\,+\,N(u)^2 \,du^2\,\,,\\
& B\,=B_t\,dt\,=\,\,e^u\,\beta(u)\,\,dt\,.
\end{align}
which is very convenient for this type of analysis. In this choice of coordinates the UV boundary is located at $u=\infty$ while the IR one at $u=-\infty$. A non trivial profile for the time component of the $B$ vector breaks Lorentz-invariance and in particular boost symmetry.\\
\begin{mdframed}[style=MyFrame3]
\begin{center}
    \textbf{Exercise \#15 : Drive through the Proca model.}
\end{center}
Consider the action defined in eq.\eqref{pp1} with the potential chosen as in \eqref{pp2}. Assume the ansatz defined in \eqref{Ansatzg}. Derive the equations of motion for the various functions $\beta,N,w$.
\end{mdframed}
\vspace{0.3cm}
The fixed points of the system of equations, following from the action \eqref{pp1}, are defined by the background functions taking constant values:
\begin{equation}
    \beta(u)\,=\,\beta_0\,,\quad w(u)\,=\,w_0\,,\quad N(u)\,=\,N_0
\end{equation}
It is straightforward to obtain the form of the most generic fixed point
\begin{equation}
  \beta_0\,=\,\sqrt{\frac{z-1}{2\,z}}\,,\quad w_0\,\equiv\,\frac{1}{z}\,,\quad m^2\,=\,\frac{36\,z}{9\,+\,z^2\,+2\,z}\,,\quad N_0\,=\,\frac{\sqrt{z^2\,+\,2\, z\,+\,9\,}}{2\, \sqrt{3}\, z}\label{sols}
\end{equation}
which is shown in fig.\ref{figdim}.
\\
Whenever $z=1$, the profile of the gauge field is trivial and Lorentz-invariance is preserved. In such case, the metric in \eqref{pp1} is simply that of AdS spacetime in the RG gauge coordinates (try to compute the Ricci scalar and convince yourself). On the contrary, whenever $z \neq 1$, Lorentz-invariance is broken by the massive vector deformation and the geometry is of the Lifshitz type. Interestingly, the solution in eq.\eqref{sols} impose a constraint on the value of the dynamics exponent, $z\leq 1$. This constraint coincide with the requirement of unitarity and causality found in \cite{Hoyos:2010at}.
\begin{mdframed}[style=MyFrame3]
\begin{center}
    \textbf{Exercise \#16 : The Null Energy Condition.}
\end{center}
Derive that a Lifshitz solution with $z<1$ would violate the Null Energy Condition (NEC). Remember that the NEC is defined as:
\begin{equation}
    T_{\mu\nu}\,v^\mu\,v^\nu\,>\,0
\end{equation}
where $v^\mu$ is an arbitrary null vector. In order to prove the statement above, you need to construct a basis for the space of null vectors and impose the condition above for each element of such basis. You can work in your favorite coordinates system.
\end{mdframed}
The next thing we want to  study is the stability of the fixed points under the deformations present in the model, in particular under the vector deformation which breaks the Lorentz group.\\
For simplicity let us start with the Lorentz invariant fixed point, corresponding to the AdS geometry. Concretely, we introduce the following bulk fields perturbations:
\begin{equation}
\beta(u)\,=\,\beta_{AdS}\,+\,\delta \beta(u)\,,\quad w(u)\,=\,w_{AdS}+\delta w(u)
\end{equation}
around the AdS solution defined by $\beta_{AdS}\,=\,0$ and $w_{AdS}=1$.\\
The linearized equations of motion for the perturbations take the following structure:
\begin{align}
&4\, \delta \beta'(u)+\frac{1}{2} \,4\, \delta \beta (u) \left(\frac{4\, V'(0)}{V(0)}+2\right)+\delta \beta ''(u)\,=\,0,\\&4 \,\delta w+\delta w'=0,
\end{align}
and they admit the simple solutions
\begin{align*}
\delta w (u) \,=\,c_2 \,e^{-\,4\,u},\quad \delta \beta (u)\,=\,c_3 \,e^{-u\,\Delta_{\beta}^{-}}\,+\,c_4\,e^{-u\,\Delta_{\beta}^{+}}
\end{align*}
where:
\begin{equation}
\Delta^{\pm}\,=\,2 \pm\sqrt{1+\,m^2}
\label{procadim}
\end{equation}
The $\delta w$ perturbation is a marginal deformation ($\delta w=d=4$) which corresponds to the injection of energy into the CFT or in other words to heating up the CFT with some non zero temperature.
The perturbation of the Proca field $B_\mu$ is more interesting and it is \textit{relevant/irrelevant} depending on the value of the vector mass $m^2$. In particular, the conformal dimension of the vector operator corresponds to $\Delta_\beta\equiv \Delta_+$. Whenever $m^2<3$, we have that $\Delta_\beta<d=4$ and therefore the vector perturbation is relevant and the system tends to flow away from the AdS fixed point. See fig.\eqref{figdim} for a visual representation of the conformal dimension of the vector operator dual to the bulk field $B^\mu$ in function of its mass.\\
Following the same kind of computations, but this time around the Lifshitz fixed point $z\neq 1$, we discover that the conformal dimension of the vector operator is in this case:
\begin{equation}
\Delta_\beta\,=\,\frac{1}{2} \,\left(\sqrt{9 \,z^2\,-\,26\, z\,+\,33}\,+\,z\,+\,3\right)
\end{equation}
The operator is relevant whenever
\begin{equation}
\Delta_\beta\,<\,3+z\,\quad\longrightarrow\,\quad z\,<\,3\equiv z_{max}
\end{equation}
At this point, we can classify the Lifshitz fixed point into two branches: what we will call the \textit{IR branch} $z>3$ and the \textit{UV branch} $z<3$. All the Lifshitz fixed points with $z<3$ have a relevant direction determined by the vector deformation; on the contrary all the Lifshitz fixed points with $z>3$ are attractive IR fixed points where such deformation is irrelevant.
\begin{figure}
    \centering
    \includegraphics[width=0.7\linewidth]{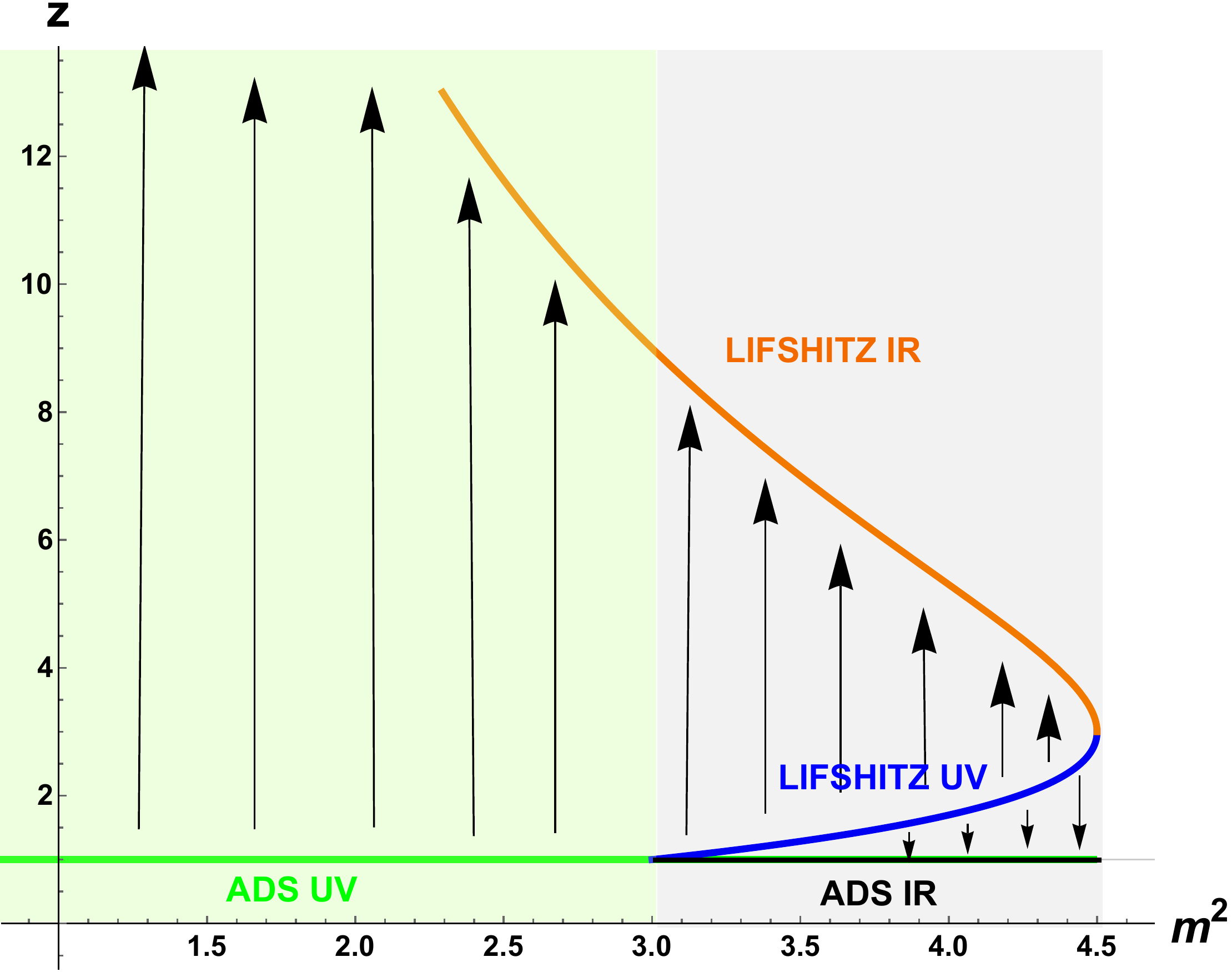}
    \caption{The RG flow structure of the theory which follows from the stability analysis. By UV fixed points we mean those where the vector deformation is relevant; in contrast, IR ones are those where the deformation is irrelevant. The arrows indicate the direction of the RG flow from high energy (UV) to low energy (IR). The different colors distinguish the different scaling symmetries of the fixed points.}
    \label{ggflow}
\end{figure}\\
From this stability analysis, we can already understand the RG flow structure of our theory, which is summarized in fig.\eqref{ggflow}.
Our final exercise consists in constructing numerically the interpolating geometries which characterize the field theory RG flows. The task is quite simple and it can be summarized in the following steps:
\begin{enumerate}
    \item Fix the value of the bulk vector field mass $m^2$. Depending on its value, different flows are expected, in agreement with fig.\ref{ggflow}.
    \item Take the equations of motion for the functions $\beta(u)$ and $w(u)$\footnote{After you have derived the equations of motion as suggested in exercise 13 you will realize that the equations of motion for the function $N(u)$ are not dynamical and once the solution in terms of $\beta(u),w(u)$ is obtained the profile of $N(u)$ follows automatically.}. Impose your boundary conditions in the IR, $u=-\infty$, to be those of the fixed point, plus a small deformation.\footnote{Alternatively you can impose the boundary conditions in the UV. For $m^2<3$ everything follows directly. In the range $m^2>3$ the flow you obtain depends on the type of deformation you impose around the UV Lifshitz fixed point. Try!}
    \item Solve numerically the equations and plots the functions $\beta(u),w(u)$.
\end{enumerate}
In fig.\ref{flow1fig} and \ref{flow2fig} an example for every flow is shown. The details of the numerical computations can be found in the complementary notebooks.
\\
You can try to play with many more complicated holographic RG flows driven by scalar and vector operators and study the corresponding physical properties. See for example \cite{Bhattacharya:2014dea,Donos:2016zpf,Gursoy:2018umf,Nitti:2017cbu,Kiritsis:2016kog,Kiritsis:2019wyk,Bea:2018whf,Anninos:2018svg}.
\begin{mdframed}[style=MyFrame3]
\begin{center}
    \textbf{Exercise \#17 : Nematic fixed points}
\end{center}
Using the same Proca model you can construct flows between fixed points with rotational invariance and fixed points where rotational invariance is broken. The latter are referred to as \textit{nematic fixed points}. The procedure is totally analogous and the details can be found in \cite{Cremonini:2014pca}. I encourage you to play with them, they are quite funny!
\end{mdframed}

%
%

\prefacedue

\hspace{0.2cm} \includegraphics[width=0.65\textwidth]{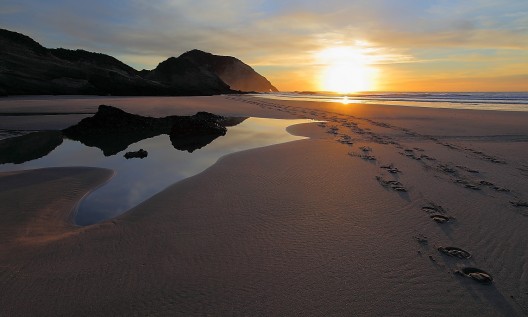}\\

This is just the beginning of the story. The idea is that, after studying these lectures, together with the available notebooks, you had a first contact with the basic methods of Applied Holography and the common types of computations which are usually involved. I believe that this is enough material to start facing more advanced and modern problems in the topic. As a matter of fact, if you confidently master all the techniques explained in these lectures you are easily been able to reproduce more than $50\%$ of my publications, meaning you are ready to contribute to the field.\\
Given the spacetime limitations, I chose a finite set of benchmark problems and I chose them in simple enough setups to allow for basic numerical computations and analytic techniques, which could be confined in few pages. Everything that comes after that, it is just an improvement of the techniques due to technical difficulties, but it will not represent a huge conceptual leap.\\

To conclude, here there are some parenting tips for you:
\begin{itemize}
    \item[(I)] Be engaged! Reading is not enough! Nobody learned Physics just by reading. It is like trying to learn to drive watching Schumacher in the TV; try by yourself!
    \item[(II)] Doubt any equation written in these notes. Try to derive them on your own. Convince yourself.
    \item[(III)] Within the book, I have suggested several exercises and checks you might want to do. Facing the problems, and most likely getting stuck with them for a bit, it is the best way to learn!
    \item[(IV)] Play with the available notebooks. You can find them at \url{https://members.ift.uam-csic.es/matteo.baggioli/lectures/}. I am definitely not an expert in the numerical techniques, therefore I am pretty sure you can improve the routines, make them faster, more automatized, and even use more advanced ones.
    \item[(V)] Use the references! I tried to be the most exhaustive possible with the references. Once you are stuck, or you want to know more, check them out!
    \item[(VI)] In several places, I indicated more advanced directions that you could take as an exercise or to learn in more details some of the basic methods which I presented. Good luck with them!
\end{itemize}
That said, I have done my best to transfer to you what I know, now it is your turn!\\

For any comment, suggestion, complaint, errata do not hesitate to contact me at \href{mailto:matteo.baggioli@uam.es}{matteo.baggioli@uam.es} .

\backmatter
\bibliographystyle{unsrt2authabbrvpp}
  \bibliography{lectures}
\end{document}